\documentclass[final]{siamltex}

\usepackage{amsmath,amsfonts,epsfig}

\newcommand{\dt}{\mbox{d}t}
\newcommand{\dsigma}{\mbox{d}\sigma}
\newcommand{\ds}{\mbox{d}s}
\newcommand{\dx}{\mbox{d}x}
\newcommand{\dy}{\mbox{d}y}
\newcommand{\dz}{\mbox{d}z}
\newcommand{\boldn}{{\mathbf n}}
\newcommand{\boldM}{{\mathbf M}}
\newcommand{\boldV}{{\mathbf V}}
\newcommand{\picturesAB}[4]{
\centerline{
\hskip #4
\raise #3 \hbox{\raise 0.9mm \hbox{(a)}}
\hskip -8mm
\epsfig{file=#1,height=#3}
\raise #3 \hbox{\raise 0.9mm \hbox{(b)}}
\hskip -8mm
\epsfig{file=#2,height=#3}
}}
\newcommand{\picturesABal}[4]{
\centerline{
\hskip #4
\raise #3 \hbox{\raise 0.9mm \hbox{(a)}}
\hskip -8mm
\epsfig{file=#1,height=#3}
\hskip 2mm
\raise #3 \hbox{\raise 0.9mm \hbox{(b)}}
\hskip -8mm
\epsfig{file=#2,height=#3}
}}

\title{A practical guide to stochastic simulations 
       of reaction-diffusion processes}
   
\author{Radek Erban\thanks{University of Oxford, 
Mathematical Institute, 24-29 St. Giles', Oxford, OX1 3LB, United Kingdom;
{\it e-mails:  erban@maths.ox.ac.uk, chapman@maths.ox.ac.uk,
maini@maths.ox.ac.uk}.}
 \and S. Jonathan Chapman\footnotemark[1]
 \and Philip K. Maini\footnotemark[1]
 }

\begin{document}

\maketitle

\begin{abstract}
A practical introduction to stochastic modelling of reaction-diffusion 
processes is presented. No prior knowledge of stochastic simulations
is assumed. The methods are explained using illustrative examples.
The article starts with the classical Gillespie algorithm 
for the stochastic modelling of chemical reactions. Then stochastic 
algorithms for modelling molecular diffusion are given. Finally, 
basic stochastic reaction-diffusion methods are presented.  
The connections between stochastic simulations and deterministic 
models are explained and basic mathematical tools (e.g. chemical master 
equation) are presented. The article concludes with an overview of 
more advanced methods and problems.
\end{abstract}

\begin{keywords} 
stochastic simulations, reaction-diffusion processes
\end{keywords}

\begin{AMS}
60G05, 92C40, 60J60, 92C15
\end{AMS}

\pagestyle{myheadings}
\thispagestyle{plain}
\markboth{RADEK ERBAN ET AL.}
{STOCHASTIC REACTION-DIFFUSION PROCESSES}

\section{Introduction}
There are two fundamental approaches to the mathematical 
modelling of chemical reactions and diffusion: 
{\it deterministic} models which are based on differential 
equations; and {\it stochastic} simulations. Stochastic
models provide a more detailed understanding of the 
reaction-diffusion processes. Such a description is often 
necessary for the modelling of biological systems where small
molecular abundances of some chemical species make 
deterministic models inaccurate or even inapplicable.
Stochastic models are also necessary when
biologically observed phenomena depend on  
stochastic fluctuations (e.g. switching between
two favourable states of the system).

In this paper, we provide an accessible introduction 
for students to the stochastic modelling of the reaction-diffusion
processes. We assume that students have a basic understanding 
of differential equations but we do not assume any prior
knowledge of advanced probability theory or stochastic 
analysis. We explain stochastic simulation methods using 
illustrative examples. We also present basic theoretical 
tools which are used for analysis of stochastic methods.
We start with a stochastic model of a single chemical 
reaction (degradation) in Section \ref{secdegradation}, 
introducing a basic stochastic simulation algorithm (SSA) 
and a mathematical equation suitable for its analysis 
(the so-called chemical master equation). Then we study 
systems of chemical reactions in the rest of 
Section \ref{sechomogen}, 
presenting the Gillespie SSA and some additional theoretical 
concepts. We introduce
new theory whenever it provides more insights into the
particular example. We believe that such an example-based
approach is more accessible for students than introducing 
the whole theory first. In Section \ref{secdiffusion}, 
we study SSAs for modelling diffusion 
of molecules. We focus on models of diffusion which are 
later used for the stochastic modelling of  
reaction-diffusion processes. Such methods are presented
in Section \ref{secRD}. We also introduce further
theoretical concepts, including the reaction-diffusion
master equation, the Smoluchowski equation and the
Fokker-Planck equation. We conclude with
Sections \ref{secextra} and \ref{secdiscussion} 
where more advanced problems,
methods and theory are discussed, giving references
suitable for further reading.

The stochastic methods and the corresponding theory 
are explained using several illustrative examples. We do not 
assume a prior knowledge of a particular computer language 
in this paper. A student might use any computer language to 
implement the examples from this paper. However, we believe 
that some students might benefit from our computer codes 
which were used to compute the illustrative results in 
this paper. The computer codes (in Matlab or Fortran) can 
be downloaded from the website 
{\tt http://www.maths.ox.ac.uk/cmb/Education/} which
is hosted by the Centre for Mathematical Biology in the
Mathematical Institute, University of Oxford. 

\section{Stochastic simulation of chemical reactions}

\label{sechomogen}

The goal of this section is to introduce stochastic
methods for the modelling of (spatially homogeneous) systems
of chemical reactions. We present the Gillespie SSA, 
the chemical master equation and
its consequences \cite{Gillespie:1977:ESS,Gillespie:1992:MPI}.
We start with the simplest case possible, that of modelling
a single chemical reaction, in Section \ref{secdegradation}.
We then study two simple systems of chemical reactions
in Sections \ref{secproddegr} and \ref{secnonlin}.

\subsection{Stochastic simulation of degradation}

\label{secdegradation}

Let us consider the single chemical reaction
\begin{equation}
A \; \mathop{\longrightarrow}^{k} \;\, \emptyset
\label{degradation}
\end{equation}
where $A$ is the chemical species of interest and $k$ is 
the rate constant of the reaction. The symbol $\emptyset$ 
denotes chemical species which are of no further 
interest in what follows. The rate constant $k$ in 
(\ref{degradation}) is defined so that $k \,\dt$ gives 
the probability that a randomly chosen molecule of chemical 
species $A$ reacts (is degraded) during the time interval 
$[t,t+\dt)$ where $t$ is time and $\dt$ an 
(infinitesimally) small time step. In particular, the 
probability that exactly one reaction (\ref{degradation}) 
occurs during the infinitesimal time interval 
$[t,t+\dt)$ is equal to $A(t) k \,\dt$ where we 
denote the number of molecules of chemical species $A$ at 
time $t$ simply as $A(t)$. This notational convention will 
be used throughout the paper. 

Let us assume that we have $n_0$ molecules of $A$ in the 
system at time $t=0$, i.e. $A(0)=n_0$. Our first goal
is to compute the number of molecules $A(t)$ for times
$t > 0.$ To do that, we need a computer routine generating 
random numbers uniformly distributed in the interval $(0,1).$ 
Such a program is included in many modern programming 
languages (e.g. function {\tt rand} in Matlab): It generates
a number $r \in (0,1)$, so that the probability that $r$ 
is in a subinterval $(a,b) \subset (0,1)$ is equal to
$b-a$ for any $a, b \in (0,1),$ $a<b$. 

The mathematical definition of the chemical reaction (\ref{degradation}) 
can be directly used to design a ``naive" SSA for simulating it.
We choose a small time step $\Delta t$. We compute the
number of molecules $A(t)$ at times $t= i \Delta t$,
$i = 1, 2, 3, \dots,$ as follows. Starting with $t=0$ and 
$A(0)=n_0$, we perform two steps at time $t$:

\leftskip 1.4cm

\medskip

{ 
\parindent -8.4mm
 
{\bf (a1)} Generate a random number $r$ uniformly distributed in 
the interval $(0,1)$.

{\bf (b1)} If $r < A(t) k \,\Delta t$, then put $A(t+\Delta t) = A(t) - 1$; 
otherwise, $A(t+\Delta t) = A(t)$. \\ Then continue with step (a1) 
for time $t+\Delta t.$

}

\leftskip 0cm

\medskip

\begin{figure}
\picturesAB{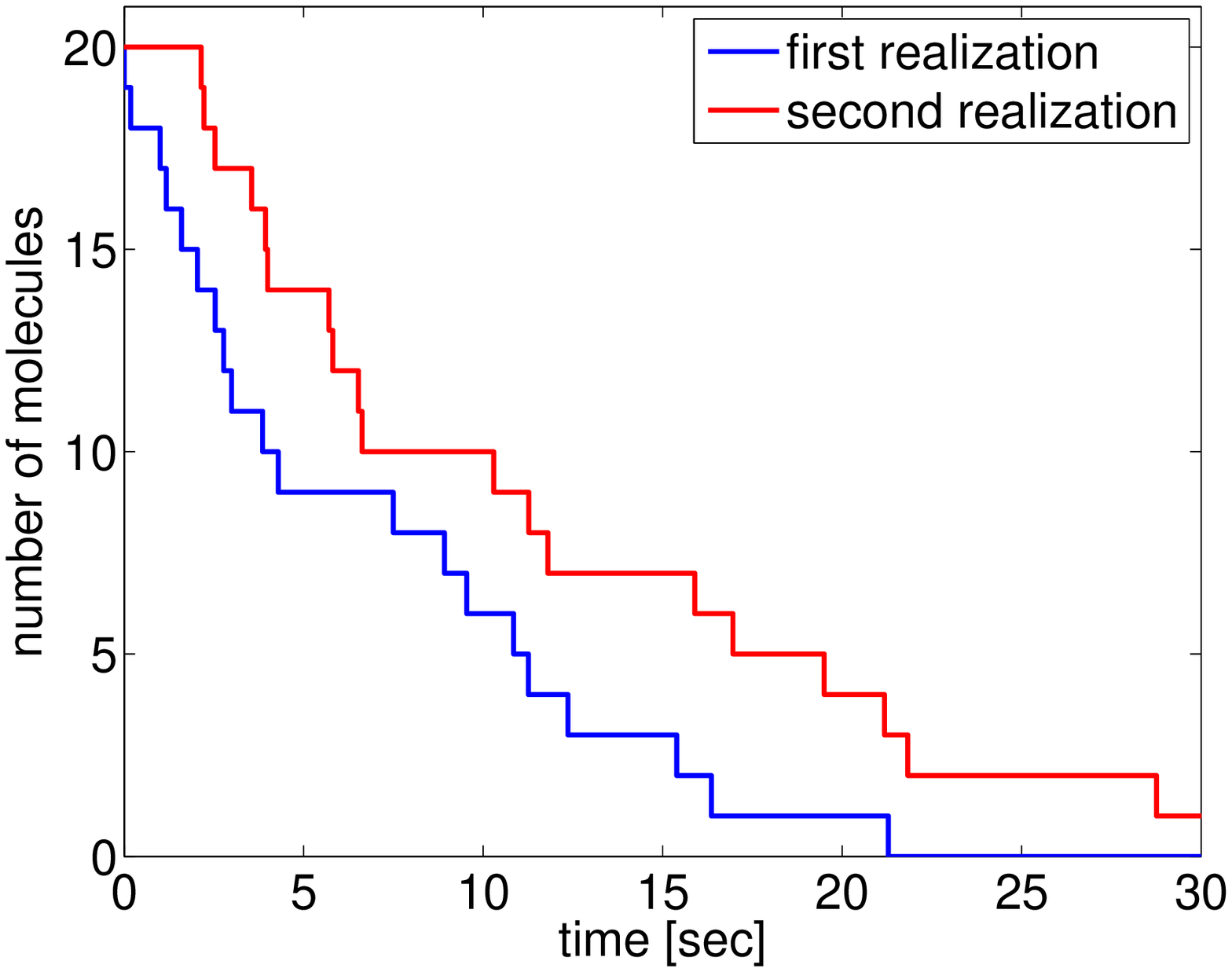}{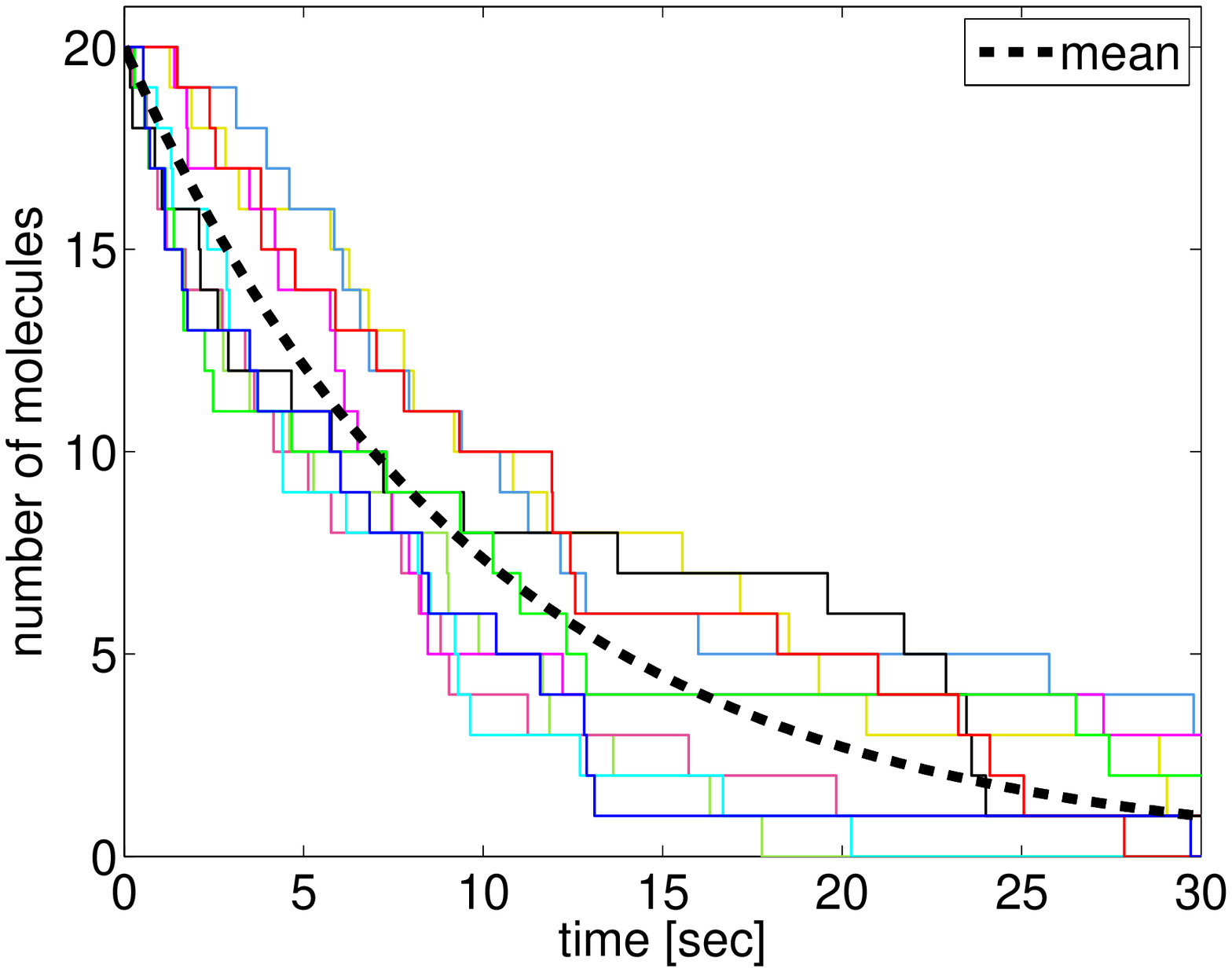}{2.05in}{5mm}
\caption{Stochastic simulation of chemical reaction
$(\ref{degradation})$ for $k= 0.1 \, \mbox{sec}^{-1}$ 
and $A(0)=20$. {\rm (a)} Number of molecules of $A$ as a function
of time for two realizations of the ``naive" SSA (a1)--(b1)
for $\Delta t=0.005 \, \mbox{sec}$; {\rm (b)} 
results of ten realizations of SSA (a2)--(c2)(solid lines;
different colours show different realizations) and stochastic 
mean $(\ref{meanAdeg})$ plotted by the dashed line.}
\label{figdegradation}
\end{figure}

\noindent
Since $r$ is a random number uniformly distributed in the interval $(0,1)$, 
the probability that $r < A(t) k \,\Delta t$ is equal to $A(t) k \,\Delta t$. 
Consequently, step (b1) says that the probability that the chemical
reaction (\ref{degradation}) occurs in the time interval $[t,t+\Delta t)$ is 
equal to $A(t) k \,\Delta t$. Thus step (b1) correctly implements
the definition of (\ref{degradation}) provided that $\Delta t$ is small.
The time evolution of $A$ obtained by the ``naive" SSA (a1)--(b1)
is given in Figure \ref{figdegradation}(a) for 
$k= 0.1 \, \mbox{sec}^{-1},$ $A(0)=20$ and $\Delta t=0.005 \, \mbox{sec}$. 
We
repeated the stochastic simulation twice and we plotted
two realizations of SSA (a1)--(b1).  We see in Figure
\ref{figdegradation}(a) that two realizations of 
SSA (a1)--(b1) give two different results. Each 
time we run the algorithm, we obtain different results. 
This is generally true for any SSA. Therefore, one might 
ask what useful and reproducible information can be 
obtained from stochastic simulations? 
This question will be addressed later in this section.

The probability that exactly one reaction (\ref{degradation}) 
occurs during the infinitesimal time interval $[t,t+\dt)$ is 
equal to $A(t) k \,\dt$. To design the SSA (a1)--(b1), 
we replaced $\dt$ by the finite time step $\Delta t$. 
In order to get reasonably accurate results, we must
ensure that $A(t) k \,\Delta t \ll 1$ during the simulation.
We used $k= 0.1 \, \mbox{sec}^{-1}$ and 
$\Delta t=0.005 \, \mbox{sec}$. Since $A(t) \le A(0) = 20$ 
for any $t \ge 0$, we have that 
$A(t) k \,\Delta t \in [0,0.01]$ for any $t \ge 0$.
Consequently, the condition $A(t) k \,\Delta t \ll 1$ is 
reasonably satisfied during the simulation. We might
further increase the accuracy of the SSA (a1)--(b1) 
by decreasing $\Delta t$. However, decreasing $\Delta t$
increases the computational intensity of the algorithm.
The probability that the reaction (\ref{degradation}) 
occurs during the time interval $[t,t+\Delta t)$ is less 
or equal to 1\% for our parameter values. During most of the 
time steps, we generate a random number $r$ in step (a1) 
to find out that no reaction occurs in step (b1). Hence,
we need to generate a lot of random numbers before
the reaction takes place. Our next task will be to design a more 
efficient method for the simulation of the chemical reaction 
(\ref{degradation}). We will need only one random
number to decide when the next reaction occurs. Moreover,
the method will be exact. There will be no approximation
in the derivation of the following SSA (a2)--(c2).

Suppose that there are $A(t)$ molecules at time $t$
in the system. Our goal is to compute time $t+\tau$
when the next reaction (\ref{degradation}) takes place.
Let us denote by $f(A(t),s) \,\ds$ the probability that,
given $A(t)$ molecules at time $t$ in the system, 
the next reaction occurs during the time interval 
$[t+s,t+s+\ds)$ where $\ds$ is an (infinitesimally) 
small time step. Let $g(A(t),s)$ 
be the probability that no reaction occurs in 
interval $[t, t+s)$. Then the probability
$f(A(t),s) \,\ds$ can be computed as a product
of $g(A(t),s)$ and the probability that a reaction 
occurs in the time interval $[t+s,t+s+\ds)$ 
which is given according to the definition of 
(\ref{degradation}) by $A(t+s) k \,\ds.$ 
Thus we have
$$
f(A(t),s) \,\ds = g(A(t),s) A(t+s) k \,\ds.
$$
Since no reaction occurs in $[t, t+s)$, we have 
$A(t+s) = A(t).$ This implies
\begin{equation}
f(A(t),s) \,\ds = g(A(t),s) A(t) k \,\ds.
\label{hgeq}
\end{equation}
To compute the probability $g(A(t),s)$, let us consider
$\sigma > 0$. The probability that no reaction
occurs in the interval $[t,t+\sigma+\dsigma)$ can be computed
as the product of the probability that no reaction
occurs in the interval $[t,t+\sigma)$ and the probability that 
no reaction occurs in the interval $[t+\sigma,t+\sigma+\dsigma)$. 
Hence
$$
g(A(t),\sigma+\dsigma) = g(A(t),\sigma) [ 1 - A(t+\sigma) k \,\dsigma].
$$
Since no reaction occurs in the interval $[t,t+\sigma)$,
we have $A(t+\sigma) = A(t)$. Consequently,
$$
\frac{g(A(t),\sigma+\dsigma) - g(A(t),\sigma)}{\dsigma} 
= - A(t) k \, g(A(t),\sigma).
$$
Passing to the limit $\dsigma \to 0$, we obtain the ordinary
differential equation (in the $\sigma$ variable)
$$
\frac{\mbox{d} g(A(t),\sigma)}{\dsigma} 
= - A(t) k \, g(A(t),\sigma).
$$
Solving this equation with initial condition $g(A(t),0)=1$,
we obtain
$$
g(A(t),\sigma) = \exp[ - A(t) k \sigma].
$$ 
Consequently, (\ref{hgeq}) can be written as
\begin{equation}
f(A(t),s) \,\ds =  A(t) k \exp[ - A(t) k s] \,\ds.
\label{heq}
\end{equation}
Our goal is to find $\tau$ such that $t+\tau$ is the time
when the next reaction occurs, provided that there are $A(t)$ 
molecules of $A$ in the system at time $t$. 
Such $\tau \in (0,\infty)$ is a random number which has to be 
generated consistently with the definition of the chemical
reaction (\ref{degradation}). To do that, we consider
the function $F(\cdot)$ defined by 
\begin{equation}
F(\tau) = \exp[ - A(t) k \tau].
\label{Heq}
\end{equation}
The function $F(\cdot)$ is monotone decreasing for $A(t)>0$. 
If $\tau$ is a random number from the interval $(0,\infty)$, then 
$F(\tau)$ is a random number from the interval $(0,1)$.
If $\tau$ is a random number chosen consistently with the
reaction (\ref{degradation}), then $F(\tau)$ is a random number 
uniformly distributed in the interval $(0,1)$ which can be
shown as follows. Let $a$, $b$, $a<b$, be chosen arbitrarily in the 
interval $(0,1)$. The probability that $F(\tau) \in (a,b)$ 
is equal to the probability that $\tau \in (F^{-1}(b),F^{-1}(a))$
which is given by the integral of $f(A(t),s)$ over $s$ in
the interval $(F^{-1}(b),F^{-1}(a))$. Using (\ref{heq}) and (\ref{Heq}), 
we obtain
$$
\int_{F^{-1}(b)}^{F^{-1}(a)} f(A(t),s) \,\ds
= 
\int_{F^{-1}(b)}^{F^{-1}(a)} A(t) k \exp[ - A(t) k s] \,\ds
$$
$$
=
- \int_{F^{-1}(b)}^{F^{-1}(a)} \frac{\mbox{d}F}{\ds} \,\ds
=
- F [ F^{-1}(a) ] + F [ F^{-1}(b) ] 
= 
b - a.
$$
Hence we have verifed that $F(\tau)$ is a random number uniformly distributed 
in $(0,1)$. Such a number can be obtained using the random number generator
(e.g. function {\tt rand} in Matlab). Let us denote it by $r$. The previous
observation implies that we can generate the time step $\tau$ by putting
$r = F(\tau).$ Using (\ref{Heq}), we obtain
$$
r = \exp[ - A(t) k \tau].
$$
Solving for $\tau$, we obtain the formula 
\begin{equation}
\tau = \frac{1}{A(t) k} \ln \left[ \frac{1}{r} \right].
\label{tauform} 
\end{equation}
Consequently, the SSA for the chemical reaction (\ref{degradation}) 
can be written as follows. Starting with $t=0$ and $A(0)=n_0$,
we perform three steps at time $t$:

\leftskip 1.4cm

\medskip

{ 
\parindent -8.4mm
 
{\bf (a2)} Generate a random number $r$ uniformly distributed in 
the interval $(0,1)$.

{\bf (b2)} Compute the time when the next reaction (\ref{degradation})
occurs as $t+\tau$ where $\tau$ is given by (\ref{tauform}).

{\bf (c2)} Compute the number of molecules at time $t+\tau$ by
$A(t+\tau) = A(t) - 1.$ \\ Then continue with step (a2) for
time $t+\tau.$

}

\leftskip 0cm

\medskip

\noindent
Steps (a2)--(c2) are repeated until we reach the time
when there is no molecule of $A$ in the system, i.e. $A=0$.
SSA (a2)--(c2) computes the time of the
next reaction $t+\tau$ using formula (\ref{tauform})
in step (b2) with the help of one random number only. 
Then the reaction is performed in step (c2)
by decreasing the number of molecules of chemical 
species $A$ by 1. The time evolution of $A$ obtained by SSA 
(a2)--(c2) is given in Figure \ref{figdegradation}(b). 
We plot ten realizations of SSA 
(a2)--(c2) for $k= 0.1 \, \mbox{sec}^{-1}$ and 
$A(0)=20$. Since the function $A(t)$ has only integer
values $\{0, 1, 2, \dots, 20\}$, it is not surprising
that some of the computed curves $A(t)$ partially 
overlap. On the other hand, all ten realizations yield 
different functions $A(t)$. Even if we made 
millions of stochastic realizations, it would be very 
unlikely (with probability zero) that there would be two 
realizations giving exactly the same results. Therefore,
the details of one realization $A(t)$ are of no special 
interest (they depend on the sequence of random numbers 
obtained by the random number generator). However,
averaging values of $A$ at time $t$ over many realizations 
(e.g. computing the stochastic mean of $A$), we obtain 
a reproducible characteristic of the system -- see the dashed 
line in Figure \ref{figdegradation}(b). The
stochastic mean of $A(t)$ over (infinitely) many realizations
can be also computed theoretically as follows.

Let us denote by $p_n(t)$ the probability that there are $n$ 
molecules of $A$ at time $t$ in the system, i.e.
$A(t)=n$. Let us consider an (infinitesimally) small time step 
$\dt$ chosen such that the probability 
that two molecules are degraded during $[t,t+\dt)$ is negligible 
compared to the probability that only one molecule is degraded 
during $[t,t+\dt).$ Then there are two possible ways for 
$A(t+\dt)$ to take the value $n$: either $A(t)=n$ and no reaction occurred
in $[t,t+\dt)$, or $A(t)=n+1$ and one molecule was degraded
in $[t,t+\dt)$. Hence
$$
p_n(t + \dt) = p_n(t) \times (1 - k n \,\dt) 
+ p_{n+1}(t) \times k (n+1) \,\dt.
$$
A simple algebraic manipulation yields
$$
\frac{p_n(t + \dt) - p_n(t)}{\dt}
=
k (n+1) \, p_{n+1}(t) - k n \, p_n(t).
$$
Passing to the limit $\dt \to 0$, we obtain the so-called
{\it chemical master equation} in the form
\begin{equation}
\frac{\mbox{d} p_n}{\dt} 
=
k (n+1) \, p_{n+1}
- k n \, p_n.
\label{cmedegradation}
\end{equation}
Equation (\ref{cmedegradation}) looks like an infinite 
system of ordinary differential equations (ODEs) for $p_n,$ 
$n=0,1,2,3, \dots$. Our initial condition $A(0)=n_0$
implies that there are never more than $n_0$ molecules
in the system. Consequently, $p_n \equiv 0$ 
for $n>n_0$ and the system (\ref{cmedegradation})
reduces to a system of $(n_0+1)$ ODEs 
for $p_n,$ $n \le n_0$. The equation
for $n=n_0$ reads as follows
$$
\frac{\mbox{d} p_{n_0}}{\dt} 
=
- k n_0 \, p_{n_0}.
$$
Solving this equation with initial condition $p_{n_0}(0)=1$, 
we get $p_{n_0}(t) = \exp[-k n_0 t].$ Using this formula 
in the chemical master equation (\ref{cmedegradation})
for $p_{n_0-1}(t)$, we obtain 
$$
\frac{\mbox{d}}{\dt} \, p_{n_0-1}(t) 
=
k n_0 \exp[-k n_0 t]
- k (n_0-1) \, p_{n_0-1}(t).
$$
Solving this equation with initial condition $p_{n_0-1}(0)=0$,
we obtain $p_{n_0-1}(t) = \exp[ -k (n_0-1) t] \, n_0 ( 1 - \exp[-kt] )$.
Using mathematical induction, it is possible to show
\begin{equation}
p_n(t) = \exp[-k n t] \binom{n_0}{n} 
\big\{ 1 - \exp[-k t] \big\}^{n_0 - n}.
\label{solcme}
\end{equation}
The formula (\ref{solcme}) provides all the information about the stochastic
process which is defined by (\ref{degradation}) and initial
condition $A(0)=n_0$. We can never say for sure that $A(t)=n$;
we can only say that $A(t)=n$ with probability $p_n(t)$. In particular, 
formula (\ref{solcme}) can be used
to derive a formula for the mean value of $A(t)$ over (infinitely)
many realizations, which is defined by
$$
M(t)
=
\sum_{n=0}^{n_0} n \, p_n(t).
$$
Using (\ref{solcme}), we deduce
\begin{eqnarray}
M(t)
& = &
\sum_{n=0}^{n_0} n \, p_n(t)
=
\sum_{n=0}^{n_0} n
\exp[-k n t] \binom{n_0}{n} \big\{ 1 - \exp[-k t] \big\}^{n_0 - n} 
\nonumber
\\
& = &
n_0  \exp[-k t]
\sum_{n=1}^{n_0}
 \binom{n_0-1}{n-1} 
 \big\{ 1 - \exp[-k t] \big\}^{(n_0-1) - (n-1)} 
 \big\{ \exp[-k t] \big\}^{n-1}
\nonumber
\\
& = &
n_0  \exp[-k t].
\label{meanAdeg}
\end{eqnarray}
The chemical master equation (\ref{cmedegradation}) and its solution
(\ref{solcme}) can be also used to quantify the stochastic
fluctuations around the mean value (\ref{meanAdeg}), i.e. how much
can an individual realization of SSA (a2)--(c2) differ
from the mean value given by (\ref{meanAdeg}). We will present
the corresponding theory and results on a more complicated 
illustrative example in Section \ref{secproddegr}. Finally,
let us note that a classical deterministic description of the
chemical reaction (\ref{degradation}) is given by the ODE 
$\mbox{d}a/\dt = - k a.$ Solving this
equation with initial condition $a(0)=n_0$, we obtain the
function (\ref{meanAdeg}), i.e. the stochastic mean 
is equal to the solution of the corresponding deterministic 
ODE. However, we should emphasize that
this is not true for general systems of chemical reactions,
as we will see in Section \ref{secnonlin} and 
Section \ref{secssr}. 

\subsection{Stochastic simulation of production and degradation}

\label{secproddegr}

We consider a system of two chemical reactions
\begin{equation}
A \; \mathop{\longrightarrow}^{k_1} \;\, \emptyset,
\qquad \qquad
\emptyset \; \mathop{\longrightarrow}^{k_2} \;\, A.
\label{degradationcreation}
\end{equation}
The first reaction describes the degradation of chemical
$A$ with the rate constant $k_1$. It was already studied 
previously as reaction (\ref{degradation}). We couple 
it with the second reaction which represents the production 
of chemical $A$ with the rate constant $k_2$. The exact 
meaning of the second chemical reaction 
in (\ref{degradationcreation}) is that one molecule 
of $A$ is created during the time interval $[t,t+\dt)$ 
with probability $k_2 \, \dt$. As before, the symbol 
$\emptyset$ denotes chemical species which are 
of no special interest to the modeller. The impact
of other chemical species on the rate of production
of $A$ is assumed to be time independent and
is already incorporated in the rate constant $k_2$.
To simulate the system of chemical reactions 
(\ref{degradationcreation}), we perform the following
four steps at time $t$ (starting with $A(0)=n_0$
at time $t=0$):

\leftskip 1.4cm

\medskip

{ 
\parindent -8.4mm
 
{\bf (a3)} Generate two random numbers $r_1$, $r_2$ uniformly distributed in 
$(0,1)$. 

{\bf (b3)} 
\!Compute $\alpha_0 = A(t) k_1 + k_2.$

{\bf (c3)} Compute the time when the next chemical reaction 
takes place as $t+\tau$ where
\begin{equation}
\tau = \frac{1}{\alpha_0} \ln \left[ \frac{1}{r_1} \right].
\label{tauform2} 
\end{equation}

{\bf (d3)} Compute the number of molecules at time $t+\tau$ by 
\begin{equation}
A(t+\tau) = \left\{ 
\begin{matrix}
A(t) + 1 & \quad \mbox{if} \; r_2 < k_2/\alpha_0; \\
A(t) - 1 & \quad \mbox{if} \; r_2 \ge k_2/\alpha_0.
\end{matrix}
\right.
\end{equation}
Then continue with step (a3) for time $t+\tau.$

}

\leftskip 0cm

\medskip

\noindent
To justify that SSA (a3)--(d3) correctly simulates
(\ref{degradationcreation}), let us note that the probability
that any of the reactions in (\ref{degradationcreation}) 
takes place in the time interval $[t, t+\dt)$ is equal
to $\alpha_0 \,\dt$. It is given as a sum of the probability 
that the first reaction occurs ($A(t) k_1 \dt$) and the
probability that the second reaction occurs ($k_2 \,\dt$).
Formula (\ref{tauform2}) gives the time $t + \tau$ 
when the next reaction takes place. It can be justified using 
the same arguments as for formula (\ref{tauform}). Once we 
know the time $t+\tau$, a molecule is produced with
probability $k_2/\alpha_0$, i.e. the second reaction in
(\ref{degradationcreation}) takes place with
probability $k_2/\alpha_0$.
Otherwise, a molecule is degraded, i.e. the first reaction
in (\ref{degradationcreation}) occurs. The decision as to
which reaction takes place is given in step (d3) with the
help of the second uniformly distributed random number
$r_2$.

\begin{figure}
\picturesAB{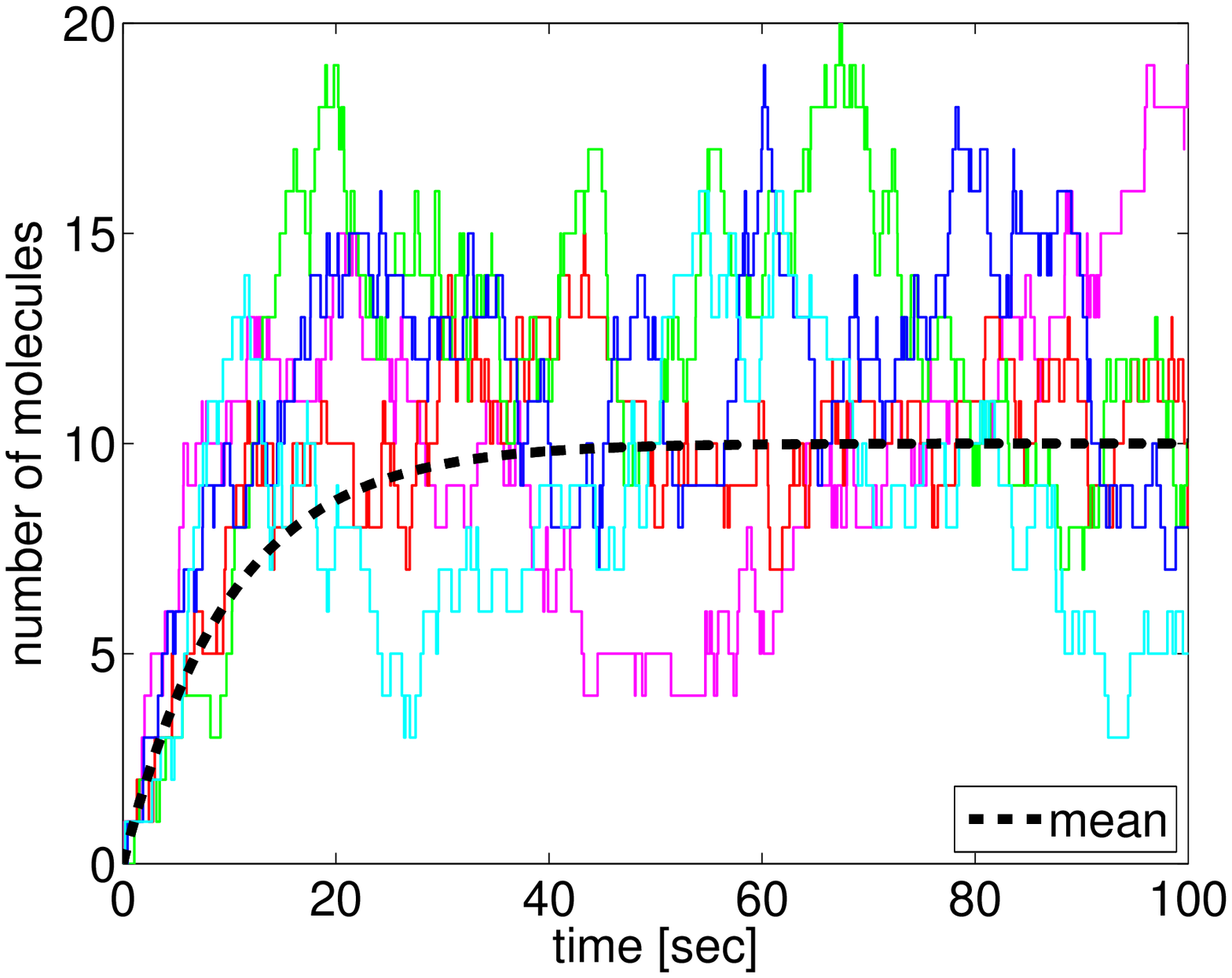}{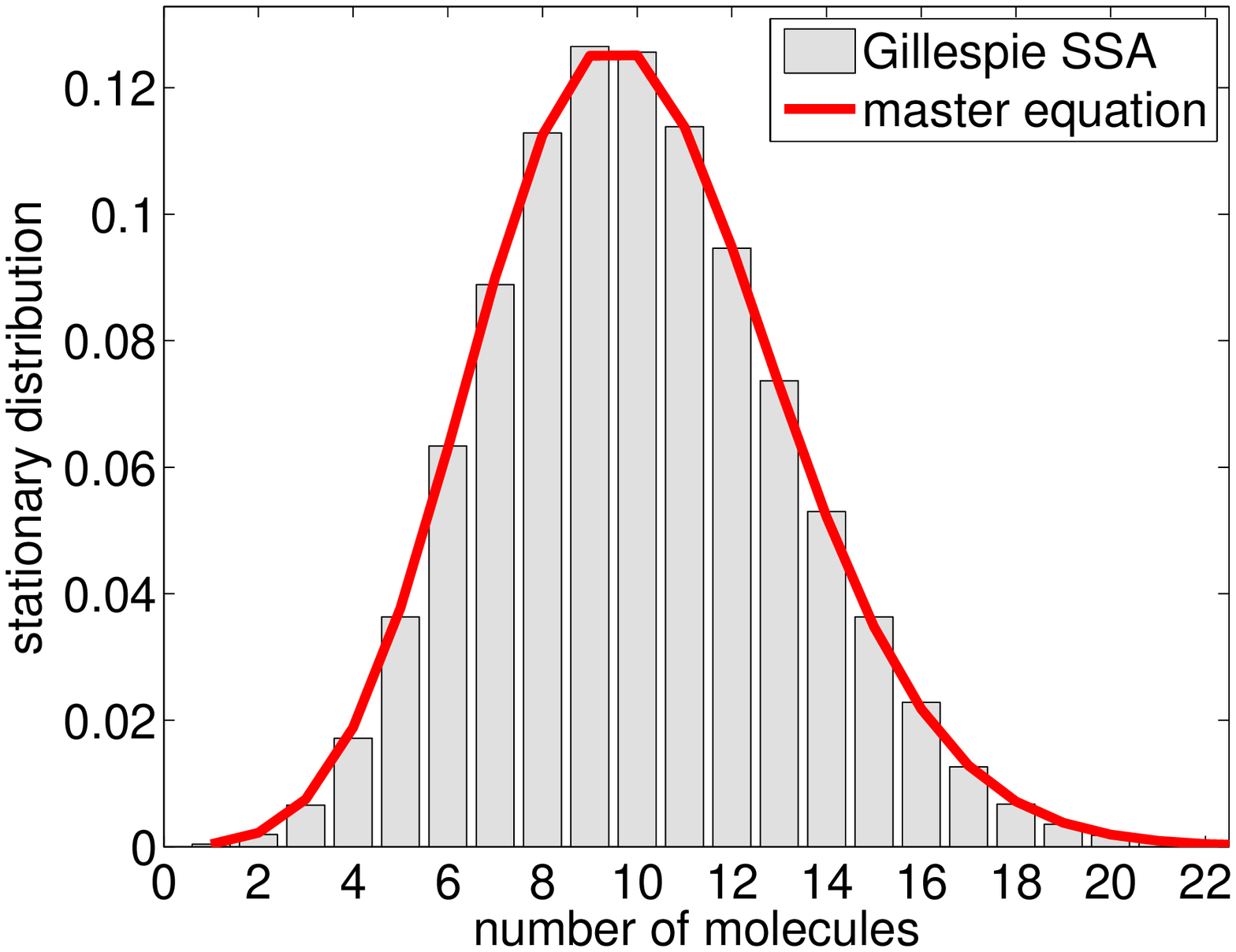}{2.05in}{5mm}
\caption{Stochastic simulation of the system of chemical
reactions $(\ref{degradationcreation})$ for 
$A(0)=0$, $k_1= 0.1 \, \mbox{sec}^{-1}$ 
and $k_2= 1 \, \mbox{sec}^{-1}.$ {\rm (a)} 
$A(t)$ given by five realizations of 
SSA (a3)--(d3) (solid lines) and
stochastic mean (dashed line). {\rm (b)} 
Stationary distribution $\phi(n)$ obtained by long time simulation 
of SSA (a3)--(d3) (gray histogram) and by
formulae $(\ref{psone})$--$(\ref{psn})$ (red solid line).}
\label{figdegradationcreation}
\end{figure}

Five realizations of SSA (a3)--(d3) 
are presented in Figure \ref{figdegradationcreation}(a)
as solid lines.
We plot the number of molecules of $A$ as a function of time
for $A(0)=0$, $k_1= 0.1 \, \mbox{sec}^{-1}$ and 
$k_2= 1 \, \mbox{sec}^{-1}.$ We see that, after an initial
transient, the number of molecules $A(t)$ fluctuates
around its mean value. To compute the stochastic mean
and quantify the stochastic fluctuations, we use
the chemical master equation which can be written
for the chemical system (\ref{degradationcreation}) in the
following form
\begin{equation}
\frac{\mbox{d} p_n}{\dt}
=
k_1 (n+1) \, p_{n+1} - k_1 n \, p_n
+
k_2 \, p_{n-1} - k_2 \, p_n 
\label{cmeproddegr}
\end{equation}
where $p_n(t)$ denotes the probability that $A(t)=n$
for $n=0,1,2,3,\dots$. Let us note that the third
term on the right hand side is missing in (\ref{cmeproddegr}) 
for $n=0$; i.e. we use the convention that
$p_{-1} \equiv 0.$ The first two terms on the right
hand side correspond to the first reaction in
(\ref{degradationcreation}). They already appeared in
equation (\ref{cmedegradation}). Production of $A$
is described by the third and fourth term on the right
hand side of (\ref{degradationcreation}). To derive the
chemical master equation (\ref{cmeproddegr}), we can
use similar arguments as in the derivation of 
(\ref{cmedegradation}). The stochastic mean $M(t)$ and
variance $V(t)$ are defined by
\begin{equation}
M(t)
=
\sum_{n=0}^{\infty} n \, p_n(t),
\qquad
V(t)
=
\sum_{n=0}^{\infty} \big( n - M(t) \big)^2 \, p_n(t).
\label{meanvariancedef}
\end{equation}
The stochastic mean $M(t)$ gives the average number of molecules
of $A$ at time $t$, while the variance $V(t)$ describes the fluctuations.
In Section \ref{secdegradation}, we first solved the chemical
master equation (\ref{cmedegradation}) and then we used
its solution (\ref{solcme}) to compute $M(t)$.
Alternatively, we could use the chemical master equation
to derive an evolution equation for $M(t)$, i.e. we could
find $M(t)$ without solving the chemical master equation.
Such an approach will be presented in this section.
Multiplying (\ref{cmeproddegr}) by $n$ and summing over $n$,
we obtain
\begin{equation*}
\frac{\mbox{d}}{\dt} \sum_{n=0}^{\infty} n p_n
 = 
k_1 \sum_{n=0}^{\infty} n (n+1) \, p_{n+1} 
- 
k_1 \sum_{n=0}^{\infty} n^2 \, p_n
 +  
k_2 \sum_{n=1}^{\infty} n \, p_{n-1} 
- 
k_2 \sum_{n=0}^{\infty} n \, p_n. 
\end{equation*}
Using definition (\ref{meanvariancedef}) on the left hand
side and changing indices $n+1 \to n$ (resp. $n-1 \to n$)
in the first (resp. third) sum on the right hand side,
we obtain
\begin{equation*}
\frac{\mbox{d} M}{\dt}
 = 
k_1 \sum_{n=0}^{\infty} (n-1) n \, p_n 
- 
k_1 \sum_{n=0}^{\infty} n^2 \, p_n
 +  
k_2 \sum_{n=0}^{\infty} (n+1) \, p_n 
- 
k_2 \sum_{n=0}^{\infty} n \, p_n. 
\end{equation*}
Adding the first and the second sum (resp. the
third and the fourth sum) on the right hand side, we
get
\begin{equation}
\frac{\mbox{d} M}{\dt}
=
- k_1 \sum_{n=0}^{\infty} n \, p_n 
+  
k_2 \sum_{n=0}^{\infty} p_n. 
\label{pomM}
\end{equation}
Since $p_n(t)$ is the probability that $A(t)=n$ and $A(t)$ is equal
to a nonnegative integer with probability 1, we have 
\begin{equation}
\sum_{n=0}^{\infty} p_n(t) = 1.
\label{probmeas}
\end{equation}
Using this fact together with the definition of $M(t)$, equation
(\ref{pomM}) implies the evolution equation for $M(t)$ 
in the form
\begin{equation}
\frac{\mbox{d} M}{\dt}
=
- k_1 M
+  
k_2. 
\label{evolM}
\end{equation}
The solution of (\ref{evolM}) with initial condition
$M(0)=0$ is plotted as a dashed line in 
Figure \ref{figdegradationcreation}(a).
To derive the evolution equation for the variance $V(t)$,
let us first observe that definition
(\ref{meanvariancedef}) implies
\begin{equation}
\sum_{n=0}^{\infty}  n^2 \, p_n(t)
=
V(t)
+
M(t)^2.
\label{secondmoment}
\end{equation}
Multiplying (\ref{cmeproddegr}) by $n^2$ and summing
over $n$, we obtain
\begin{equation*}
\frac{\mbox{d}}{\dt} \sum_{n=0}^{\infty} n^2 p_n
 = 
k_1 \sum_{n=0}^{\infty} n^2 (n+1) \, p_{n+1} 
- 
k_1 \sum_{n=0}^{\infty} n^3 \, p_n
 +  
k_2 \sum_{n=1}^{\infty} n^2 \, p_{n-1} 
- 
k_2 \sum_{n=0}^{\infty} n^2 \, p_n. 
\end{equation*}
Changing indices $n+1 \to n$ (resp. $n-1 \to n$) in the first 
(resp. third) sum on the right hand side
and adding the first and the second sum (resp. the
third and the fourth sum) on the right hand side, we
get
$$
\frac{\mbox{d}}{\dt} \sum_{n=0}^{\infty} n^2 p_n
=
k_1 \sum_{n=0}^{\infty} (- 2n^2 + n) \, p_n
+  
k_2 \sum_{n=0}^{\infty} (2n + 1) \, p_n. 
$$
Using (\ref{secondmoment}), (\ref{probmeas}) and
(\ref{meanvariancedef}), we obtain
$$
\frac{\mbox{d} V}{\dt}  
+
2 M \frac{\mbox{d} M}{\dt} 
=
- 
2 k_1 \big[ V + M^2 \big]
+ 
k_1 M
+  
2 k_2 M
+ 
k_2. 
$$
Substituting (\ref{evolM}) for $\mbox{d}M/\dt$, we derive the evolution
equation for the variance $V(t)$ in the following form
\begin{equation}
\frac{\mbox{d} V}{\dt}  
=
- 
2 k_1 V
+ 
k_1 M 
+ 
k_2. 
\label{evolV}
\end{equation}
The time evolution of $M(t)$ and $V(t)$ is described by
(\ref{evolM}) and (\ref{evolV}). Let us define the
stationary values of $M(t)$ and $V(t)$ by
\begin{equation}
M_s = \lim_{t \to \infty} M(t),
\qquad
V_s = \lim_{t \to \infty} V(t).
\label{MsVs}
\end{equation}
The values of $M_s$ and $V_s$ can be computed using
the steady state equations corresponding to
(\ref{evolM}) and (\ref{evolV}), namely by solving
$$
0
=
- k_1 M_s
+  
k_2, 
\qquad \mbox{and} \qquad
0 
=
- 
2 k_1 V_s
+ 
k_1 M_s
+ 
k_2. 
$$
Consequently,
$$
M_s = V_s = \frac{k_2}{k_1}.
$$
For our parameter values $k_1= 0.1 \, \mbox{sec}^{-1}$ and 
$k_2= 1 \, \mbox{sec}^{-1},$ we obtain $M_s = V_s = 10.$
We see in Figure \ref{figdegradationcreation}(a)
that $A(t)$ fluctuates after a sufficiently long time around
the mean value $M_s=10$. To quantify the fluctuations, one often
uses the square root of $V_s$, the so-called mean standard deviation
which is equal to $\sqrt{10}$. 

More detailed information about the fluctuations is given by 
the so-called {\it stationary distribution} $\phi(n)$, 
$n=0,1,2,3,\dots$, which is defined as
\begin{equation}
\phi(n) = \lim_{t \to \infty} p_n(t).
\label{statdistrp}
\end{equation}
This means that $\phi(n)$ is the probability that $A(t)=n$ after
an (infinitely) long time. One way to compute $\phi(n)$ is to
run SSA (a3)--(d3) for a long
time and create a histogram of values of $A(t)$ at given
time intervals. Using $k_1= 0.1 \, \mbox{sec}^{-1}$ and 
$k_2= 1 \, \mbox{sec}^{-1},$ the results of such a long 
time computation are presented in 
Figure \ref{figdegradationcreation}(b) as a gray histogram.
To compute it, we ran SSA (a3)--(d3) for $10^{5}$ seconds,
recording the value of $A(t)$ every second and then dividing
the whole histogram by the number of recordings, i.e. by 
$10^{5}$. An alternative way to compute $\phi(n)$ is
to use the steady state version of the chemical master
equation (\ref{cmeproddegr}), namely
\begin{eqnarray*}
0
& = &
k_1 \, \phi(1) - k_2 \, \phi(0)
\\ 
0
& = &
k_1 (n+1) \, \phi(n+1) - k_1 n \, \phi(n)
+
k_2 \, \phi(n-1) - k_2 \, \phi(n),
\qquad \mbox{for} \; n \ge 1, 
\end{eqnarray*}
which implies
\begin{eqnarray}
\phi(1)
& = &
\frac{k_2}{k_1} \, \phi(0),
\label{psone}
\\
\phi(n+1)
& = &
\frac{1}{k_1 (n+1)}
\big[ k_1 n \, \phi(n) + k_2 \, \phi(n) - k_2 \, \phi(n-1) \big],
\quad \mbox{for} \; n \ge 1.
\label{psn}
\end{eqnarray}
Consequently, we can express $\phi(n)$ for any $n \ge 1$ in terms 
of $\phi(0)$. The formulae (\ref{psone})--(\ref{psn}) yield 
an alternative way to compute $\phi(n)$. 
We put $\phi(0)=1$ and we compute $\phi(n)$, for sufficiently
many $n$, by (\ref{psone})--(\ref{psn}). Then we divide 
$\phi(n)$, $n \ge 0$, by $\sum \phi(n)$. The results obtained by 
(\ref{psone})--(\ref{psn}) are plotted in 
Figure \ref{figdegradationcreation}(b) as a (red) 
solid line. As expected, the results compare well
with the results obtained by the long time stochastic 
simulation.

We can also find the formula for $\phi(n)$ directly. 
We let a reader to verify that the solution of the 
recurrence formula (\ref{psone})--(\ref{psn}) can be written as
\begin{equation}
\phi(n) = \frac{C}{n!} \left( \frac{k_2}{k_1} \right)^{\!n} 
\label{solrecform}
\end{equation}
where $C$ is a real constant. Using 
(\ref{probmeas}) and (\ref{statdistrp}), we have
\begin{equation}
\sum_{n=0}^{\infty} \phi(n) = 1.
\label{probmeasphi}
\end{equation}
Substituting (\ref{solrecform}) into the normalization
condition (\ref{probmeasphi}), we get
$$
1 
= 
\sum_{n=0}^{\infty} \frac{C}{n!}  \left( \frac{k_2}{k_1} \right)^{\!n}
= 
C \sum_{n=0}^{\infty} \frac{1}{n!} \left( \frac{k_2}{k_1} \right)^{\!n} 
= 
C \exp\left[ \frac{k_2}{k_1} \right]
$$
where we used the Taylor series for the exponential function to get the 
last equality. Consequently, $C = \exp[-k_2/k_1]$
which, together with (\ref{solrecform}), implies that the
stationary distribution $\phi(n)$ is the Poisson distribution
\begin{equation}
\phi(n) = \frac{1}{n!} \left( \frac{k_2}{k_1} \right)^{\!n} 
\exp\left[ - \frac{k_2}{k_1} \right].
\label{phindistr}
\end{equation}
Thus the red solid line in Figure \ref{figdegradationcreation}(b)
which was obtained numerically by the recurrence formula 
(\ref{psone})--(\ref{psn}) can be also viewed as the stationary 
distribution $\phi(n)$ given by the explicit exact formula 
(\ref{phindistr}).

\subsection{Gillespie algorithm}

\label{secnonlin}

SSAs (a2)--(c2) and (a3)--(d3)
were special forms of the so-called Gillespie SSA. In this section, 
we present this algorithm for a more complicated illustrative example 
which will also involve second-order chemical reactions.
Such chemical reactions are of the following form 
\begin{equation}
A + A \; \mathop{\longrightarrow}^{k_1} \;\, C,
\qquad \qquad
A + B \; \mathop{\longrightarrow}^{k_2} \;\, D.
\label{nonlinearmodelintro}
\end{equation}
In the first equation, two molecules of $A$ react
with rate constant $k_1$ to produce $C$. The probability 
that the reaction takes place in the time interval 
$[t,t+\dt)$ is equal to $A(t) (A(t)-1) k_1 \dt$.
We define the {\it propensity function} of the first reaction
as $\alpha_1(t) = A(t) (A(t)-1) k_1$. Then the probability
that the first reaction occurs in the time interval 
$[t,t+\dt)$ is equal to $\alpha_1(t) \,\dt$. 
The propensity function which corresponds
to the second equation in (\ref{nonlinearmodelintro})
is defined as $\alpha_2(t) = A(t) B(t) k_1$ and 
the probability that the second reaction occurs 
in the time interval $[t,t+\dt)$ is equal to 
$\alpha_2(t) \,\dt$. In such a case, one molecule
of $A$ and one molecule of $B$ react to form 
a molecule of $D$. In general, the propensity 
function can be defined for any chemical reaction 
so that its product with $\dt$ gives the probability 
that the given reaction occurs in the infinitesimally
small time interval $[t,t+\dt).$

We consider that $A$ and $B$ can react according
to (\ref{nonlinearmodelintro}). Moreover, we assume
that they are also produced with constant rates,
that is, we consider a system of four chemical
equations:
\begin{equation}
A + A \; \mathop{\longrightarrow}^{k_1} \;\, \emptyset
\qquad \qquad \quad
A + B \; \mathop{\longrightarrow}^{k_2} \;\, \emptyset
\label{nonlinearmodel1}
\end{equation}
\begin{equation}
\emptyset \; \mathop{\longrightarrow}^{k_3} \;\, A
\qquad \qquad \quad
\emptyset \; \mathop{\longrightarrow}^{k_4} \;\, B.
\label{nonlinearmodel2}
\end{equation}
Let us note that we are not interested in chemical 
species $C$ and $D$. Hence, we replaced them by 
$\emptyset$, consistent with our previous notation
of unimportant chemical species. 
To simulate the system of chemical reactions 
(\ref{nonlinearmodel1})--(\ref{nonlinearmodel2}),
we perform the following four steps at time $t$ 
(starting with $A(0)=n_0$, $B(0)=m_0$ at time $t=0$):

\leftskip 1.4cm

\medskip

{ 
\parindent -8.4mm
 
{\bf (a4)} Generate two random numbers $r_1$, $r_2$ uniformly distributed in 
$(0,1)$. 

{\bf (b4)} Compute the propensity functions of each reaction by
$\alpha_1 = A(t) (A(t)-1) k_1$, $\alpha_2 = A(t) B(t) k_2$, $\alpha_3 = k_3$
and $\alpha_4 = k_4$. 
Compute $\alpha_0 = \alpha_1 + \alpha_2 + \alpha_3 + \alpha_4.$

{\bf (c4)} Compute the time when the next chemical reaction 
takes place as $t+\tau$ where
\begin{equation}
\tau = \frac{1}{\alpha_0} \ln \left[ \frac{1}{r_1} \right].
\label{tauform3} 
\end{equation}

{\bf (d4)} Compute the number of molecules at time $t+\tau$ by 
\begin{equation}
A(t+\tau) = \left\{ 
\begin{matrix}
A(t) - 2 & 
\quad \mbox{if} \; 0 \le r_2 < \alpha_1/\alpha_0; \\
A(t) - 1 & 
\quad \mbox{if} \; \alpha_1/\alpha_0 \le r_2 < (\alpha_1+\alpha_2)/\alpha_0; \\
A(t) + 1 & 
\quad \mbox{if} \; (\alpha_1+\alpha_2)/\alpha_0 \le r_2 
                   < (\alpha_1+\alpha_2+\alpha_3)/\alpha_0; \\
A(t)     & 
\quad \mbox{if} \; (\alpha_1+\alpha_2+\alpha_3)/\alpha_0 \le r_2 < 1; 
\end{matrix}
\right.
\label{Aform}
\end{equation}
\begin{equation}
B(t+\tau) = \left\{ 
\begin{matrix}
B(t) & 
\quad \mbox{if} \; 0 \le r_2 < \alpha_1/\alpha_0; \\
B(t) - 1 & 
\quad \mbox{if} \; \alpha_1/\alpha_0 \le r_2 < (\alpha_1+\alpha_2)/\alpha_0; \\
B(t) & 
\quad \mbox{if} \; (\alpha_1+\alpha_2)/\alpha_0 \le r_2 
                    < (\alpha_1+\alpha_2+\alpha_3)/\alpha_0; \\
B(t) + 1  & 
\quad \mbox{if} \; (\alpha_1+\alpha_2+\alpha_3)/\alpha_0 \le r_2 < 1; 
\end{matrix}
\right.
\label{Bform}
\end{equation}
Then continue with step (a4) for time $t+\tau.$

}

\leftskip 0cm

\medskip

\begin{figure}
\picturesAB{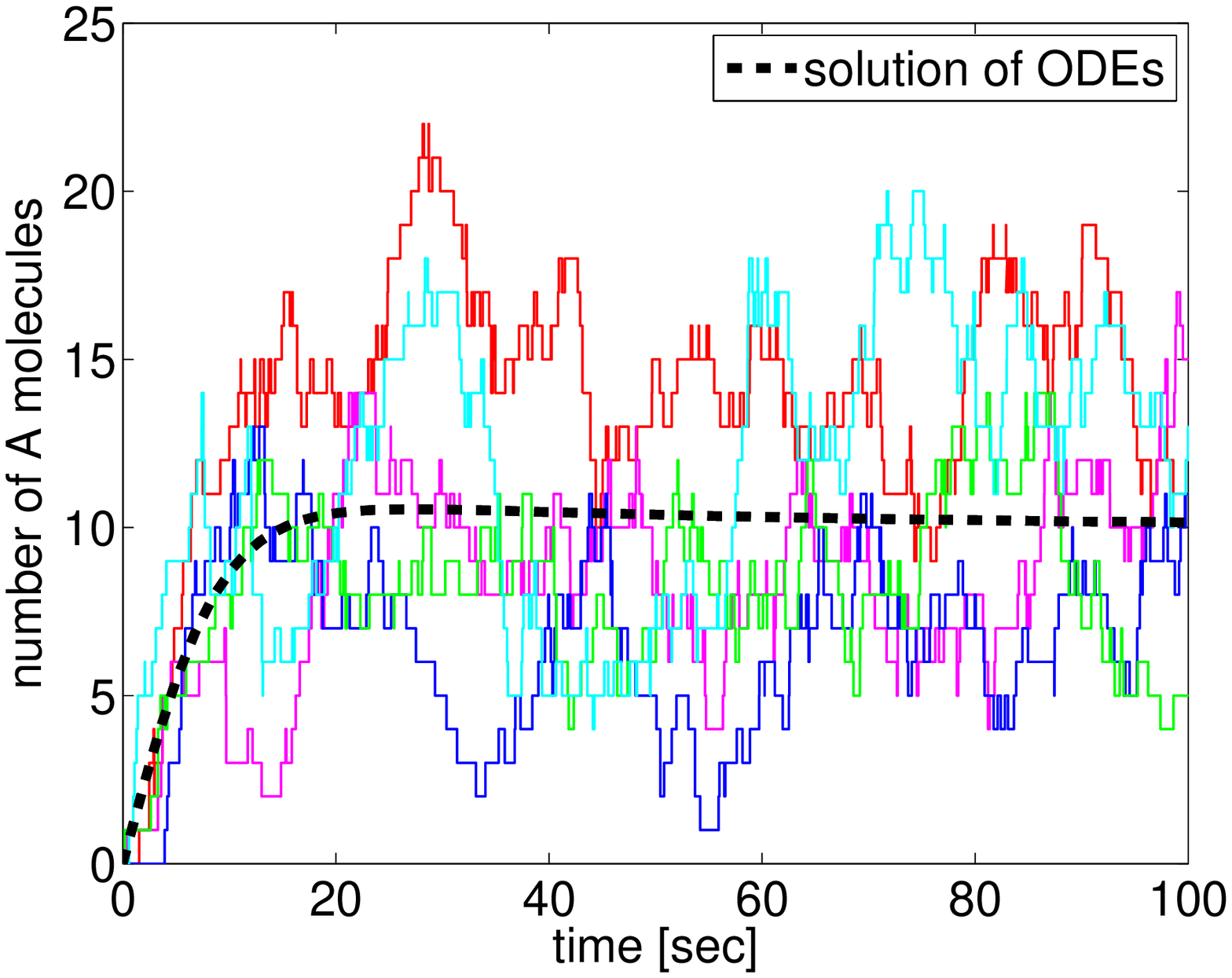}{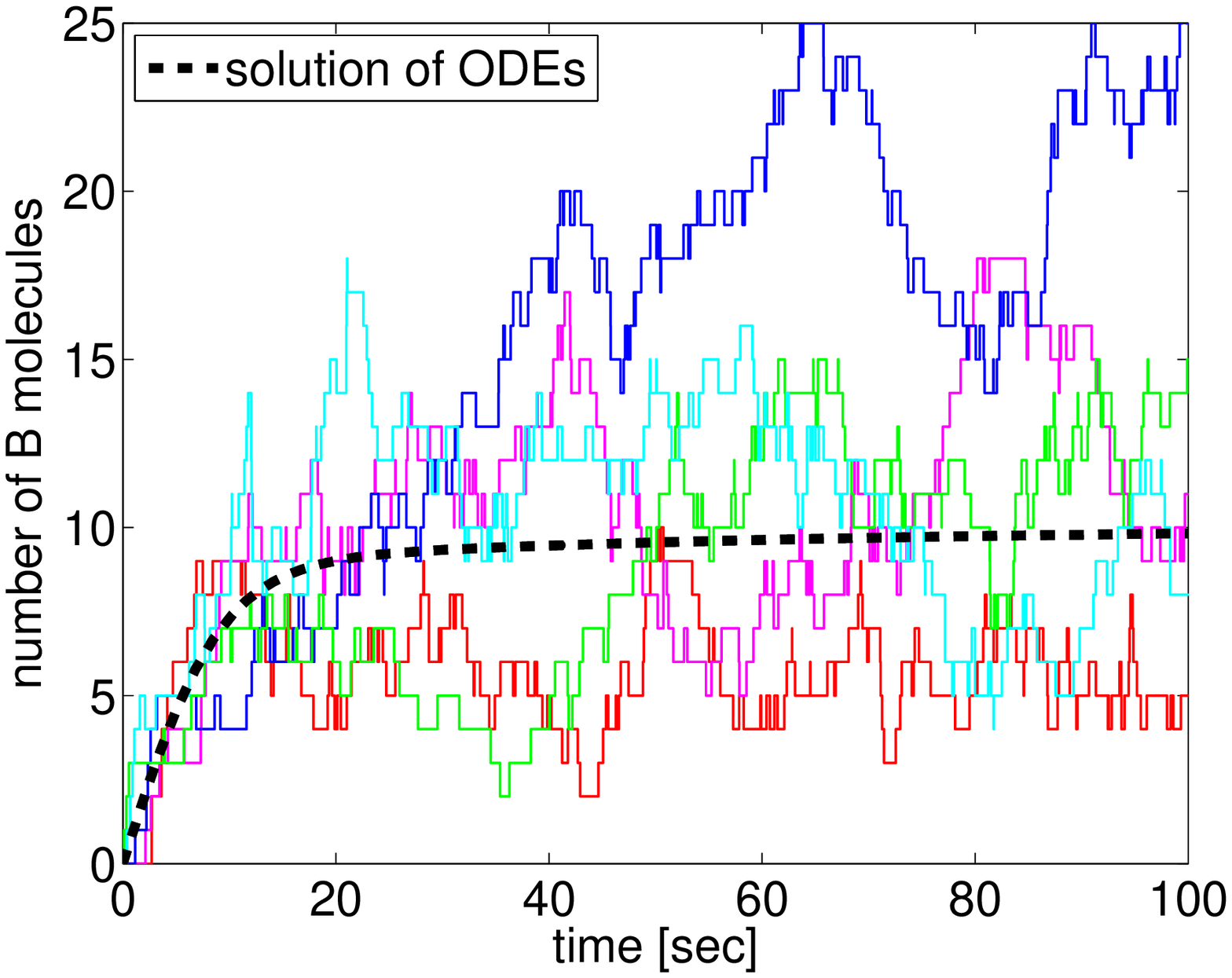}{2.05in}{5mm}
\caption{Five realizations of SSA (a4)--(d4). Number of molecules 
of chemical species
$A$ (left panel) and $B$ (right panel) are plotted as functions 
of time as solid lines. Different colours correspond to different
realizations. The solution of $(\ref{OdeA})$--$(\ref{OdeB})$ is
given by the dashed line. We use $A(0)=0$, $B(0)=0$, 
$k_1= 10^{-3} \, \mbox{sec}^{-1}$, $k_2= 10^{-2} \, \mbox{sec}^{-1},$
$k_3= 1.2 \; \mbox{sec}^{-1}$ and $k_4= 1 \, \mbox{sec}^{-1}.$}
\label{fignonlintime}
\end{figure}

\noindent
SSA (a4)--(d4) is a direct generalisation of SSA (a3)--(d3). At each 
time step, we first ask the
question when will the next reaction occur? The answer is given
by formula (\ref{tauform3}) which can be  justified using 
the same arguments as formulae (\ref{tauform}) or
(\ref{tauform2}). Then we ask the question which reaction
takes place. The probability that the $i$-th chemical
reaction occurs is given by $\alpha_i/\alpha_0$. The decision
which reaction takes place is given in step (d4) with the
help of the second uniformly distributed random number
$r_2$. Knowing that the $i$-th reaction took place,
we update the number of reactants and products accordingly.
Results of five realizations of SSA (a4)--(d4) are plotted 
in Figure \ref{fignonlintime} as solid lines.
We use $A(0)=0$, $B(0)=0$, $k_1= 10^{-3} \, \mbox{sec}^{-1}$, 
$k_2= 10^{-2} \, \mbox{sec}^{-1},$
$k_3= 1.2 \; \mbox{sec}^{-1}$ and
$k_4= 1 \, \mbox{sec}^{-1}.$ We plot the number of molecules 
of chemical species $A$ and $B$ as functions of time.
We see that, after initial transients, $A(t)$ and $B(t)$
fluctuate around their average values. They can be estimated
from long time stochastic simulations as 9.6 for $A$ and
12.2 for $B$.

Let $p_{n,m}(t)$ be the probability that $A(t)=n$ and $B(t)=m$.
The chemical master equation can be written in the following
form
\begin{eqnarray}
\frac{\mbox{d} p_{n,m}}{\dt} 
& = &
k_1 (n+2) (n+1) \, p_{n+2,m} - k_1 n (n-1) \, p_{n,m}
\nonumber
\\
& + &
k_2 (n+1) (m+1) \, p_{n+1,m+1} - k_2 n m \, p_{n,m}
\nonumber
\\
& + &
k_3 \, p_{n-1,m} - k_3 \, p_{n,m}
+ 
k_4 \, p_{n,m-1} - k_4 \, p_{n,m}
\label{cmenonlin}
\end{eqnarray}
for $n$, $m \ge 0$, with the convention that 
$p_{n,m} \equiv 0$ if $n<0$ or $m<0.$ The first
difference between (\ref{cmenonlin}) and the chemical
master equations from the previous sections is that 
equation (\ref{cmenonlin}) is parametrised by two
indices $n$ and $m$. The second important difference
is that (\ref{cmenonlin}) cannot be solved analytically
as we did with (\ref{cmedegradation}). Moreover, it does
not lead to closed evolution equations for stochastic means
and variances; i.e. we cannot follow the same technique
as in the case of equation (\ref{cmeproddegr}). 
The approach from the previous section does 
not work. Let us note that the probability $p_{n,m}(t)$
is sometimes denoted by $p(n,m,t)$; such a notational
convention is often used when we consider systems 
of many chemical species. We will use it in the 
following sections to avoid long subscripts.

The classical deterministic description of the chemical system
(\ref{nonlinearmodel1})--(\ref{nonlinearmodel2}) is given
by the system of ODEs
\begin{eqnarray}
\frac{\mbox{d}a}{\dt} & = & - 2 k_1 a^2 - k_2 \, a b + k_3, 
\label{OdeA}
\\
\frac{\mbox{d}b}{\dt} & = & - k_2 \, a b + k_4.
\label{OdeB}
\end{eqnarray}
The solution of (\ref{OdeA})--(\ref{OdeB}) with initial
conditions $a(0)=0$ and $b(0)=0$ is plotted as a dashed
line in Figure \ref{fignonlintime}. Let us note that
the equations (\ref{OdeA})--(\ref{OdeB}) do not describe
the stochastic means of $A(t)$ and $B(t)$. For example,
the steady state values of (\ref{OdeA})--(\ref{OdeB})
are (for the parameter values of Figure \ref{fignonlintime})
equal to $a_s = b_s = 10$.
On the other hand, the average values estimated
from long time stochastic simulations are 9.6 for $A$ 
and 12.2 for $B$. We will see later in Section \ref{secssr}
that the difference between the results of stochastic
simulations and the corresponding ODEs can be even more significant.

\begin{figure}
\picturesAB{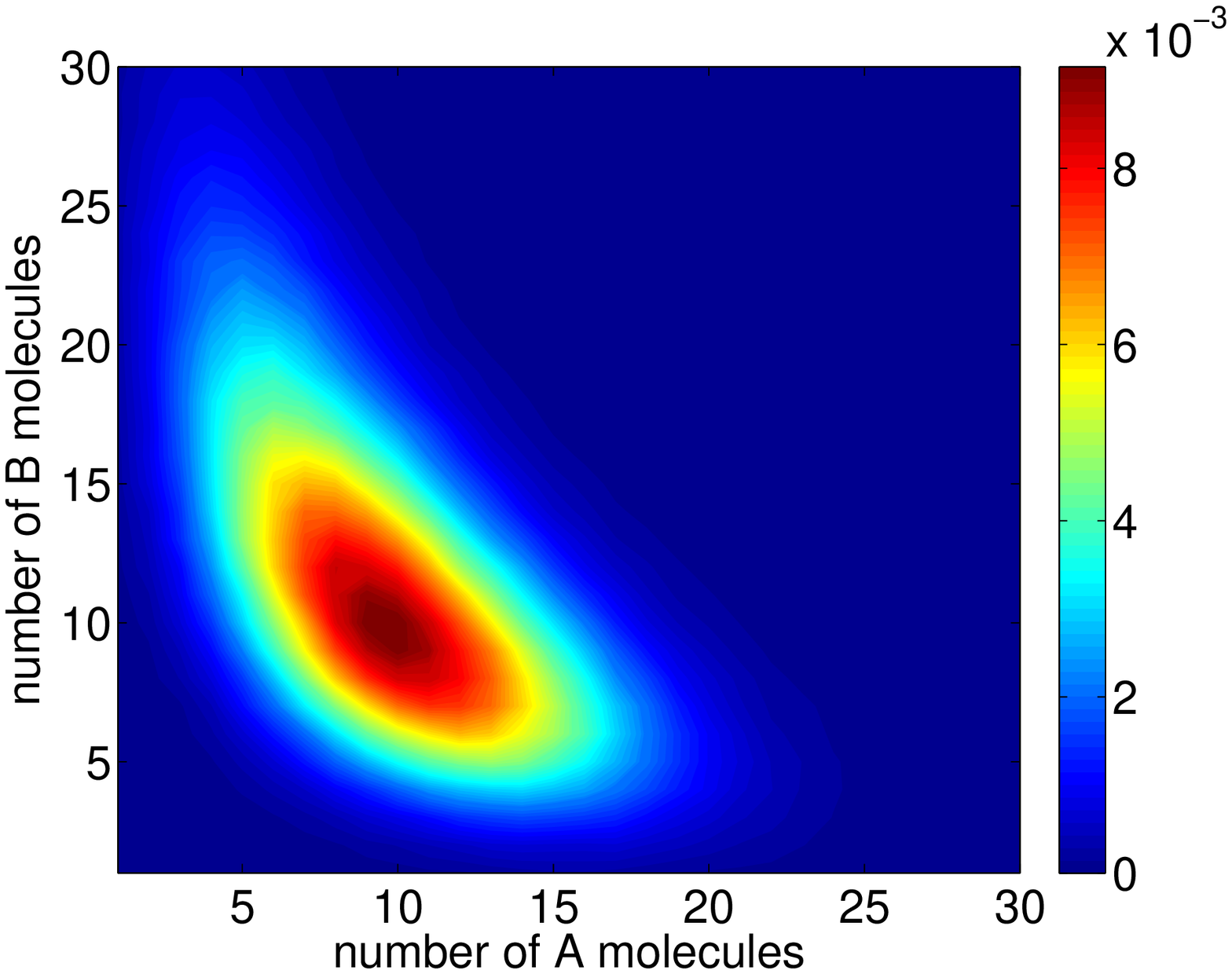}{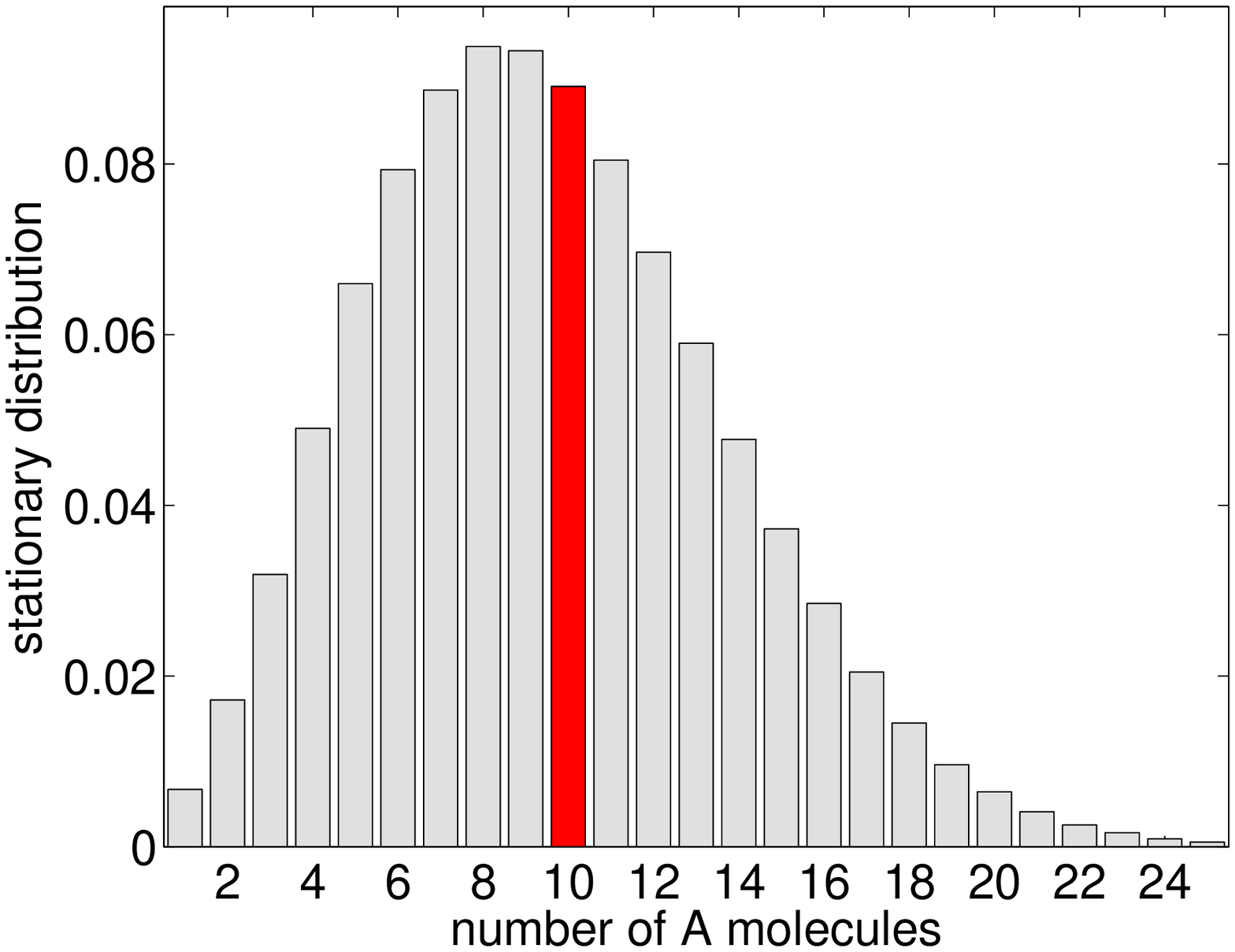}{2.05in}{5mm}
\caption{{\rm (a)} Stationary distribution $\phi(n,m)$ obtained
by long time simulation of (a4)--(d4) for 
$k_1= 10^{-3} \, \mbox{sec}^{-1}$, 
$k_2= 10^{-2} \, \mbox{sec}^{-1},$
$k_3= 1.2 \; \mbox{sec}^{-1}$ and
$k_4= 1 \, \mbox{sec}^{-1}$. {\rm (b)}
Stationary distribution of $A$ obtained by $(\ref{psnnonlin})$.}
\label{fignonlindistr}
\end{figure}

The stationary distribution is defined by 
$$
\phi(n,m) = \lim_{t \to \infty} p_{n,m}(t).
$$
This can be computed by long time simulations of 
SSA (a4)--(d4) and is plotted in Figure
\ref{fignonlindistr}(a).
We see that there is a correlation between 
the values of $A$ and $B$. This can also be 
observed in Figure \ref{fignonlintime}. Looking
at the blue realizations, we see that the values of
$A(t)$ are below the average and the values of $B(t)$
are above the average, similarly for other 
realizations. One can also define the stationary distribution
of $A$ only by 
\begin{equation}
\phi(n) = \sum_{m=0}^\infty \phi(n,m).
\label{psnnonlin}
\end{equation}
Summing the results of Figure
\ref{fignonlindistr}(a) over $m$, we obtain $\phi(n)$
which is plotted in Figure
\ref{fignonlindistr}(b) as a gray histogram. The red bar
highlights the steady state value $a_s = 10$ of
system (\ref{OdeA})--(\ref{OdeB}).

SSAs (a3)--(d3) and (a4)--(d4) were special forms of the
so-called Gillespie SSA. To conclude this section, we 
formulate the Gillespie SSA in its full generality.
Let us consider that we have a system of $q$ chemical 
reactions. Let $\alpha_i(t)$ be the propensity function 
of the $i$-th reaction, $i=1,2, \dots, q,$ at time $t$,
that is, $\alpha_i(t) \, \dt$ is the probability 
that the $i$-th reaction occurs in the time interval $[t,t+\dt)$. Then
the Gillespie SSA consists of the following four steps 
at time $t$.

\leftskip 1.4cm

\medskip

{ 
\parindent -8.4mm
 
{\bf (a5)} Generate two random numbers $r_1$, $r_2$ uniformly distributed in 
 $(0,1)$. 

{\bf (b5)} Compute the propensity function $\alpha_i(t)$ of each reaction. 
Compute 
\begin{equation}
\alpha_0 = \sum_{i=1}^{q} \alpha_i(t).
\label{compalphaRD}
\end{equation}

{\bf (c5)} Compute the time when the next chemical reaction 
takes place as $t+\tau$ where $\tau$ is given by (\ref{tauform3}).

{\bf (d5)} Compute which reaction occurs at time $t+\tau$. Find $j$ such that
$$
r_2 \ge \frac{1}{\alpha_0} \sum_{i=1}^{j-1} \alpha_i 
\qquad \mbox{and} \qquad 
r_2 < \frac{1}{\alpha_0} \sum_{i=1}^{j} \alpha_i.
$$ 
Then the $j$-th reaction takes place, i.e. update numbers of reactants
and products of the $j$-th reaction. \\
Continue with step (a5) for time $t+\tau.$

}

\leftskip 0cm

\medskip

\noindent
The Gillespie SSA (a5)--(d5) provides an exact method 
for the stochastic simulation of systems of chemical 
reactions. It was applied previously as SSA (a2)--(c2) 
for the chemical reaction (\ref{degradation}), 
as SSA (a3)--(d3) for the chemical system 
(\ref{degradationcreation}) and as SSA (a4)--(d4) 
for the chemical system (\ref{nonlinearmodel1})--(\ref{nonlinearmodel2}).
Our simple examples can be simulated quickly in Matlab
(in less than a second on present-day computers).
If one considers systems of many chemical reactions
and many chemical species, then SSA (a5)--(d5) might be 
computationally intensive. In such a case, there are
ways to make the Gillespie SSA more efficient.
For example, it would be a waste of time to recompute all the
propensity functions at each time step (step (b5)). We simulate 
one reaction per one time step. Therefore, it makes sense 
to update only those propensity functions which are changed by the 
chemical reaction which was selected in step (d5) of
SSA (a5)--(d5). A more detailed discussion about the
efficient computer implementation of the Gillespie SSA
can be found e.g. in \cite{Gibson:2000:EES}.

\section{Diffusion}

\label{secdiffusion}

\begin{figure}
\picturesAB{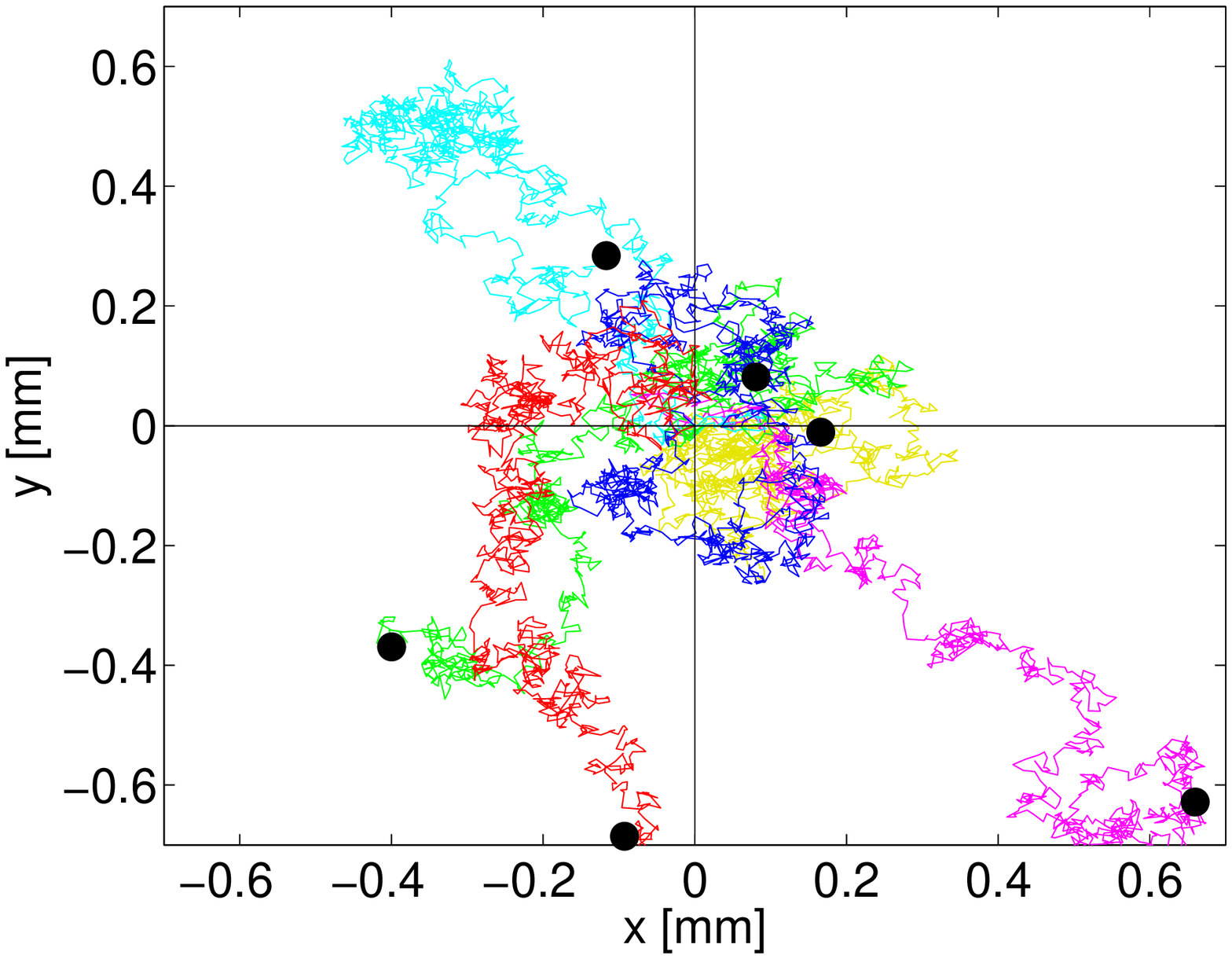}{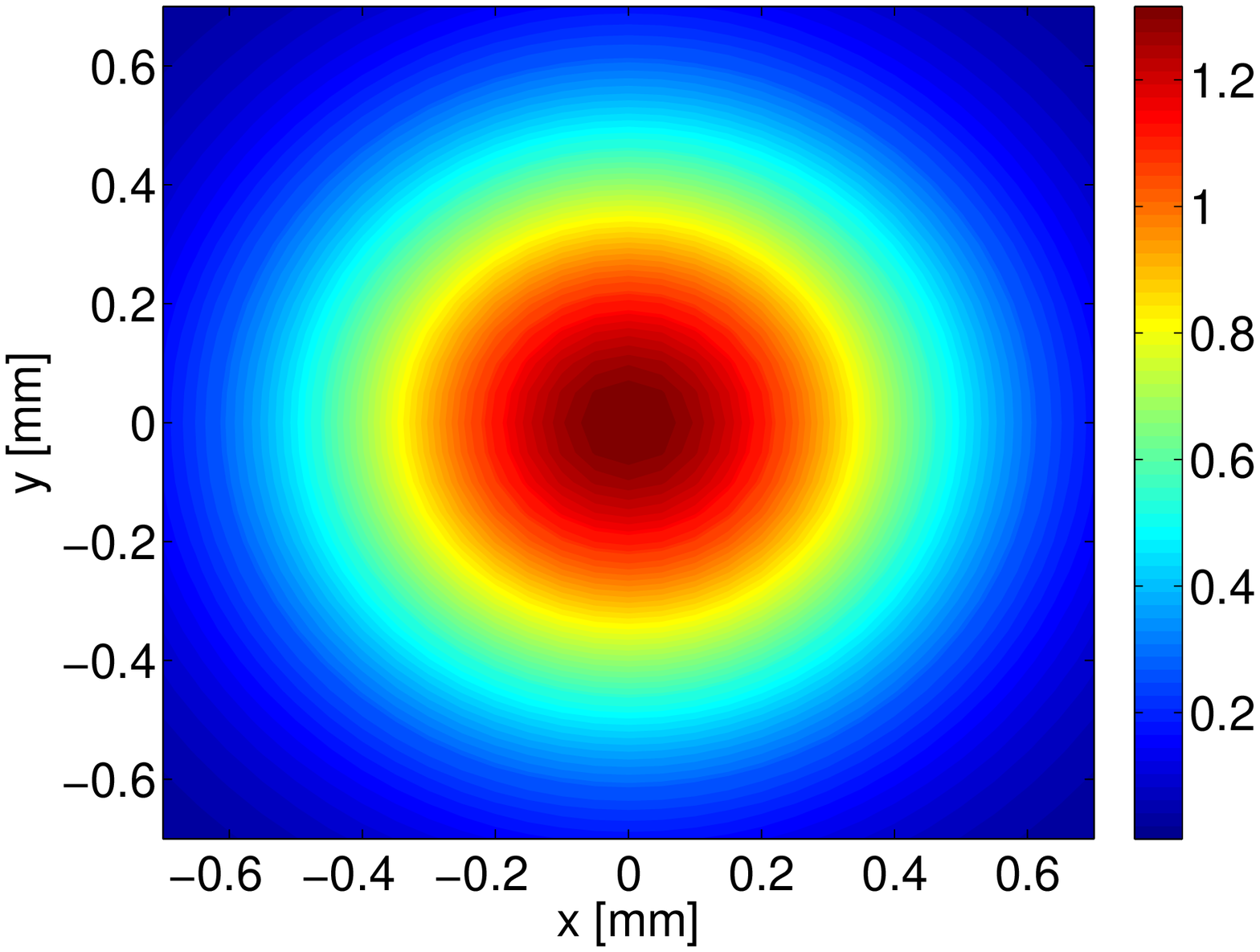}{2in}{5mm}
\caption{{\rm (a)} Six trajectories obtained by SSA (a6)--(b6)
for $D=10^{-4} \; \mbox{mm}^2 \, \mbox{sec}^{-1}$ 
and $\Delta t = 0.1 \; \mbox{sec}$. Trajectories were started at the origin
and followed for 10 minutes. {\rm (b)} Probability distribution
function $\psi(x,y,t)$ given by $(\ref{psixy})$
at time $t=10 \; \mbox{min}$.}
\label{figdiffusion2D}
\end{figure}

Diffusion is the random migration of molecules (or
small particles) arising from motion due to thermal 
energy \cite{Berg:1983:RWB}. As shown by Einstein,
the kinetic energy of a molecule (e.g. protein)
is proportional to the absolute temperature. 
In particular, the protein molecule has a non-zero 
instantaneous speed at, for example, room temperature 
or at the temperature of the human body. A typical 
protein molecule is immersed 
in the aqueous medium of a living cell. Consequently,
it cannot travel too far before it bumps into 
other molecules (e.g. water molecules) in the solution. 
As a result, the trajectory of the molecule
is not straight but it executes a random walk as
shown in Figure \ref{figdiffusion2D}(a).
We plot six possible trajectories of the protein
molecule with six different colours. All trajectories
start at the origin and are followed for 10 minutes. We
will provide more details about this figure together
with the methods for simulating molecular 
diffusion in the rest of this section. Stochastic
models of diffusion which are based on the Smoluchowski 
equation are introduced in Section \ref{secdiffSmoluchowski}. 
In Section \ref{secdiffGillespie}, we introduce a
model which is suitable for coupling with the
Gillespie SSA. Both modelling approaches will be used 
later in Section \ref{secRD} for the stochastic
modelling of reaction-diffusion processes. Let
us note that there exist other models of 
molecular diffusion -- they will be discussed
in Section \ref{secdiscussion}.

\subsection{Smoluchowski equation and diffusion}

\label{secdiffSmoluchowski}

Let $[X(t),Y(t),Z(t)] \in {\mathbb R}^3$ be the position
of a diffusing molecule at time $t$. Starting with 
$[X(0),Y(0),Z(0)]=[x_0,y_0,z_0]$, we want to
compute the time evolution of $[X(t),Y(t),Z(t)]$.
To do that, we make use of a generator of random
numbers which are normally distributed with zero mean
and unit variance. Such a generator is part of many
modern computer languages (e.g. function {\tt randn}
in Matlab). Diffusive spreading of molecules
is characterised by a single {\it diffusion constant} $D$
which depends on the size of the molecule, absolute
temperature and viscosity of the solution \cite{Berg:1983:RWB}. 
Choosing time step $\Delta t$, we compute the time evolution
of the position of the diffusing molecule by performing 
two steps at time $t$:

\leftskip 1.4cm

\medskip

{ 
\parindent -8.4mm

{\bf (a6)} \!Generate three normally distributed  
(with zero mean and unit variance) random numbers 
$\xi_x$, $\xi_y$ and $\xi_z$.

{\bf (b6)} Compute the position of the molecule at time
$t+\Delta t$ by
\begin{eqnarray} 
X(t + \Delta t) & = & X(t) + \sqrt{2 D \, \Delta t} \; \xi_x, 
\label{xequation}
\\
Y(t + \Delta t) & = & Y(t) + \sqrt{2 D \, \Delta t} \; \xi_y, 
\label{yequation}
\\
Z(t + \Delta t) & = & Z(t) + \sqrt{2 D \, \Delta t} \; \xi_z, 
\label{zequation}
\end{eqnarray}
Then continue with step (a6) for time $t+\Delta t.$

}

\leftskip 0cm

\medskip

\noindent
Choosing $D=10^{-4} \; \mbox{mm}^2 \, \mbox{sec}^{-1}$
(diffusion constant of a typical protein molecule),
$[X(0),Y(0),Z(0)]=[0,0,0]$ and $\Delta t = 0.1 \; \mbox{sec}$,
we plot six realizations of SSA (a6)--(b6)
in Figure \ref{figdiffusion2D}(a). We plot
only the $x$ and $y$ coordinates. We follow the diffusing 
molecule for 10 minutes. The position of the molecule at
time $t = 10$ min is denoted as a black circle for
each trajectory.

Equations (\ref{xequation})--(\ref{zequation}) are discretized
versions of stochastic differential equations (SDEs) which are 
sometimes called Smoluchowski equations. Some basic facts
about SDEs can be found e.g. in 
\cite{Arnold:1974:SDE,Gardiner:1985:HSM}. A more accessible 
introduction to SDEs can be found in \cite{Higham:2001:AIN}
which has a similar philosophy as our paper. Reference 
\cite{Higham:2001:AIN} is a nice algorithmic introduction
to SDEs for students who do not have a prior knowledge of
advanced probability theory or stochastic analysis.
We will not go into details of SDEs in this paper, 
but only highlight some interesting facts which will be useful 
later. 

First, equations (\ref{xequation})--(\ref{zequation}) are not 
coupled. To compute the time evolution of $X(t)$, we do not need 
to know the time evolution of $Y(t)$ or $Z(t)$. We will later focus 
only on the time evolution of the $x$-th coordinate, effectively 
studying one-dimensional problems. Two-dimensional
or three-dimensional problems can be treated similarly.
Second, we see that different realizations of SSA
(a6)--(b6) give different results. To get more reproducible
quantities, we will shortly study the behaviour of several
molecules. However, even in the case of a single diffusing
molecule, there are still quantities whose evolution is
deterministic. Let $\varphi(x,y,t) \,\dx \dy \dz$ be the 
probability that  $X(t) \in [x,x+\dx),$ 
$Y(t) \in [y,y+\dy)$ and $Z(t) \in [z,z+\dz)$ at time $t$. 
It can be shown that $\varphi$ evolves according to the 
partial differential equation 
\begin{equation}
\frac{\partial \varphi}{\partial t} 
=
D \left( 
\frac{\partial^2 \varphi}{\partial x^2} 
+
\frac{\partial^2 \varphi}{\partial y^2} 
+
\frac{\partial^2 \varphi}{\partial z^2} 
\right),
\label{FokPladiff}
\end{equation}
which is a special form of the so-called Fokker-Planck equation.
Since our random walk starts at the origin, we can solve
(\ref{FokPladiff}) with initial condition
$\varphi(x,y,z,0)=\delta(x,y,z)$ where $\delta$ is the Dirac
distribution at the origin. We obtain
\begin{equation*}
\varphi(x,y,z,t)
=
\frac{1}{(4 D \pi t)^{3/2}}
\exp \left[ - \frac{x^2+y^2+z^2}{4Dt} \right].
\end{equation*}
In order to visualise this probability distribution
function, we integrate it over $z$ to get probability 
distribution function
\begin{equation}
\psi(x,y,t) = \int_{\mathbb R} \varphi(x,y,z,t) \dz
=
\frac{1}{4 D \pi t} \exp \left[ - \frac{x^2+y^2}{4Dt} \right].
\label{psixy}
\end{equation}
This means that $\psi(x,y,t) \,\dx \dy$ is the 
probability that  $X(t) \in [x,x+\dx)$ and $Y(t) \in [y,y+\dy)$ 
at time $t$. The function $\psi(x,y,t)$ at time $t=10$ min
is plotted in Figure \ref{figdiffusion2D}(b).
It can be obtained also by computing many realizations
of SSA (a6)--(b6) and plotting the histogram of positions
of a molecule at time $10$ min; such positions were
denoted as black circles for the six illustrative trajectories
in Figure \ref{figdiffusion2D}(a).

One important issue which was not addressed previously 
is that molecules diffuse in bounded volumes, i.e. the 
domain of interest has boundaries and suitable boundary 
conditions must be implemented. In the rest of this paper, 
we focus on one-dimensional problems to avoid technicalities.
Hence, we effectively study diffusion of molecules
in the one-dimensional interval $[0,L]$. Then the SSA can 
be formulated as follows:

\leftskip 1.4cm

\medskip

{ 
\parindent -8.4mm

{\bf (a7)} Generate a normally distributed  
(with zero mean and unit variance) random number $\xi$.

{\bf (b7)} Compute the position of the molecule at time
$t+\Delta t$ by
\begin{equation} 
X(t + \Delta t) = X(t) + \sqrt{2 D \, \Delta t} \; \xi. 
\label{x5equation}
\end{equation}

{\bf (c7)} If $X(t+\Delta t)$ computed by (\ref{x5equation}) is less
than $0$, then \\
$X(t+\Delta t)= - X(t) - \sqrt{2 D \, \Delta t} \; \xi$. \\
If $X(t+\Delta t)$ computed by (\ref{x5equation}) is greater
than $L$, then \\
$X(t+\Delta t)=2 L - X(t) - \sqrt{2 D \, \Delta t} \; \xi$. \\ 
Then continue with step (a7) for time $t+\Delta t.$

}

\leftskip 0cm

\medskip

\noindent
The boundary condition implemented in step (c7) is the so-called
{\it reflective boundary condition} or zero flux boundary condition.
It can be used when there is no chemical interaction between 
the boundary and diffusing molecules. More complicated boundary
conditions are discussed in \cite{Erban:2007:RBC,Erban:2007:TSR}.

\begin{figure}
\picturesAB{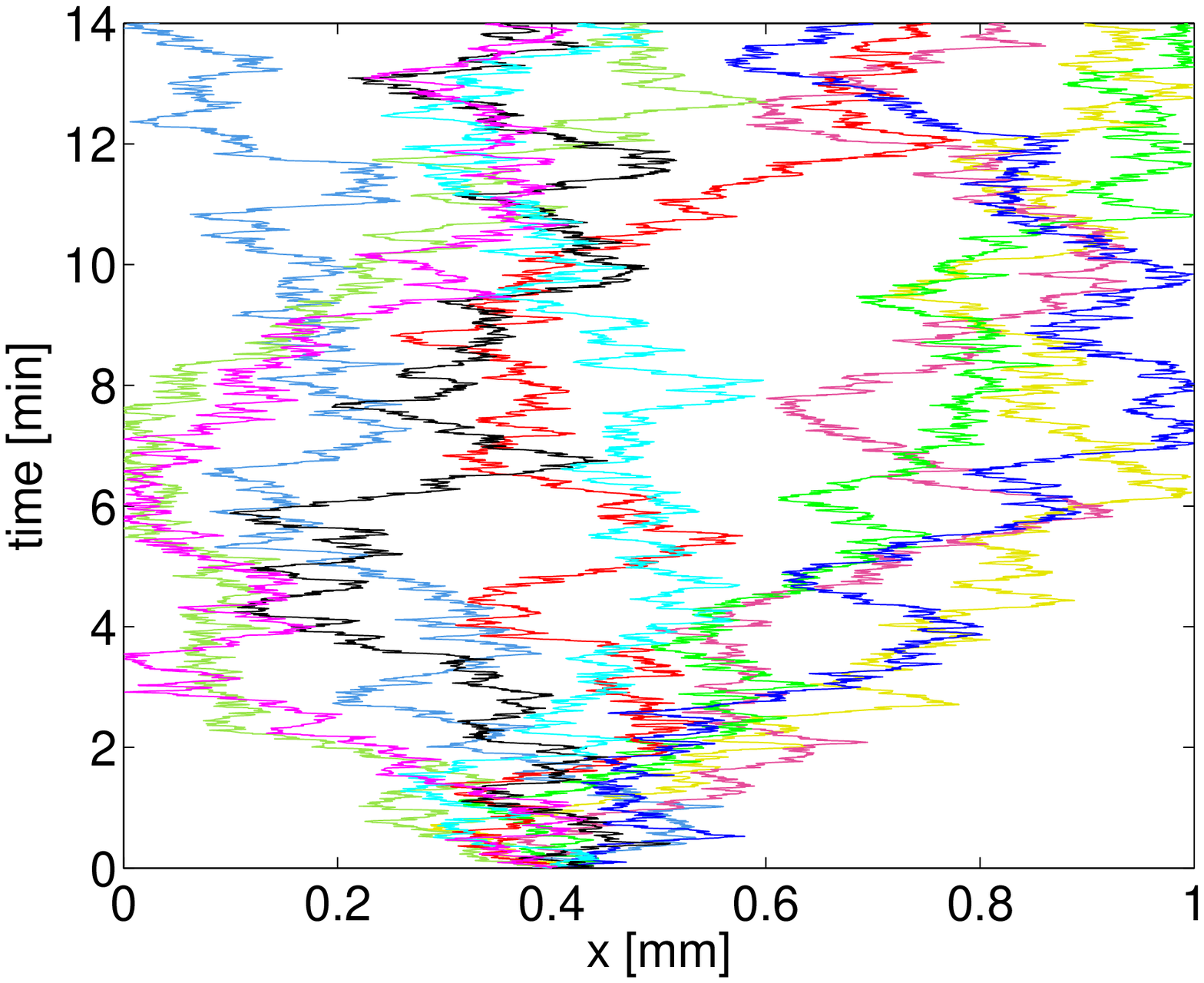}{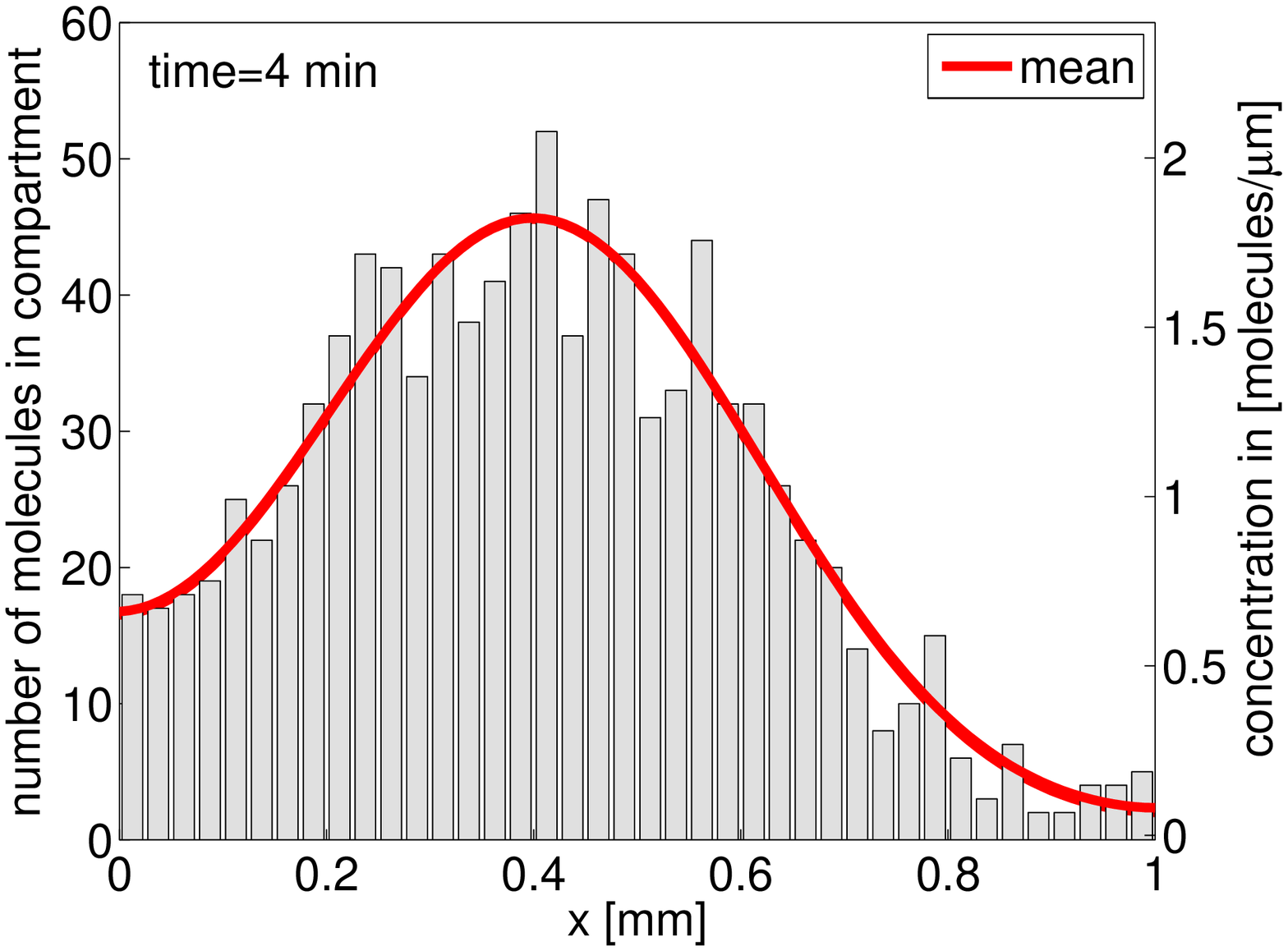}{2in}{0mm}
\caption{{\rm (a)} Ten trajectories computed by SSA (a7)--(c7)
for $D=10^{-4} \; \mbox{mm}^2 \, \mbox{sec}^{-1}$,
$L=1$ mm, $X(0)=0.4$ mm and $\Delta t = 0.1 \; \mbox{sec}$. 
{\rm (b)} Numbers of molecules in bins of length 
$h = 25 \;\mu$m at time $t=4 \; \mbox{min}$.}
\label{figdiffusion1DSmol}
\end{figure}

Choosing $D=10^{-4} \; \mbox{mm}^2 \, \mbox{sec}^{-1}$,
$L=1$ mm, $X(0)=0.4$ mm and $\Delta t = 0.1 \; \mbox{sec}$,
we plot ten realizations of SSA (a7)--(c7) 
in Figure \ref{figdiffusion1DSmol}(a). 
Let us assume that we have a system of 1000 molecules
which are released at position $x=0.4$ mm at time $t=0$.
Then Figure \ref{figdiffusion1DSmol}(a)
can be viewed as a plot of the trajectories of ten representative
molecules. Considering 1000 molecules, the trajectories
of individual molecules are of no special interest.
We are rather interested in spatial histograms (density 
of molecules). An example of such a plot is given
in Figure \ref{figdiffusion1DSmol}(b).
We simulate 1000 molecules, each following SSA
(a7)--(c7). At time $t=4$ min, we divided the domain
of interest $[0,L]$ into 40 bins of length 
$h = L/40 = 25 \;\mu$m. We calculated the number of molecules
in each bin $[(i-1)h,ih)$, $i=1,2, \dots, 40$, 
at time $t=4$ min and plotted them as a histogram.
 
Let us note that the deterministic counterpart to the stochastic
simulation is a solution of the corresponding Fokker-Planck
equation (diffusion equation in our case) which, in one
dimension with zero flux boundary conditions, reads as follows
\begin{equation}
\frac{\partial \varphi}{\partial t} 
=
D \frac{\partial^2 \varphi}{\partial x^2}
\qquad \mbox{where} \qquad 
\frac{\partial \varphi}{\partial x}(0)
= 
\frac{\partial \varphi}{\partial x}(L)
=
0. 
\label{FokPladiff1D}
\end{equation}
The solution of (\ref{FokPladiff1D}) with the Dirac-like initial condition
at $x=0.4$ mm is plotted as a red solid line 
in Figure \ref{figdiffusion1DSmol}(b) for comparison. 

\subsection{Compartment-based approach to diffusion}

\label{secdiffGillespie}

In Section \ref{secdiffSmoluchowski}, we simulated the behaviour of
1000 molecules by computing the individual trajectories of
every molecule (using SSA (a7)--(c7)). At the end 
of the simulation, we divided the computational domain
$[0,L]$ into $K=40$ compartments and we plotted numbers
of molecules in each compartment 
in Figure \ref{figdiffusion1DSmol}(b). In particular, most
of the computed information (1000 trajectories) was not used 
for the final result -- the spatial histogram. We visualised
only 40 numbers (numbers of molecules in compartments) instead
of 1000 computed positions of molecules. 
In this section, we present a different SSA for the simulation
of molecular diffusion. We redo the example from 
Section \ref{secdiffSmoluchowski} but instead of simulating
1000 positions of the individual molecules, we are going to
simulate directly the time evolution of 40 compartments.

To do that, we divide the computational domain
$[0,L]$ into $K=40$ compartments of length $h = L/K=25 \,\mu$m.
We denote the number of molecules of chemical species $A$ 
in the $i$-th compartment $[(i-1)h,ih)$ by $A_i$, 
$i=1,\dots,K$. We apply the Gillespie SSA to the following chain 
of ``chemical reactions":
\begin{equation}
A_1
\;
\mathop{\stackrel{\displaystyle\longrightarrow}\longleftarrow}^{d}_{d}
\;
A_2
\;
\mathop{\stackrel{\displaystyle\longrightarrow}\longleftarrow}^{d}_{d}
\;
A_3
\;
\mathop{\stackrel{\displaystyle\longrightarrow}\longleftarrow}^{d}_{d}
\;
\dots
\;
\mathop{\stackrel{\displaystyle\longrightarrow}\longleftarrow}^{d}_{d}
\;
A_K
\label{diffGill}
\end{equation}
where
$$
A_i
\;
\mathop{\stackrel{\displaystyle\longrightarrow}\longleftarrow}^{d}_{d}
\;
A_{i+1}
\qquad
\mbox{means that}
\qquad
A_i \; \mathop{\longrightarrow}^{d} \;\, A_{i+1}
\;\;\;
\mbox{and}
\;\;\;
A_{i+1} \; \mathop{\longrightarrow}^{d} \;\, A_{i}.
$$
We will shortly show that the Gillespie SSA of (\ref{diffGill}) provides
a correct model of diffusion provided that the rate constant $d$ 
in (\ref{diffGill}) is chosen as $d = D/h^2$ where $D$ is the
diffusion constant and $h$ is the compartment length. The
compartment-based SSA can be described as follows. Starting
with initial condition $A_i(t)=a_{0,i},$ $i=1,2, \dots, K$,
we perform six steps at time $t$:

\leftskip 1.4cm

\medskip

{ 
\parindent -8.4mm
 
{\bf (a8)} Generate two random numbers $r_1$, $r_2$ uniformly distributed in 
 $(0,1)$. 

{\bf (b8)} Compute propensity functions of reactions by
$\alpha_i = A_i(t) d$ for $i = 1, 2, \dots, K$. Compute 
\begin{equation}
\alpha_0 = \sum_{i=1}^{K-1} \alpha_i + \sum_{i=2}^{K} \alpha_i.
\label{compalpha}
\end{equation}

{\bf (c8)} Compute the time at which the next chemical reaction 
takes place as $t+\tau$ where $\tau$ is given by (\ref{tauform3}).

{\bf (d8)} If $r_2 < \sum_{i=1}^{K-1} \alpha_i/\alpha_0$, then find 
$j \in \{ 1, 2, \dots, K-1\}$ such that 
$$
r_2 \ge \frac{1}{\alpha_0} \sum_{i=1}^{j-1} \alpha_i 
\qquad \mbox{and} \qquad 
r_2 < \frac{1}{\alpha_0} \sum_{i=1}^{j} \alpha_i.
$$ 
Then compute the number of molecules at time $t+\tau$ by 
\begin{eqnarray}
A_j(t+\tau) & = & A_j(t) - 1, \\
A_{j+1}(t+\tau) & = & A_{j+1}(t) + 1, \\
A_i(t+\tau) & = & A_{i}(t), \; \; \; \mbox{for} \; i \ne j, i \ne j+1.
\label{AiAip1}
\end{eqnarray}

{\bf (e8)} If $r_2 \ge \sum_{i=1}^{K-1} \alpha_i/\alpha_0$, then find 
$j \in \{ 2, 3, \dots, K \}$ such that 
$$
r_2 \ge \frac{1}{\alpha_0} 
\left( \sum_{i=1}^{K-1} \alpha_i +\sum_{i=2}^{j-1} \alpha_i \right)
\qquad \mbox{and} \qquad 
r_2 < \frac{1}{\alpha_0} 
\left( \sum_{i=1}^{K-1} \alpha_i +\sum_{i=2}^{j} \alpha_i \right). 
$$ Then compute the number of molecules at time $t+\tau$ by 
\begin{eqnarray}
A_j(t+\tau) & = & A_j(t) - 1, \\
A_{j-1}(t+\tau) & = & A_{j-1}(t) + 1, \\
A_i(t+\tau) & = & A_{i}(t), \; \; \; \mbox{for} \; i \ne j, i \ne j-1.
\label{AiAim1}
\end{eqnarray}

{\bf (f8)} Continue with step (a8) for time $t+\tau.$

}

\leftskip 0cm

\medskip

\noindent
The first term on the right hand side of (\ref{compalpha})
corresponds to reactions $A_i \to A_{i+1}$ (jumps to the right)
and the second term corresponds to 
reactions $A_i \to A_{i-1}$ (jumps to the left).
The time of the next chemical reaction is computed in the
step (c8) using formula (\ref{tauform3}) derived previously. 
The decision about which reaction takes place is done in 
steps (d8)--(e8) with the help of random number $r_2$. 
Jumps to the right are implemented in step (d8) and jumps
to the left in step (e8).

\begin{figure}
\picturesABal{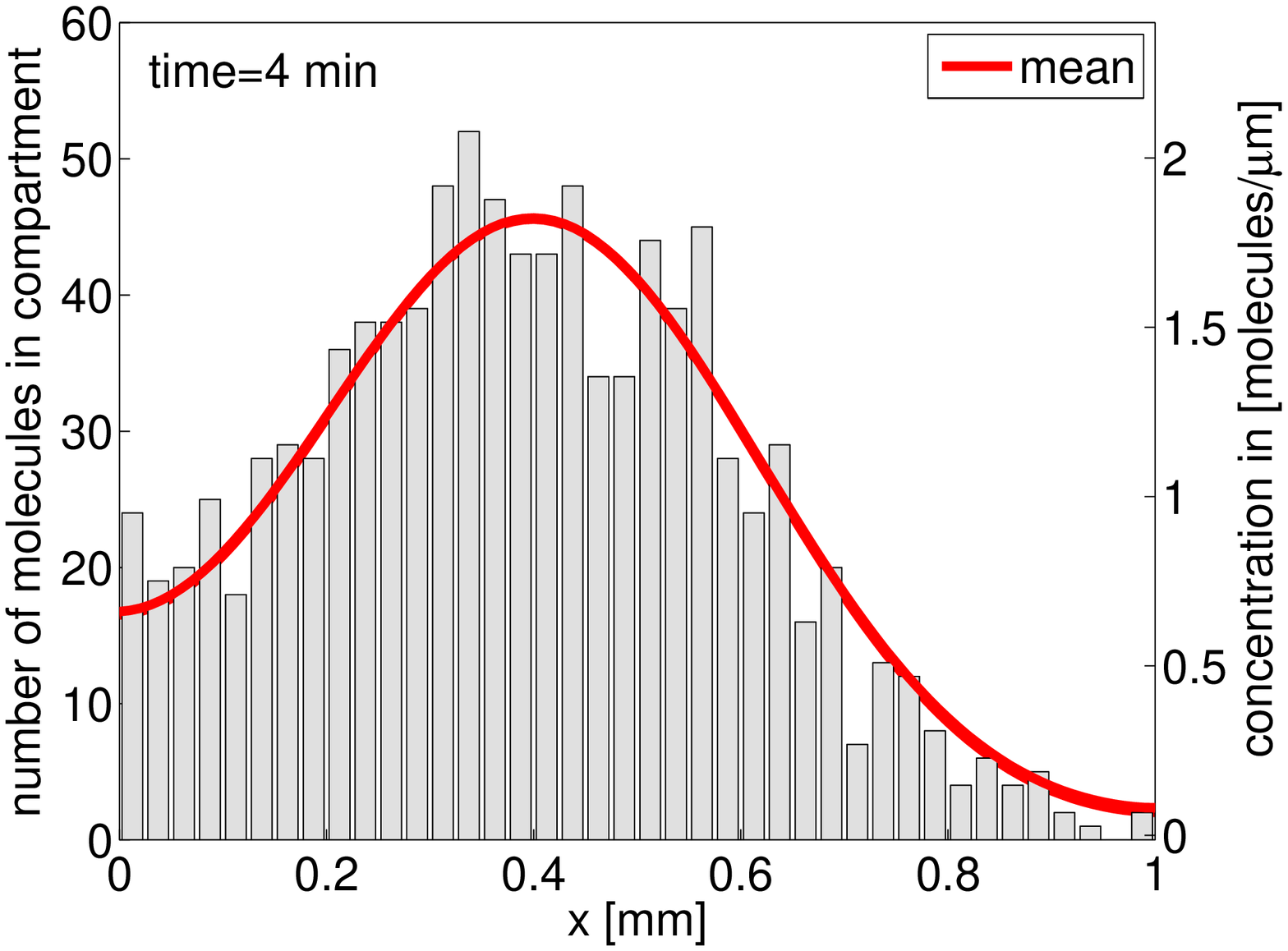}{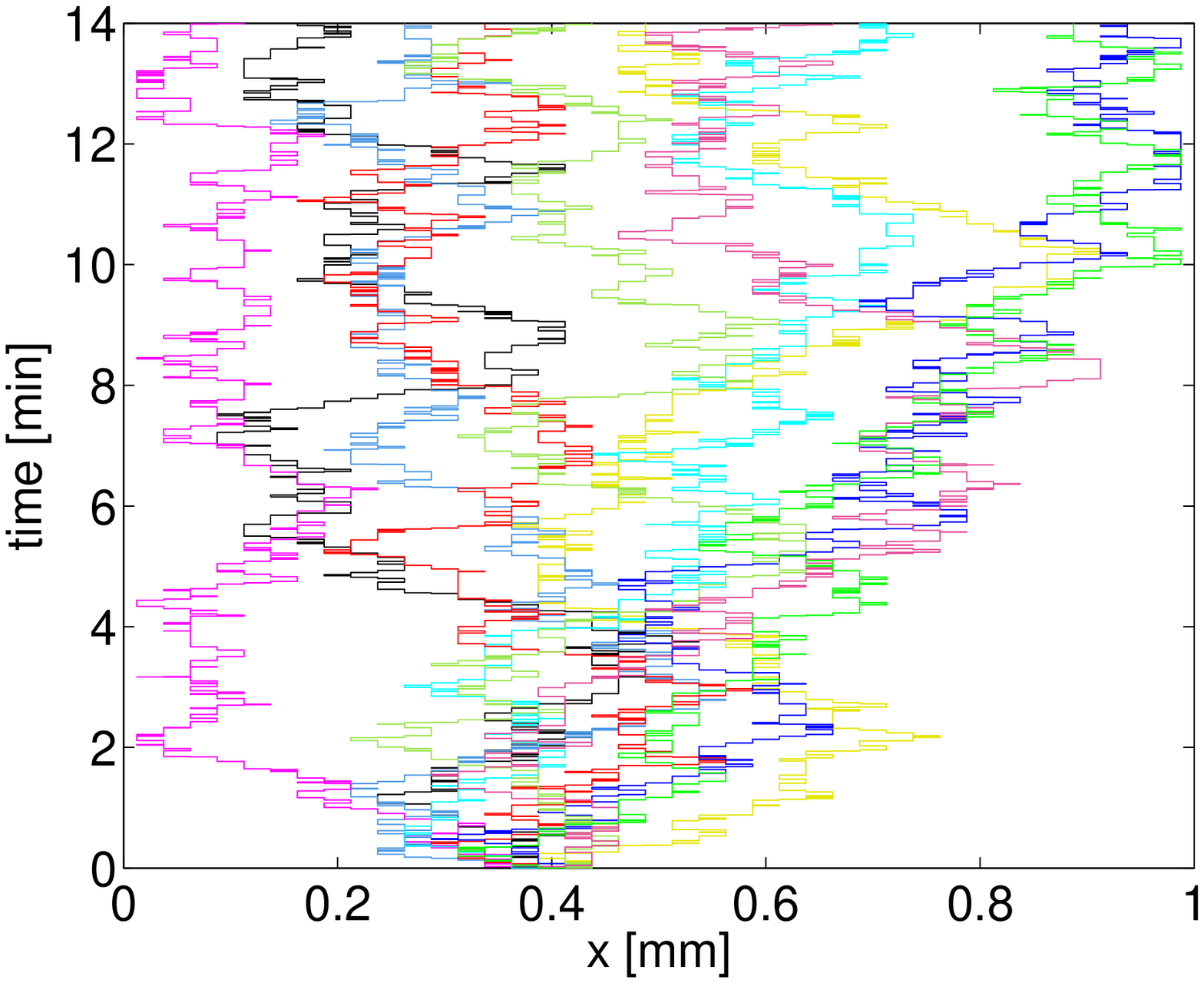}{2in}{4mm}
\caption{Compartment-based SSA model of diffusion. {\rm (a)} 
Numbers $A_i(t)$, $i=1, 2, \dots, K$, at time $t=4 \; \mbox{min}$
obtained by SSA (a8)--(f8). We use $d = D/h^2 = 0.16 \, \mbox{sec}^{-1}$,
$K=40$ and 
initial condition $A_{16}(0) = 500,$ $A_{17}(0)=500$ and
$A_i(0)=0$ for $i \ne 16,$ $i \ne 17.$
{\rm (b)} Ten realizations of the simulation of an individual molecule
by SSA (a8)--(f8).}
\label{figdiffusion1DGill}
\end{figure}

We want to redo the example from Section \ref{secdiffSmoluchowski},
i.e. simulate 1000 molecules starting from position 0.4 mm
in the interval $[0,L]$ for $L=1$ mm. We use $K=40$. 
Since 0.4 mm is exactly
a boundary between the 16th and 17th compartment, the initial
condition is given by $A_{16}(0) = 500,$ $A_{17}(0)=500$ and
$A_i(0)=0$ for $i \ne 16,$ $i \ne 17.$ As 
$D=10^{-4} \; \mbox{mm}^2 \, \mbox{sec}^{-1}$,
we have $d = D/h^2 = 0.16 \; \mbox{sec}^{-1}.$ 
The numbers $A_i(t)$, $i=1, \dots, K$, at time $t=4$ min, are plotted
in Figure \ref{figdiffusion1DGill}(a) as a histogram. 
This panel can be directly compared with Figure 
\ref{figdiffusion1DSmol}(b). The computational intensity 
of SSA (a8)--(f8) can be decreased using the appropriate
way to implement it in the computer. For example, only one 
chemical reaction occurs per time step. Consequently,
only two propensity functions change and need to
be updated in step (b8). Moreover, the formula 
(\ref{compalpha}) can be simplifed as follows
$$
\alpha_0 
= 
\sum_{i=1}^{K-1} \alpha_i + \sum_{i=2}^{K} \alpha_i
=
2 \sum_{i=1}^{K} \alpha_i 
-
\alpha_1 
-
\alpha_K
=
2 d \sum_{i=1}^{K} A_i(t) 
-
\alpha_1 
-
\alpha_K
=
2 d N 
-
\alpha_1 
-
\alpha_K,
$$
where $N=1000$ is the total number of molecules in the simulation
(this number is conserved because there is no creation or
degradation of the molecules in the system). Hence, we need
to recompute $\alpha_0$ only when there is a change in
$\alpha_1$ or $\alpha_K$, i.e. whenever the boundary compartments
were involved in the previous reaction.

SSA (a8)--(f8) does not compute the trajectories of individual
molecules. However, we can still compute a plot comparable 
with Figure \ref{figdiffusion1DSmol}(a). To do that,
we repeat the simulation with 1 molecule instead of 1000.
Then, at given time $t$, exactly one of numbers $A_i(t)$,
$i=1,2, \dots, K$, is non-zero and equal to 1. This is a
position of the molecule at time $t$. Ten realizations
of SSA (a8)--(f8) with one molecule released at 0.4 mm
at $t=0$ are plotted in Figure \ref{figdiffusion1DGill}(b).
This panel can be directly compared with Figure 
\ref{figdiffusion1DSmol}(a).

Let $p(\boldn,t)$ be the joint probability that $A_i(t)=n_i$,
$i=1, \dots, K$, where we denoted $\boldn=[n_1,n_2, \dots, n_K].$
Let us define operators $R_i, L_i: {\mathbb N}^K \to {\mathbb N}^K$ by
\begin{equation}
R_i: [n_1, \dots, n_i, n_{i+1}, \dots, n_K] 
\to [n_1, \dots, n_i+1, n_{i+1}-1, \dots,n_K],
\quad
i=1, \dots, K-1,
\label{Ridef}
\end{equation}
\begin{equation}
L_i: [n_1, \dots, n_{i-1}, n_i, \dots, n_K] 
\to [n_1, \dots, n_{i-1}-1, n_i+1, \dots,n_K],
\quad
i=2, \dots, K.
\label{Lidef}
\end{equation}
Then the chemical master equation, which corresponds to the system
of chemical reactions given by (\ref{diffGill}), can be written as follows
\begin{equation}
\frac{\partial P(\boldn)}{\partial t} 
= 
d
\sum_{j=1}^{K-1}
\Big\{
(n_j+1)
 \, P (R_j \boldn)
- 
n_j \, P (\boldn)
\Big\}
+ 
d
\sum_{j=2}^{K}
\Big\{
(n_j+1)
 \, P (L_j \boldn)
- 
n_j \, P (\boldn)
\Big\}.
\label{cmediff}
\end{equation}
The stochastic mean is defined as the vector $\boldM(t)
\equiv [M_1,M_2, \dots, M_K]$ where
\begin{equation}
M_i(t)
=
\sum_{\boldn} n_i \, P(\boldn,t)
\equiv
\sum_{n_1=0}^\infty
\sum_{n_2=0}^\infty
\dots
\sum_{n_K=0}^\infty
 n_i \, P(\boldn,t)
\label{Midef}
\end{equation}
gives the mean number of molecules in the $i$-th compartment,
$i=1,2, \dots, K$. To derive an evolution equation for
the stochastic mean vector $\boldM(t)$, we can follow the 
method from Section \ref{secproddegr} -- see derivation 
of (\ref{evolM}) from chemical master equation
(\ref{cmeproddegr}). Multiplying (\ref{cmediff}) by $n_i$ 
and summing over $\boldn$, we obtain
(leaving the details to the student) a system of equations for 
$M_i$ of the form
\begin{equation}
\frac{\partial M_i}{\partial t} 
= 
d (M_{i+1} + M_{i-1} - 2 M_i),
\qquad
i = 2, \dots, K-1,
\label{meaneq1}
\end{equation}
\begin{equation}
\frac{\partial M_1}{\partial t} 
=
d (M_{2} - M_1), \qquad
\frac{\partial M_K}{\partial t} 
= 
d (M_{K-1} - M_K).
\label{meaneq2}
\end{equation}
System (\ref{meaneq1})--(\ref{meaneq2}) is equivalent
to a discretization of (\ref{FokPladiff1D}) provided that
$d = D/h^2$. Hence, we have derived the relation between
the rate constant $d$ in (\ref{diffGill}), diffusion constant $D$ 
and compartment length $h$. 
The solution of (\ref{FokPladiff1D}) with the Dirac-like initial 
condition at $x=0.4$ mm is plotted for comparison
as a red solid line in Figure \ref{figdiffusion1DGill}(a). 
 
The noise is described by the variance vector $\boldV(t)
\equiv [V_1,V_2, \dots, V_K]$ where
\begin{equation}
V_i(t)
=
\sum_{\boldn} (n_i - M_i(t))^2 \, P(\boldn,t)
\equiv
\sum_{n_1=0}^\infty
\sum_{n_2=0}^\infty
\dots
\sum_{n_K=0}^\infty
 (n_i - M_i(t))^2 \, P(\boldn,t)
\label{Videf}
\end{equation}
gives the variance of number of molecules in the $i$-th compartment,
$i=1,2, \dots, K$. To derive the evolution equation for the
vector $\boldV(t)$, we define the matrix $\{V_{i,j}\}$ by
$$
V_{ij} = \sum_{\boldn} n_i n_j \, P(\boldn,t) - M_iM_j,
\qquad \mbox{for} \; i,j = 1,2, \dots, K.
$$
Using (\ref{Videf}), we obtain $V_i = V_{ii}$ for
$i = 1,2, \dots, K$.  Multiplying (\ref{cmediff}) by $n_i^2$ 
and summing over $\boldn$, we obtain
\begin{eqnarray}
\frac{\partial}{\partial t} 
\sum_{\boldn} n_i^2 P(\boldn)
& = & 
d
\sum_{j=1}^{K-1}
\Big\{
\sum_{\boldn} n_i^2 (n_j+1)
 \, P (R_j \boldn)
- 
\sum_{\boldn} n_i^2 n_j \, P (\boldn)
\Big\}
\nonumber
\\
& + & 
d
\sum_{j=2}^{K}
\Big\{
\sum_{\boldn} n_i^2 (n_j+1)
 \, P (L_j \boldn)
- 
\sum_{\boldn} n_i^2 n_j \, P (\boldn)
\Big\}.
\label{cmediffpom}
\end{eqnarray}
Let us assume that $i=2, \dots, K-1$. Let us consider the term corresponding 
to $j=i$ in the first sum on the right hand side. We get
$$
\sum_{\boldn} n_i^2 (n_i+1)
 \, P (R_i \boldn)
- 
\sum_{\boldn} n_i^2 n_i \, P (\boldn)
=
\sum_{\boldn} (n_i-1)^2 n_i
 \, P (\boldn)
- 
\sum_{\boldn} n_i^2 n_i \, P (\boldn)
$$
$$
=
\sum_{\boldn} (-2 n_i^2 + n_i) 
\, P (\boldn)
=
- 2 V_{i} - 2 M_i^2 + M_i.
$$
First, we changed indices in the first sum $R_i \boldn \to \boldn$
and then we used definitions (\ref{Midef}) and (\ref{Videf}). 
Similarly, the term corresponding to $j=i-1$ in the first sum on 
the right hand side of (\ref{cmediffpom}) can be rewritten as
$$
\sum_{\boldn} n_i^2 (n_{i-1}+1)
 \, P (R_{i-1} \boldn)
- 
\sum_{\boldn} n_i^2 n_{i-1} \, P (\boldn)
=
\sum_{\boldn} (2 n_i n_{i-1} + n_{i-1})
 \, P (\boldn)
$$
$$
=
2 V_{i,i-1} + 2 M_i M_{i-1} + M_{i-1}.
$$
Other terms (corresponding to $j \ne i, i-1$) in the first sum
on the right hand side of (\ref{cmediffpom}) are equal to zero.
The second sum can be handled analogously. We obtain
\begin{eqnarray}
\frac{\partial}{\partial t} 
\sum_{\boldn} n_i^2 P(\boldn)
& = & 
d
\Big\{
2 V_{i,i-1} + 2 M_i M_{i-1} + M_{i-1}
- 2 V_{i} - 2 M_i^2 + M_i
\Big\}
\nonumber
\\
& + & 
d
\Big\{
2 V_{i,i+1} + 2 M_i M_{i+1} + M_{i+1}
- 2 V_{i} - 2 M_i^2 + M_i
\Big\}.
\label{cmediffpom2}
\end{eqnarray}
Using (\ref{Videf}) and (\ref{meaneq1})
on the left hand side of (\ref{cmediffpom2}), we obtain
$$
\frac{\partial}{\partial t} 
\sum_{\boldn} n_i^2 P(\boldn)
=
\frac{\partial V_i}{\partial t} 
+ 
2
M_i
\frac{\partial M_i}{\partial t} 
= 
\frac{\partial V_i}{\partial t} 
+ 
d (2 M_i M_{i+1} + 2 M_i M_{i-1} - 4 M_i^2).
$$
Substituting this into (\ref{cmediffpom2}), we get
\begin{eqnarray}
\frac{\partial V_i}{\partial t} 
= 
2 d
\Big\{V_{i,i+1} + V_{i,i-1}  - 2 V_{i} \Big\}
+ 
d \Big\{ M_{i+1}+ M_{i-1} + 2 M_i \Big\}
\label{Viequation}
\end{eqnarray}
for $i=2, \dots, K-1$. Similarly, we get
\begin{eqnarray}
\frac{\partial V_1}{\partial t} 
= 
2 d
\Big\{V_{1,2}  - V_{1} \Big\}
+ 
d \Big\{ M_2 + M_1 \Big\},
\label{Viequation1Ka}
\end{eqnarray}
\begin{eqnarray}
\frac{\partial V_K}{\partial t} 
= 
2 d
\Big\{V_{K,K-1}  - V_{K} \Big\}
+ 
d \Big\{M_{K-1} + M_K \Big\}.
\label{Viequation1Kb}
\end{eqnarray}
We see that the evolution equation for the variance vector $\boldV(t)$
depends on the mean $\boldM$, variance $\boldV$ and on non-diagonal 
terms of the matrix $V_{i,j}$. To get a closed system of equations, we have
to derive evolution equations for $V_{i,j}$ too. This can be done 
by multiplying (\ref{cmediff}) by $n_i n_j$, summing over 
$\boldn$ and following the same arguments as before. We conclude
this section with some consequences of (\ref{meaneq1})--(\ref{meaneq2})
and (\ref{Viequation})--(\ref{Viequation1Kb}). Looking at the steady
states of equations (\ref{meaneq1})--(\ref{meaneq2}), we obtain
$M_{i} = N/K$, $i=1,2, \dots, K$, where $N$ is the total number 
of diffusing molecules. Moreover, the variance equations imply that 
$V_{i} = N/K$, $i=1,2, \dots, K$, at the steady state. 

\section{Stochastic reaction-diffusion models}

\label{secRD}

In this section, we add chemical reactions to both models of 
molecular diffusion which were presented in Section 
\ref{secdiffusion}. We introduce two methods for the stochastic
modelling of reaction-diffusion processes. The first one is 
based on the diffusion model from Section \ref{secdiffGillespie}, 
the second one on the diffusion model from Section 
\ref{secdiffSmoluchowski}. We explain both methods using the
same example. Namely, we consider molecules (e.g. protein) 
which diffuse in the domain $[0,L]$ with diffusion
constant $D$ as we considered in Section  \ref{secdiffusion}. Moreover,
we assume that protein molecules are degraded (in the whole domain)
and produced in part of the domain, i.e. we consider 
the chemical reactions from Sections \ref{secdegradation} and 
\ref{secproddegr} in our illustrative reaction-diffusion
model. The model has a realistic motivation which is discussed 
in more detail later in Section \ref{secpattern}. 
In Section \ref{secRDnonlin}, we present another illustrative 
example of a reaction-diffusion process incorporating the nonlinear 
model (\ref{nonlinearmodel1})--(\ref{nonlinearmodel2}).

\subsection{Compartment-based reaction-diffusion SSA}

\label{seccompartmentRD}

We consider molecules of chemical species $A$ which are 
diffusing in the domain $[0,L]$, where $L=1$ mm, with diffusion constant 
$D=10^{-4} \; \mbox{mm}^2 \, \mbox{sec}^{-1}$. Initially,
there are no molecules in the system. Molecules are produced
in the part of the domain $[0,L/5]$ with rate
$k_p = 0.012 \; \mbox{$\mu$m}^{-1} \, \mbox{sec}^{-1}$.
This means that the probability that a molecule is created 
in the subinterval of the length 1 $\mu$m is equal
to $k_p \, \dt$. Consequently, the probability that a molecule 
is created somewhere in the interval $[0,L/5]$ is equal to 
$k_p L/5 \, \dt$. Molecules are degraded with rate 
$k_1 = 10^{-3} \, \mbox{sec}^{-1}$
according to the chemical reaction (\ref{degradation}).

Following Section \ref{secdiffGillespie}, we divide the computational 
domain $[0,L]$ into $K=40$ compartments of length $h = L/K=25 \,\mu$m.
We denote the number of molecules of chemical species $A$ 
in the $i$-th compartment $[(i-1)h,ih)$ by $A_i$, 
$i=1,\dots,K$. Then our reaction-diffusion process
is described by the system of chemical reactions
\begin{equation}
A_1
\;
\mathop{\stackrel{\displaystyle\longrightarrow}\longleftarrow}^{d}_{d}
\;
A_2
\;
\mathop{\stackrel{\displaystyle\longrightarrow}\longleftarrow}^{d}_{d}
\;
A_3
\;
\mathop{\stackrel{\displaystyle\longrightarrow}\longleftarrow}^{d}_{d}
\;
\dots
\;
\mathop{\stackrel{\displaystyle\longrightarrow}\longleftarrow}^{d}_{d}
\;
A_K,
\label{diffGillRD}
\end{equation}
\begin{equation}
A_i \; \mathop{\longrightarrow}^{k_1} \;\, \emptyset,
\qquad \mbox{for} \; i = 1, 2, \dots, K,
\label{degradationRD}
\end{equation}
\begin{equation}
\emptyset \; \mathop{\longrightarrow}^{k_2} \;\, A_i,
\qquad \mbox{for} \; i = 1, 2, \dots, K/5.
\label{creationRD}
\end{equation}
Equation (\ref{diffGillRD}) describes diffusion and is identical
to (\ref{diffGill}). In particular, the rate constant $d$ is given by
$d = D/h^2.$ Equation (\ref{degradationRD}) describes the degradation
of $A$ and is, in fact, equation (\ref{degradation}) applied to every 
compartment. Equation (\ref{creationRD}) describes the production
of $A$ in the first $K/5$ compartments (e.g. in part $[0,L/5]$ of
the computational domain). The rate constant $k_2$ describes the
rate of production per compartment. Since each compartment has length
$h$, we have $k_2 = k_p h$. 

\begin{figure}
\picturesAB{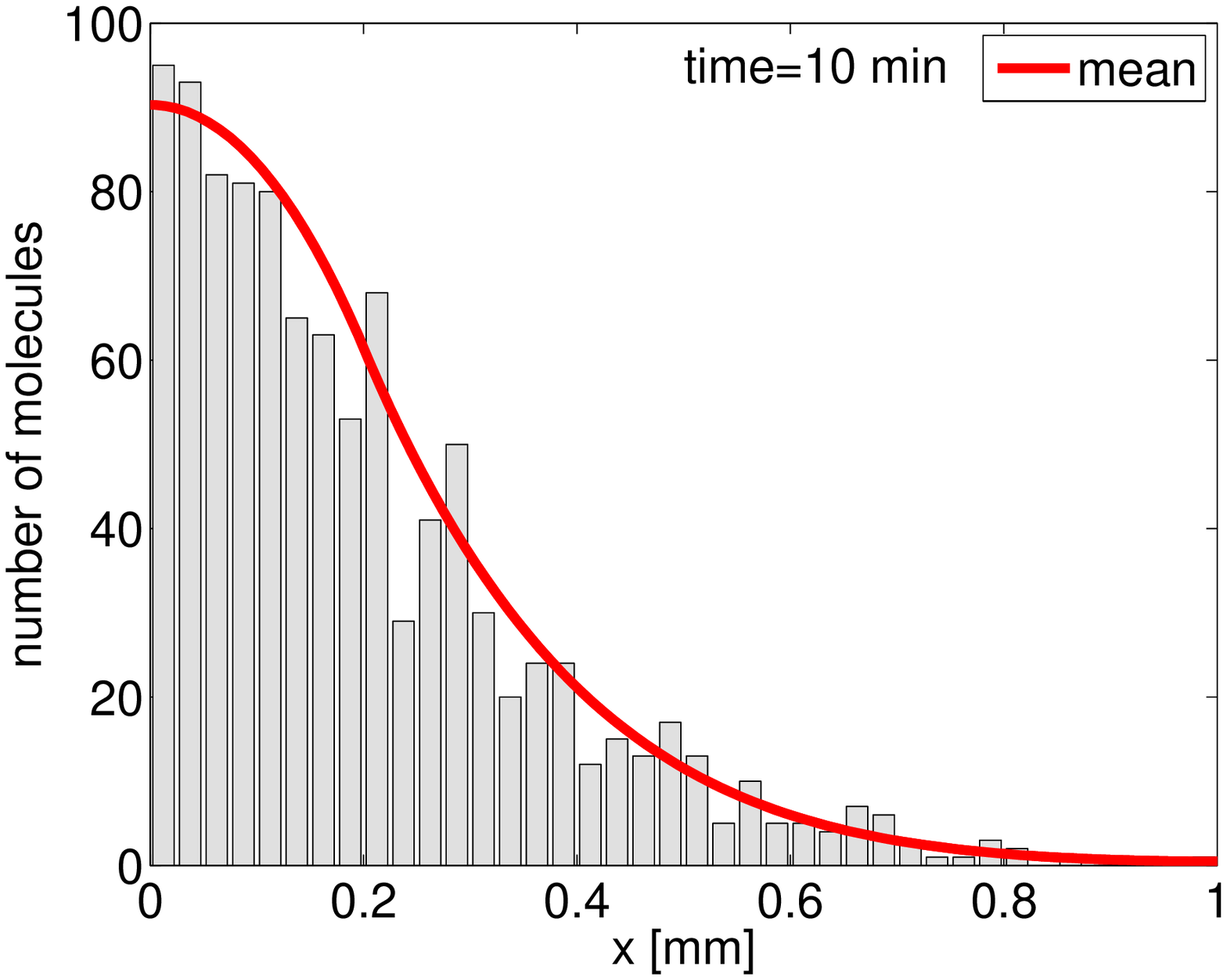}{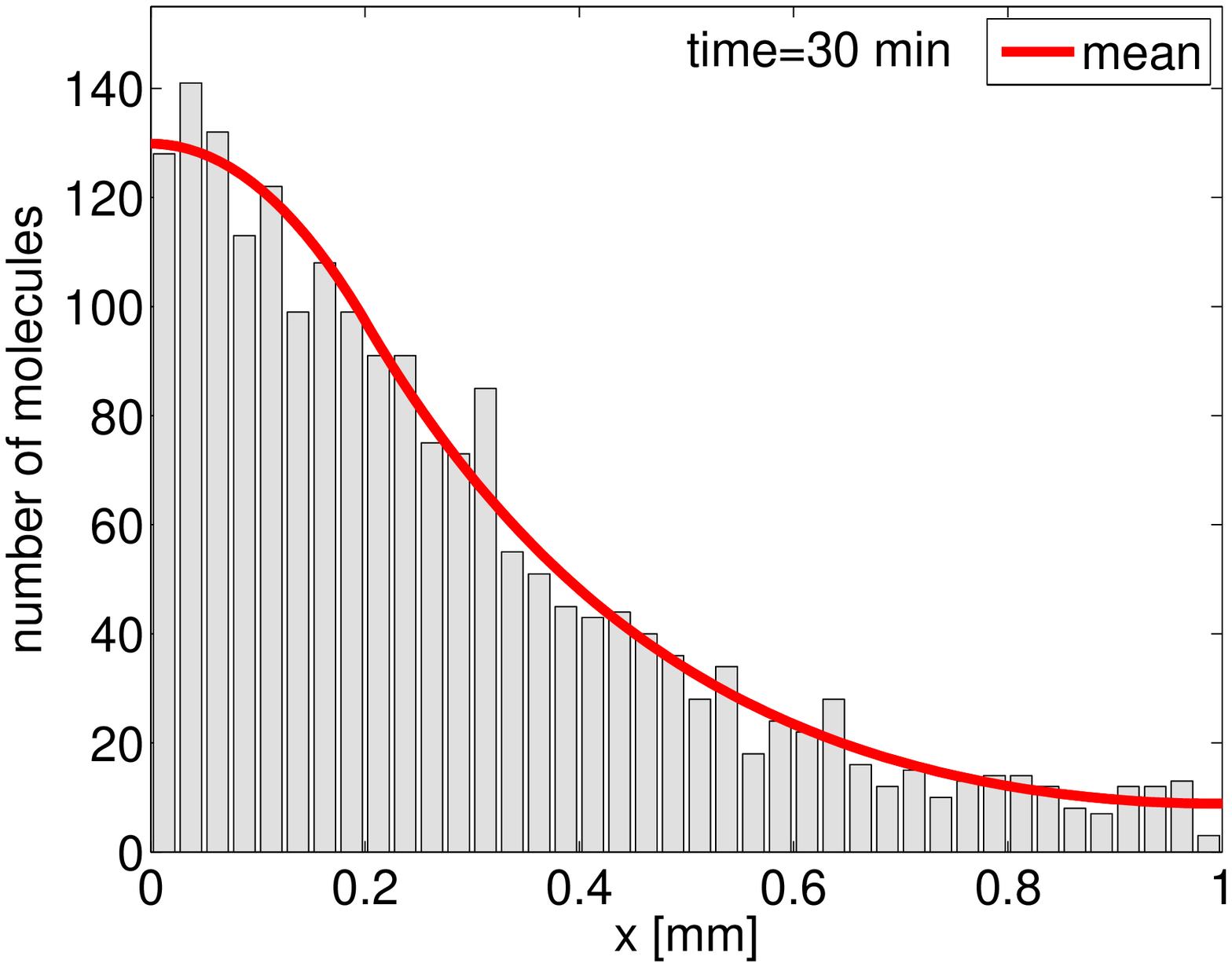}{2in}{5mm}
\caption{One realization of the Gillespie SSA (a5)--(d5)
for the system of chemical reactions 
$(\ref{diffGillRD})$--$(\ref{creationRD})$. Gray histograms
show numbers of molecules in compartments at
time:  {\rm (a)} $t=10$ min; {\rm (b)} $t=30$ min.
Solution of $(\ref{rdPrDeDi})$--$(\ref{boPrDeDi})$ is plotted
as the red solid line.}
\label{figrdgill}
\end{figure}

The system of chemical reactions (\ref{diffGillRD})--(\ref{creationRD})
is simulated using the Gillespie SSA (a5)--(d5). In our case, the
propensity functions of reactions in (\ref{diffGillRD})
are given as $A_i(t) d$, the propensity functions of reactions
in (\ref{degradationRD}) are given as $A_i(t) k_1$ and
propensity functions of reactions in (\ref{creationRD})
are equal to $k_2.$ Starting with no molecules of $A$ in the 
system, we compute one realization of SSA (a5)--(d5)
for the system of reactions (\ref{diffGillRD})--(\ref{creationRD}).
We plot the numbers of molecules in compartments at two different
times in Figure \ref{figrdgill}.

Let $p(\boldn,t)$ be the joint probability that $A_i(t)=n_i$,
$i=1, \dots, K$, where we use the notation $\boldn=[n_1,n_2, \dots, n_K].$
Let us define operators $R_i, L_i: {\mathbb N}^K \to {\mathbb N}^K$ by
(\ref{Ridef})--(\ref{Lidef}). Then the chemical master equation,
which corresponds to the system of chemical reactions 
(\ref{diffGillRD})--(\ref{creationRD}), can be written as follows
\begin{eqnarray}
\frac{\partial p(\boldn)}{\partial t} 
& = & 
d
\sum_{i=1}^{K-1}
\Big\{
(n_i+1)
 \, p (R_i \boldn)
- 
n_i \, p (\boldn)
\Big\}
+ 
d
\sum_{i=2}^{K}
\Big\{
(n_i+1)
 \, p (L_i \boldn)
- 
n_i \, p (\boldn)
\Big\} 
\nonumber
\\
& + &
k_1
\sum_{i=1}^{K}
\Big\{
(n_i+1)
 \, p (n_1,\dots,n_i+1,\dots,n_K)
- 
n_i \, p (\boldn)
\Big\}
\nonumber
\\
& + &
k_2
\sum_{i=1}^{K/5}
\Big\{
p (n_1,\dots,n_i-1,\dots,n_K)
- 
p (\boldn)
\Big\}.
\label{cmeRD}
\end{eqnarray}
The first two sums correspond to diffusion (\ref{diffGillRD}),
the third sum to degradation (\ref{degradationRD})
and the fourth sum to production (\ref{creationRD}).
The stochastic mean is defined as the vector $\boldM(t)
\equiv [M_1,M_2, \dots, M_K]$ where $M_i$
is given by (\ref{Midef}). This gives the mean number 
of molecules in the $i$-th compartment, $i=1,2, \dots, K$, 
at time $t$ (averaged over many realizations of SSA 
(a5)--(d5)). To derive the evolution equation for
the stochastic mean vector $\boldM(t)$, we can follow the 
method from Section \ref{secproddegr} -- see derivation 
of (\ref{evolM}) from the chemical master equation
(\ref{cmeproddegr}). Multiplying (\ref{cmeRD}) by $n_i$ 
and summing over all $n_j$, $j=1, \dots, K$, we obtain
(leaving the details to the student) a system of equations for 
$M_i$ in the form
\begin{eqnarray}
\frac{\partial M_1}{\partial t} 
& = &
d (M_{2} - M_1) + k_2 - k_1 M_1, 
\label{meaneq1RD}
\\
\frac{\partial M_i}{\partial t} 
& = & 
d (M_{i+1} + M_{i-1} - 2 M_i) + k_2 - k_1 M_i,
\qquad
i = 2, \dots, K/5,
\label{meaneq2RD}
\\
\frac{\partial M_i}{\partial t} 
& = & 
d (M_{i+1} + M_{i-1} - 2 M_i) - k_1 M_i,
\qquad
i = K/5+1, \dots, K-1,
\label{meaneq3RD}
\\
\frac{\partial M_K}{\partial t} 
& = & 
d (M_{K-1} - M_K) - k_1 M_K.
\label{meaneq4RD}
\end{eqnarray}
System (\ref{meaneq1RD})--(\ref{meaneq4RD}) is a discretized
version of the reaction-diffusion equation
\begin{equation}
\frac{\partial a}{\partial t}
=
D \, \frac{\partial^2 a}{\partial x^2}
+ k_2 \chi_{[0,L/5]}- k_1 a
\label{rdPrDeDi}
\end{equation}
with zero-flux boundary conditions
\begin{equation}
\frac{\partial a}{\partial x}(0)
=
\frac{\partial a}{\partial x}(L)
=
0.
\label{boPrDeDi}
\end{equation}
Here, $\chi_{[0,L/5]}$ is the characteristic function of the interval
$[0,L/5]$. Using initial condition $a(\cdot,0) \equiv 0$,
we computed the solution of (\ref{rdPrDeDi})--(\ref{boPrDeDi}) 
numerically. It is plotted as a red solid line in Figure 
\ref{figrdgill} for comparison. 

The concentration of molecules
in the $i$-th compartment is defined as $\overline{M}_i=M_i/h$,
$i = 1, \dots, K$. Dividing (\ref{meaneq1RD})--(\ref{meaneq4RD}) 
by $h$, we can write a system of ODEs for $\overline{M}_i$.
It is a discretized version of the reaction-diffusion equation
\begin{equation}
\frac{\partial \overline{a}}{\partial t}
=
D \, \frac{\partial^2 \overline{a}}{\partial x^2}
+ k_p \chi_{[0,L/5]}- k_1 \overline{a}
\label{concentrationEq}
\end{equation}
where $\overline{a} \equiv \overline{a}(x,t)$ is the
concentration of molecules of $A$ at point $x$ and time
$t.$ The equation (\ref{concentrationEq}) provides 
a classical deterministic description of the 
reaction-diffusion process. Its parameters $D$, $k_p$
and $k_1$ are independent of $h$. Solving (\ref{concentrationEq})
is equivalent to solving (\ref{rdPrDeDi}). Consequently,
the red solid line in Figure \ref{figrdgill} can
be also viewed as a plot of $\overline{a} h$ where
$\overline{a}$ is the solution of the classical deterministic
model (\ref{concentrationEq}) with the zero-flux boundary conditions.

\subsection{Reaction-diffusion SSA based on the Smoluchowski equation}

\label{secRDsmol}

In this section, we present a SSA which implements the Smoluchowski
model of diffusion from Section \ref{secdiffSmoluchowski},
that is, we follow the trajectories of individual molecules.
Diffusion of each molecule is modelled according to the
model (a7)--(c7). We explain the SSA using the reaction-diffusion
example from Section \ref{seccompartmentRD}. Choosing a small time 
step $\Delta t$, we perform the following three steps at time
$t$:

\leftskip 1.4cm

\medskip

{ 
\parindent -8.4mm
 
{\bf (a9)} For each molecule, compute its position at time $t+\Delta t$
according to steps (a7)--(c7). 

{\bf (b9)} For each molecule, generate a random number $r_1$ uniformly 
distributed in the interval $(0,1)$. If $r_1 < k_1 \, \Delta t$, then remove 
the molecule from the system.

{\bf (c9)} Generate a random number $r_2$ uniformly distributed 
in the interval $(0,1)$. If $r_2 < k_p L/5 \, \Delta t$, then generate 
another random number $r_3$ uniformly distributed 
in the interval $(0,1)$ and introduce a new molecule at position 
$r_3 L/5.$ \\
Continue with step (a9) for time $t+\Delta t.$

}

\leftskip 0cm

\medskip

\begin{figure}
\picturesAB{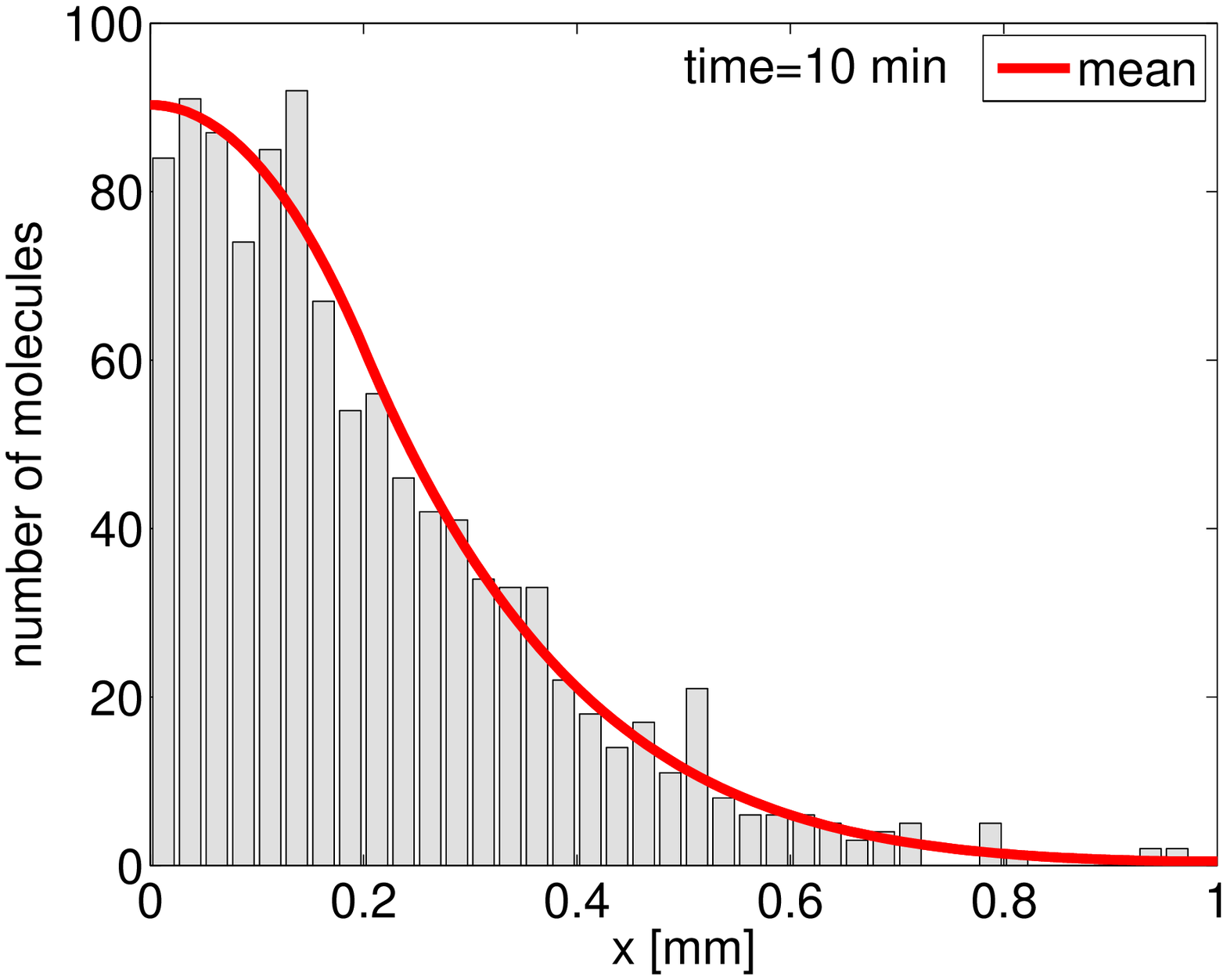}{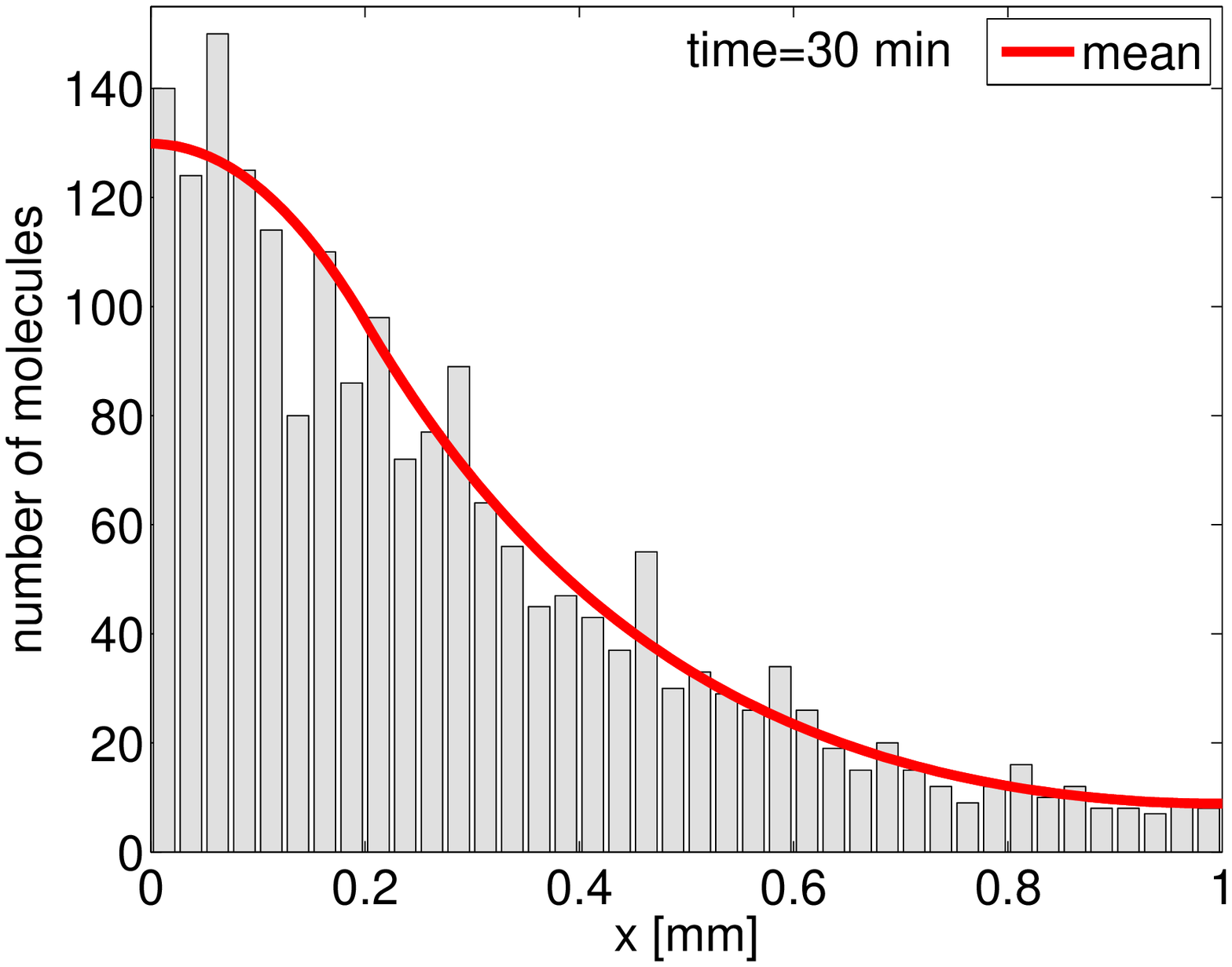}{2in}{5mm}
\caption{One realization of SSA (a9)--(c9).
Dividing domain $[0,L]$ into 40 bins, we plot the number of molecules
in each bin at time: {\rm (a)} $t=10$ min; {\rm (b)} $t=30$ min. 
Solution of $(\ref{rdPrDeDi})$--$(\ref{boPrDeDi})$ is plotted
as the red solid line.}
\label{figrdsmol}
\end{figure}

\noindent
The degradation of molecules is modelled by step (b9). 
Equation (\ref{degradation}) implies that 
$k_1 \, \dt$ is the probability that a molecule is degraded 
in the time interval $[t,t+\dt)$ for infinitesimally small $\dt$. 
SSA (a9)--(c9) replaces $\dt$ by the finite time step
$\Delta t$ (compare with SSA (a1)--(b1)) which has to be chosen 
sufficiently small so that $k_1 \, \Delta t \ll 1.$ Similarly, 
the probability
that a molecule is created in $[0,L/5]$ in time interval $[t,t+\dt)$
is equal to $k_p L/5 \, \dt$. Consequently, we have to 
choose $\Delta t$ so small that $k_p L/5 \, \Delta t$ is 
significantly less than 1. We choose 
$\Delta t = 10^{-2} \, \mbox{sec}$. Then 
$k_1 \, \Delta t = 10^{-5}$ and 
$k_p L/5 \, \Delta t = 2.4 \times 10^{-2}$
for our parameter values $k_1= 10^{-3} \, \mbox{sec}^{-1}$,
$k_p = 0.012 \; \mbox{$\mu$m}^{-1} \, \mbox{sec}^{-1}$ and $L=1$ 
mm. Starting with no molecules of $A$ in the 
system, we compute one realization of SSA (a9)--(c9).
To visualise the results, we divide the interval $[0,L]$ into
40 bins and we plot the numbers of molecules in bins 
at time 10 minutes in Figure \ref{figrdsmol}(a). The same
plot at time 30 minutes is given in Figure \ref{figrdsmol}(b).
We used the same number of bins to visualise the results
of SSA (a9)--(c9) as we used previously in the compartment-based
model. Thus Figure \ref{figrdsmol} is directly comparable
with Figure \ref{figrdgill}. We also plot the solution of 
$(\ref{rdPrDeDi})$--$(\ref{boPrDeDi})$ as a red solid line
for comparison.

\subsection{Reaction-diffusion models of nonlinear chemical kinetics}

\label{secRDnonlin}
 
In the previous sections, we studied an example of a reaction-diffusion
model which did not include the second-order chemical reactions 
(\ref{nonlinearmodelintro}). We considered only production and 
degradation, i.e. we considered chemical reactions from Sections 
\ref{secdegradation} and \ref{secproddegr}. In this section, we discuss 
generalisations of our approaches to models which involve second-order 
chemical reactions too. Our illustrative example is a reaction-diffusion 
process incorporating the nonlinear model 
(\ref{nonlinearmodel1})--(\ref{nonlinearmodel2}).
The second-order chemical reactions (\ref{nonlinearmodelintro})
require that two molecules collide (be close to each other) 
before the reaction can take place. The generalisation of 
SSA (a9)--(c9) to such a case is nontrivial and we will not
present it in this paper (it can be found in 
\cite{Andrews:2004:SSC}). Application of the Gillespie SSA
(a5)--(d5) is more straightforward and is presented below.

\begin{figure}
\picturesAB{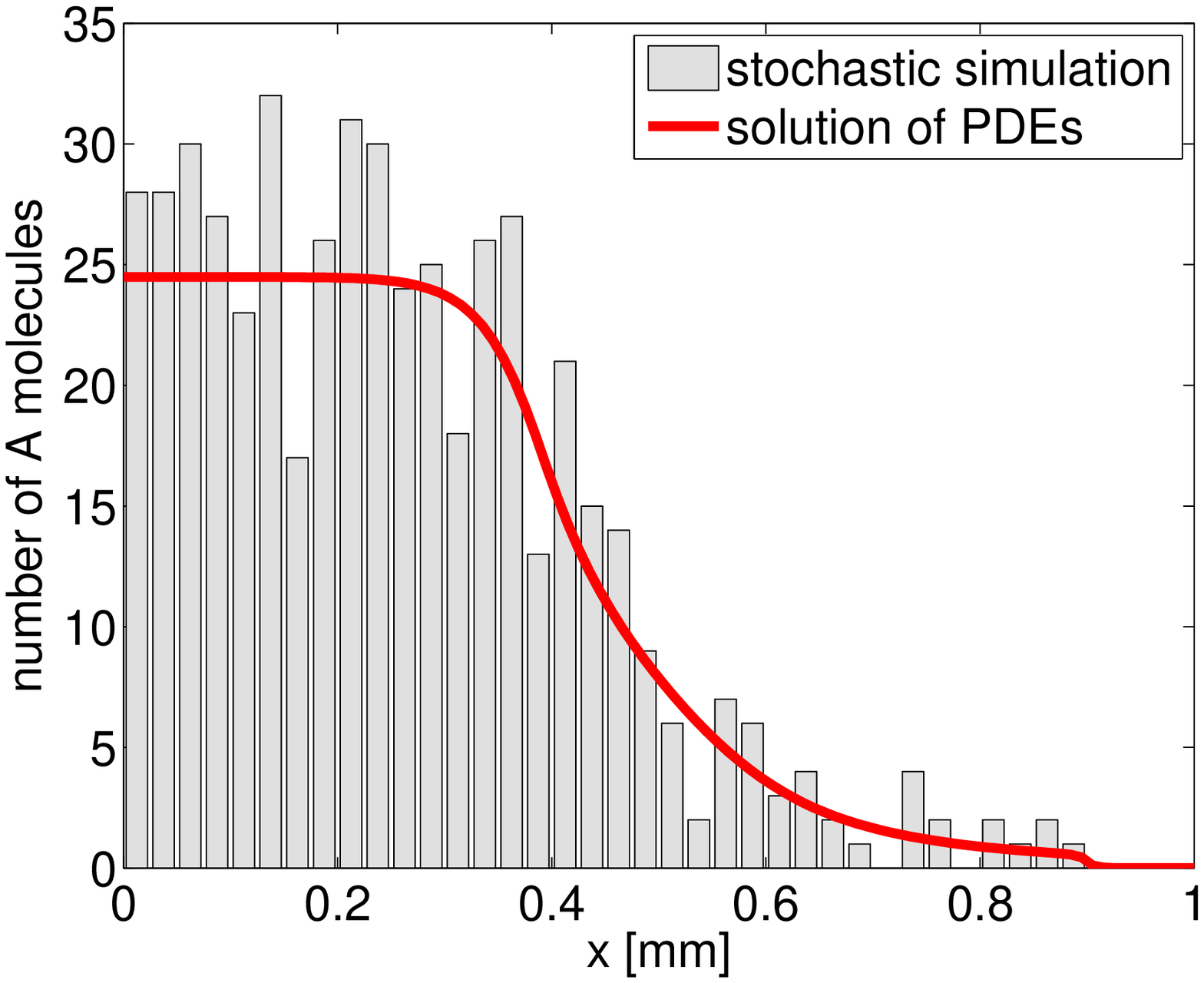}{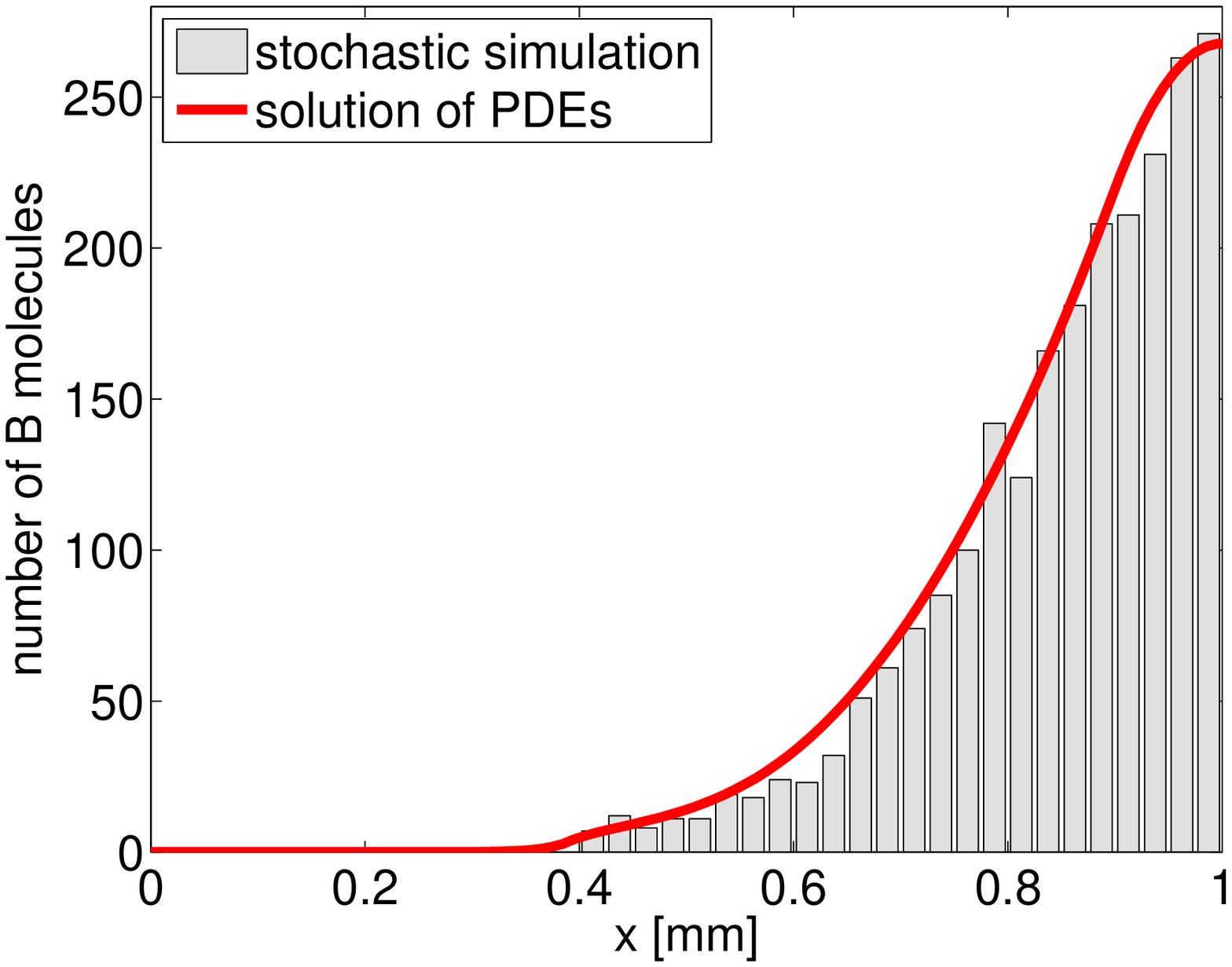}{2in}{5mm}
\caption{One realization of the Gillespie SSA (a5)--(d5) 
for the system of chemical reactions
$(\ref{diffGillRDnonlinA})$--$(\ref{prodBi})$. Numbers of molecules
of chemical species $A$ (left panel) and $B$ (right panel)
in compartments at time $30$ minutes (gray histograms).
Solution of $(\ref{PdeA})$--$(\ref{boPdeAB})$ is plotted as
the red solid line.}
\label{figrdnonlin}
\end{figure}

We consider that both chemical species $A$ and $B$ diffuse 
in the domain $[0,L]$, where $L=1$ mm, with diffusion constant 
$D=10^{-4} \; \mbox{mm}^2 \, \mbox{sec}^{-1}$. Following
the method of Section \ref{seccompartmentRD}, we divide 
the computational domain $[0,L]$ into $K=40$ compartments 
of length $h = L/K=25 \,\mu$m. We denote the number of molecules 
of chemical species $A$ (resp. $B$) in the $i$-th compartment 
$[(i-1)h,ih)$ by $A_i$ (resp. $B_i$), $i=1,\dots,K$.
Diffusion corresponds to two chains of ``chemical reactions":
\begin{equation}
A_1
\;
\mathop{\stackrel{\displaystyle\longrightarrow}\longleftarrow}^{d}_{d}
\;
A_2
\;
\mathop{\stackrel{\displaystyle\longrightarrow}\longleftarrow}^{d}_{d}
\;
A_3
\;
\mathop{\stackrel{\displaystyle\longrightarrow}\longleftarrow}^{d}_{d}
\;
\dots
\;
\mathop{\stackrel{\displaystyle\longrightarrow}\longleftarrow}^{d}_{d}
\;
A_K
\label{diffGillRDnonlinA}
\end{equation}
\begin{equation}
B_1
\;
\mathop{\stackrel{\displaystyle\longrightarrow}\longleftarrow}^{d}_{d}
\;
B_2
\;
\mathop{\stackrel{\displaystyle\longrightarrow}\longleftarrow}^{d}_{d}
\;
B_3
\;
\mathop{\stackrel{\displaystyle\longrightarrow}\longleftarrow}^{d}_{d}
\;
\dots
\;
\mathop{\stackrel{\displaystyle\longrightarrow}\longleftarrow}^{d}_{d}
\;
B_K
\label{diffGillRDnonlinB}
\end{equation}
Molecules of $A$ and $B$ are assumed to react according to chemical 
reactions (\ref{nonlinearmodel1}) in the whole domain with
rate constants  $k_1= 10^{-3} \, \mbox{sec}^{-1}$ and 
$k_2= 10^{-2} \, \mbox{sec}^{-1}$ per one compartment, that is,
\begin{equation}
A_i + A_i \; \mathop{\longrightarrow}^{k_1} \;\, \emptyset,
\qquad \qquad
A_i + B_i \; \mathop{\longrightarrow}^{k_2} \;\, \emptyset,
\qquad
\mbox{for} \; i=1,2,\dots,K.
\label{nonlinearmodel1RD}
\end{equation}
Production of chemical species (\ref{nonlinearmodel2}) is
assumed to take place only in parts of the computational
domain $[0,L]$. Molecules of chemical species $A$ (resp. $B$) 
are assumed to be produced in subinterval $[0,9L/10]$ 
(resp. $[2L/5,L]$) with rate $k_{3} = 1.2 \; \mbox{sec}^{-1}$
(resp. $k_{4} = 1 \; \mbox{sec}^{-1}$) per one compartment
of length $h$, that is,
\begin{equation}
\emptyset \; \mathop{\longrightarrow}^{k_3} \;\, A_i,
\qquad \mbox{for} \; i = 1,2, \dots, 9K/10,
\label{prodAi}
\end{equation}
\begin{equation}
\emptyset \; \mathop{\longrightarrow}^{k_4} \;\, B_i,
\qquad \mbox{for} \; i = 2K/5, \dots, K.
\label{prodBi}
\end{equation}
Starting with no molecules in the system at time $t=0$,
we present one realization of the Gillespie SSA 
(a5)--(d5) applied to the chemical system 
(\ref{diffGillRDnonlinA})--(\ref{prodBi}) in Figure
\ref{figrdnonlin}. We plot the numbers of molecules of
$A$ and $B$ at time 30 minutes.

We already observed in Section \ref{secnonlin} that the analysis
of the master equation for chemical systems involving the second
order reactions is not trivial. It is not possible to derive
the equation for stochastic means as was done in Section 
\ref{seccompartmentRD} for the linear model. Hence, we will
not attempt such an approach here. We also observed
in Section \ref{seccompartmentRD} that the equation for the mean
vector (\ref{meaneq1RD})--(\ref{meaneq4RD}) was actually
equal to a discretized version of the reaction-diffusion equation
(\ref{rdPrDeDi})--(\ref{boPrDeDi}) which would be used
as a traditional deterministic description. When considering the
nonlinear chemical model (\ref{nonlinearmodel1RD})--(\ref{prodBi}),
we cannot derive the equation for the mean vector but we can still
write a deterministic system of partial differential equations
(PDEs).
We simply add diffusion to the system of ODEs (\ref{OdeA})--(\ref{OdeB})
to obtain
\begin{eqnarray}
\frac{\partial a}{\partial t}
& = & 
D \, \frac{\partial^2 a}{\partial x^2}
- 2 k_1 a^2 - k_2 \, a b + k_3 \chi_{[0,9L/10]}, 
\label{PdeA}
\\
\frac{\partial b}{\partial t}
& = & 
D \, \frac{\partial^2 b}{\partial x^2}
 - k_2 \, a b + k_4 \chi_{[2L/5,L]},
\label{PdeB}
\end{eqnarray}
and couple it with zero-flux boundary conditions
\begin{equation}
\frac{\partial a}{\partial x}(0)
=
\frac{\partial a}{\partial x}(L)
=
\frac{\partial b}{\partial x}(0)
=
\frac{\partial b}{\partial x}(L)
=
0.
\label{boPdeAB}
\end{equation}
Using initial condition $a(\cdot,0) \equiv 0$ and
$b(\cdot,0) \equiv 0$, we can compute the solution of 
(\ref{PdeA})--(\ref{boPdeAB}) numerically. It is plotted 
as a red solid line in Figure \ref{figrdnonlin} for comparison. 
We see that (\ref{PdeA})--(\ref{boPdeAB}) gives
a reasonable description of the system when comparing
with one realization of SSA (a5)--(d5). However, let us note
that solution of (\ref{PdeA})--(\ref{boPdeAB}) is not
equal to the stochastic mean. 

The equations (\ref{PdeA})--(\ref{boPdeAB}) can be also rewritten
in terms of concentrations $\overline{a} = a/h$ and $\overline{b} = b/h$ 
as we did in the case of equations (\ref{rdPrDeDi}) and 
(\ref{concentrationEq}). Let us note that the rate constants
scale with $h$ as $k_1=\overline{k_1}/h$, 
$k_2=\overline{k_2}/h$, $k_3=\overline{k_3} h$, $k_4=\overline{k_4} h$
where $\overline{k_1},$ $\overline{k_2},$ $\overline{k_3},$
$\overline{k_4}$ are independent of $h$. Consequently, the 
equations for concentrations $\overline{a}$ and $\overline{b}$ 
are independent of $h$. They can be written in terms of the
parameters $D$, $\overline{k_1}$, $\overline{k_2},$ 
$\overline{k_3}$ and $\overline{k_4}$ only (compare
with (\ref{concentrationEq})).

Finally, let us discuss the choice of the compartment length $h$. 
In Sections \ref{secdiffGillespie} and \ref{seccompartmentRD},
we considered linear models and we were able to derive
the equations for the mean vectors (e.g. (\ref{meaneq1RD})--(\ref{meaneq4RD})).
Dividing (\ref{meaneq1RD})--(\ref{meaneq4RD}) by $h$
and passing to the limit $h \to 0$,
we derive the corresponding deterministic reaction-diffusion 
PDE (\ref{concentrationEq}) which can be viewed
(for linear models) as an equation for the probability distribution
function of a single molecule (i.e. the exact description which
we want to approximate by the compartment-based SSA). 
Consequently, we can increase the accuracy of the SSA 
by decreasing $h$. Considering the nonlinear model from this 
section, the continuum limit $h \to 0$ is not well-defined. The 
compartment-based SSA is generally considered valid only 
for a range of $h$ values (i.e. the length 
$h$ cannot be chosen arbitrarily small); conditions 
which the length $h$ has to satisfy are subject of current 
research  -- see e.g. \cite[Section 3.5]{Isaacson:2006:IDC}.

\section{Two important remarks}

\label{secextra}

We explained SSAs for chemical reactions, molecular diffusion
and reaction-diffusion processes in the previous sections. 
This final section is devoted to two important questions:

(a) Why do we care about stochastic modelling? The answer
is given in Section \ref{secssr} where we discuss connections 
between stochastic and deterministic modelling. In particular, 
we present examples where deterministic modelling fails 
and a stochastic approach is necessary. We start with 
a simple example of stochastic switching between favourable
states of the system, a phenomenon which cannot be fully
understood without stochastic modelling. Then we illustrate 
the fact that the stochastic model might have qualitatively 
different properties than its  deterministic limit, i.e. 
the stochastic model is not just ``equal" to the ``noisy 
solution" of  the corresponding deterministic equations. 
We present a simple system of chemical reactions for
which the deterministic description converges to 
a steady state. On the other hand, the stochastic model 
of the same system of chemical reactions has oscillatory 
solutions. Finally, let us note that stochasticity plays 
important roles in biological applications, see e.g. 
\cite{Rao:2002:CET,Elowitz:2002:SGE,Paulsson:2000:SFF}.

(b) Why do we care about reaction-diffusion processes?
The answer is given in Section \ref{secpattern} where we
discuss biological pattern formation. Reaction-diffusion 
models are key components of models in developmental biology. 
We present stochastic analogues of two classical pattern 
forming models. The first one is the so-called French flag
problem where we re-interpret the illustrative example 
from Sections \ref{seccompartmentRD} and \ref{secRDsmol}.
Then we present the reaction-diffusion pattern forming model 
based on the so-called Turing instability.

\subsection{Deterministic vs. stochastic modelling}

\label{secssr}

The models presented so far have one thing in common. One could use
the deterministic description (given by ODEs or PDEs) and one
would obtain a reasonable description of the system. In Sections
\ref{secdegradation}, \ref{secproddegr}, \ref{secdiffSmoluchowski},
\ref{secdiffGillespie}, 
\ref{seccompartmentRD} and \ref{secRDsmol}, we studied linear
models. We showed that the evolution equations for the stochastic mean 
are equal to (the discretized versions of) the corresponding 
deterministic differential equations. In Sections \ref{secnonlin} and
\ref{secRDnonlin}, we presented nonlinear models.  
We were not able to derive equations for the stochastic mean.
However, we solved numerically the corresponding systems of 
deterministic equations (ODEs (\ref{OdeA})--(\ref{OdeB}) in 
Section \ref{secnonlin} and PDEs (\ref{PdeA})--(\ref{boPdeAB}) 
in Section \ref{secRDnonlin}) and we obtained results
comparable with the SSAs, i.e. results of the SSAs looked like ``noisy
solutions" of the corresponding differential equations. Here,
we discuss examples of problems when SSAs give results which 
cannot be obtained by corresponding deterministic models.

\begin{figure}
\picturesAB{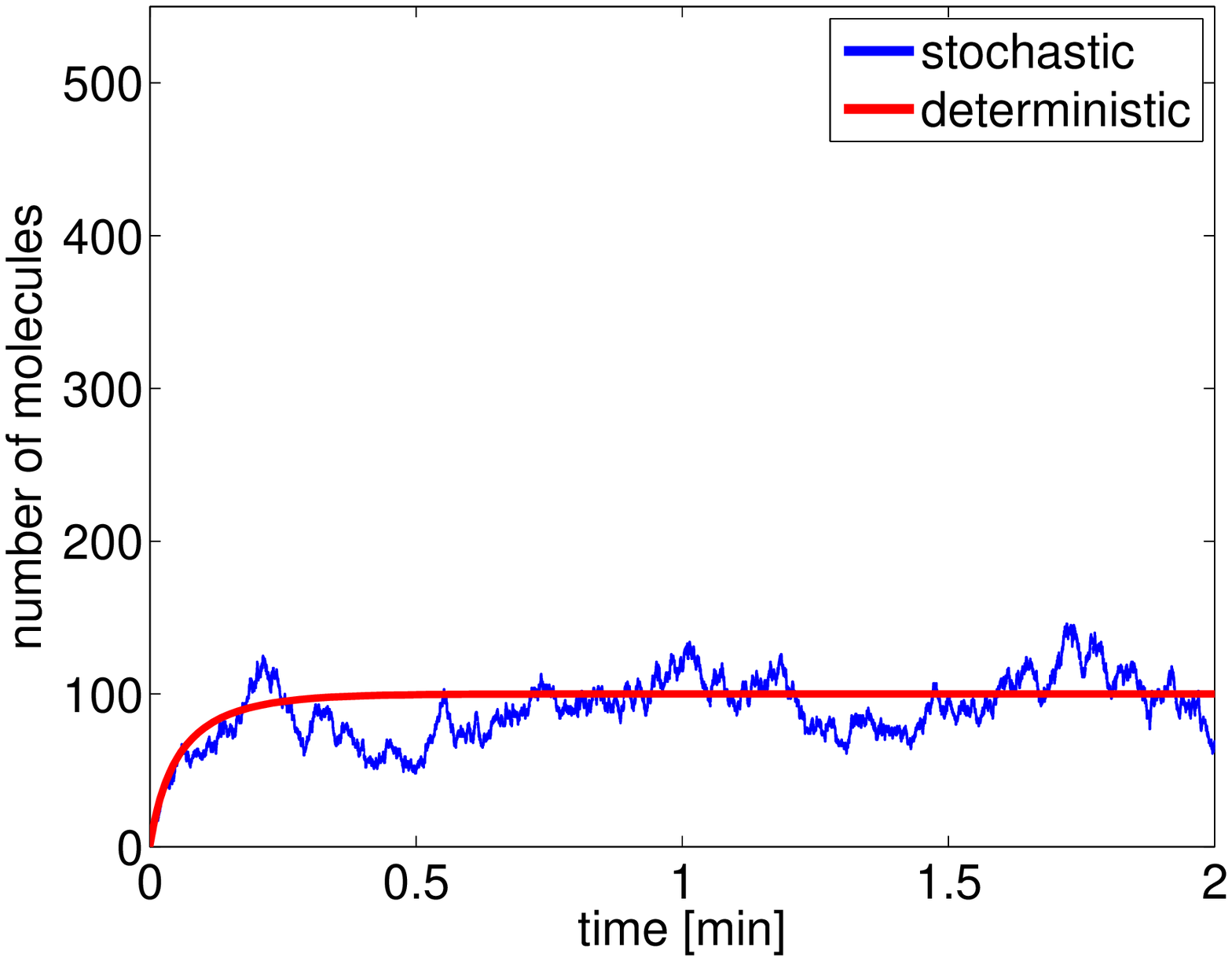}{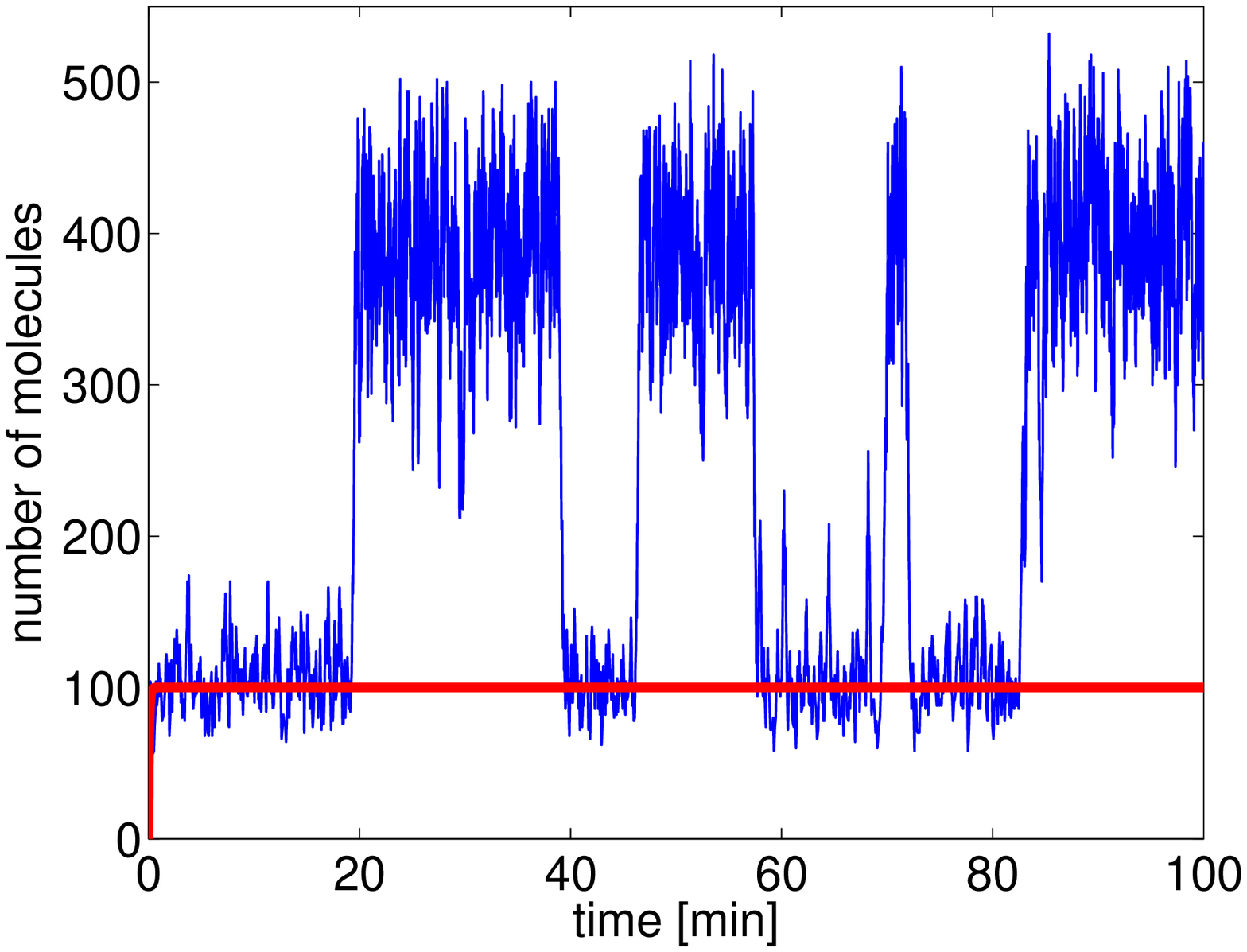}{2in}{5mm}
\caption{Simulation of $(\ref{schlogl})$. One realization 
of SSA (a5)--(d5) for the system of chemical reactions 
$(\ref{schlogl})$ (blue line) 
and the solution of the deterministic ODE $(\ref{odeschlogl})$ 
(red line). {\rm (a)} The number of molecules of $A$ as
a function of time over the first two minutes of simulation.
{\rm (b)} Time evolution over 100 minutes.}
\label{figswitches}
\end{figure}

Let us consider the model from Section \ref{secnonlin}. Its 
deterministic description is given by the system of ODEs 
(\ref{OdeA})--(\ref{OdeB}). Such a system
has only one nonnegative (stable) steady state for our parameter
values, namely $a_s = b_s = 10$. It can be observed from 
Figure \ref{fignonlintime} that solutions of 
(\ref{OdeA})--(\ref{OdeB}) converge to $a_s$ and $b_s$ as
$t \to \infty$. This is true for any nonnegative initial 
condition. The results of SSAs show fluctuation about the 
means, which are roughly equal to $a_s$ and $b_s$ (they are
9.6 for $A$ and 12.2 for $B$). However, there are
chemical systems which have two or more favourable 
states, so that the corresponding ODEs have 
more than one nonnegative stable steady state. 
For example, let us consider the system of chemical reactions
for chemical $A$ introduced by Schl\"ogl \cite{Schlogl:1972:CRM}
\begin{equation}
{\mbox{ \raise 0.851 mm \hbox{$2 A$}}}
\;
\mathop{\stackrel{\displaystyle\longrightarrow}\longleftarrow}^{k_1}_{k_2}
\;
{\mbox{\raise 0.851 mm\hbox{$3 A$,}}}
\qquad\qquad\qquad
{\mbox{ \raise 0.851 mm \hbox{$\emptyset$}}}
\;
\mathop{\stackrel{\displaystyle\longrightarrow}\longleftarrow}^{k_3}_{k_4}
\;
{\mbox{\raise 0.851 mm\hbox{$A$.}}}
\label{schlogl}
\end{equation}
The corresponding ODE is given as follows
\begin{equation}
\frac{\mbox{d}a}{\dt} =  - \, k_2 \, a^3 + k_1 \, a^2 - k_4 \, a + k_3.
\label{odeschlogl}
\end{equation}
We choose the rate constants as follows:
$k_1 = 0.18 \; \mbox{min}^{-1},$ 
$k_2=2.5 \times 10^{-4} \; \mbox{min}^{-1},$ 
$k_3 = 2200 \; \mbox{min}^{-1}$ 
and 
$k_4=37.5 \; \mbox{min}^{-1}.$ 
Then the ODE (\ref{odeschlogl}) has two stable
steady states $a_{s1} = 100$ and $a_{s2} = 400$ and one
unstable steady state $a_u = 220$. The solution of (\ref{odeschlogl})
converges to one of the steady states with the choice of the 
steady state dependent on the initial condition. Let us consider
that there are initially no molecules of $A$ in the system,
i.e. $A(0)=0$. The solution of (\ref{odeschlogl}) is plotted
in Figure \ref{figswitches}(a) as a red line.
We see that the solution of (\ref{odeschlogl}) converges
to the stable steady state $a_{s1} = 100.$ This is true
for any initial condition $A(0) \in [0,a_u)$. If
$A(0)>a_u$, then the solution of (\ref{odeschlogl})
converges to the second stable steady state $a_{s2} = 400.$
Next, we use the Gillespie SSA (a5)--(d5) to simulate
the chemical system (\ref{schlogl}). Starting with no
molecules of $A$ in the system, we plot one realization
of SSA (a5)--(d5) in Figure \ref{figswitches}(a) as a blue line. 
We see that the time evolution of $A$ given by SSA (a5)--(d5) initially
(over the first 2 minutes) looks like the noisy solution 
of (\ref{odeschlogl}). However, we can find significant
differences between the stochastic and deterministic
model if we observe both models over sufficiently
large times -- see Figure \ref{figswitches}(b) where
we plot the time evolution of $A$ over the first 100 minutes.
As expected, the solution of the deterministic model 
(\ref{odeschlogl}) stays forever close
to the stable steady state $a_{s1} = 100$. The number of molecules 
given by the stochastic model initially fluctuates around one of the 
favourable states of the system (which is close to $a_{s1} = 100$). 
However, the fluctuations 
are sometimes so strong that the system spontaneously switches 
to another steady state (which is close to $a_{s2} = 400$).
This random switching is missed by the deterministic description. 
If one wants to find the mean switching time between favourable 
states of the system, then it is necessary to implement SSAs. Random 
switching between states has been found in gene regulatory
networks \cite{Gardner:2000:CGT,Hasty:2002:EGC}. 
Theoretical or computational approaches for the analysis
of suitable stochastic models are given
in \cite{Kepler:2001:STR,Erban:2006:GRN}.

Our next example is a nonlinear system of chemical 
equations for which the stochastic model has qualitatively 
different behaviour than its deterministic counterpart
in some parameter regimes. 
\begin{figure}
\picturesAB{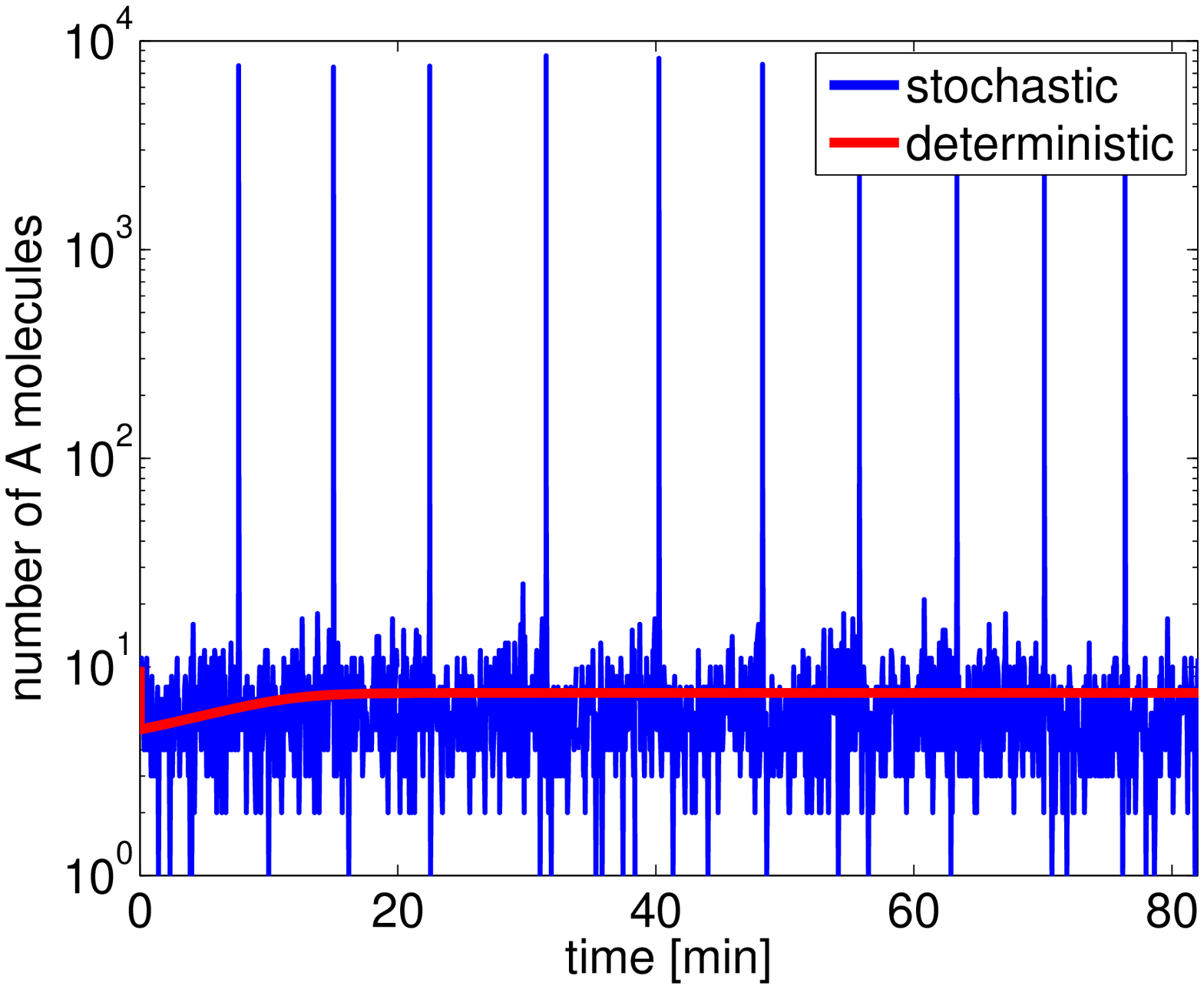}{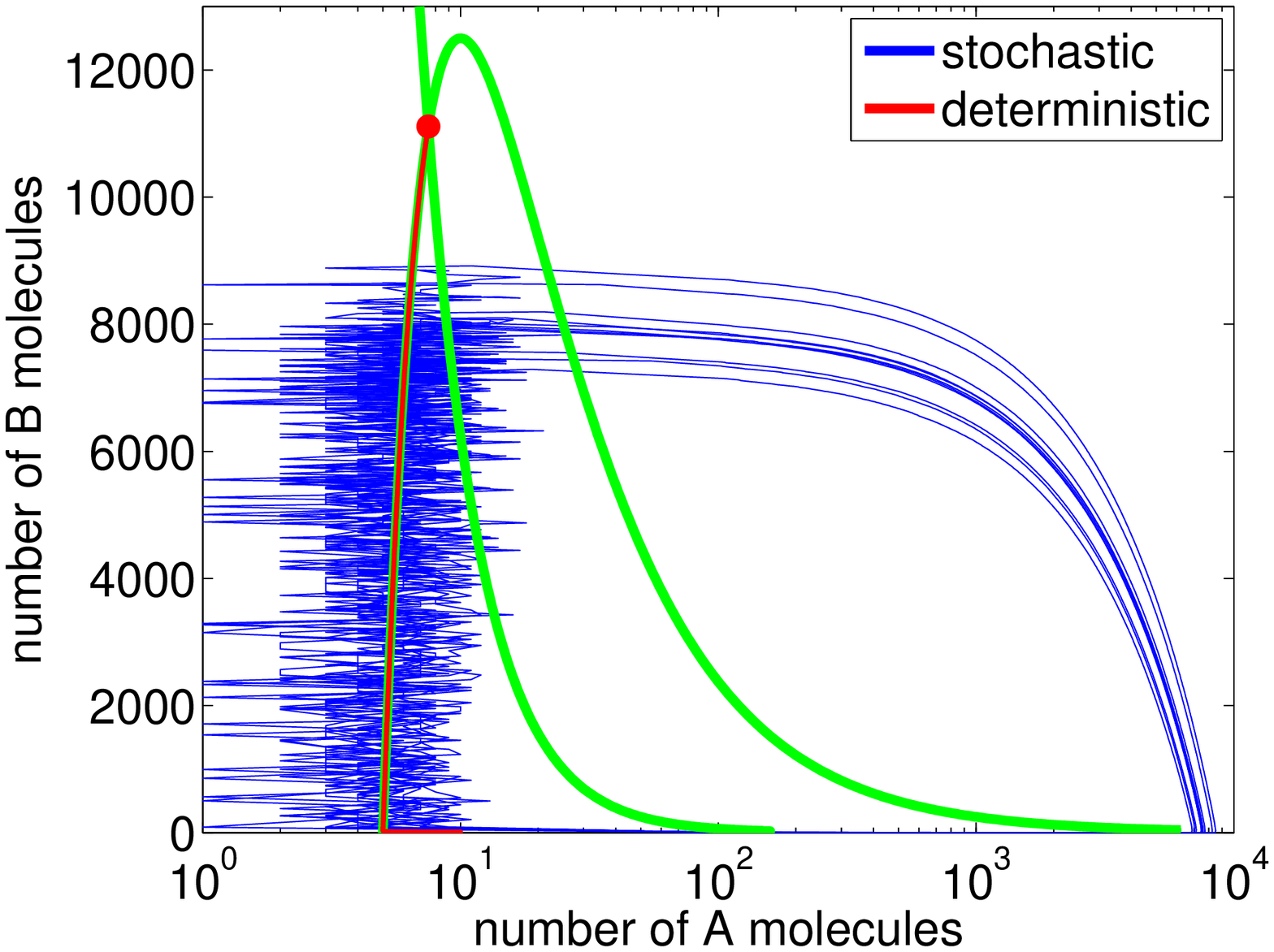}{2in}{5mm}
\caption{Self-induced stochastic resonance.
{\rm (a)} One realization of SSA (a5)--(d5) for 
the system of chemical reactions $(\ref{schnak})$ (blue line) 
and solution of the deterministic ODEs (red line). 
{\rm (b)} Comparison of the stochastic 
and deterministic trajectories in the $(A,B)$-plane.
Nullclines of the deterministic ODEs are plotted as green lines.}
\label{figstores}
\end{figure}
The presented phenomenon is sometimes called 
self-induced stochastic resonance \cite{Muratov:2005:SSR}.
Following an example from \cite{DeVille:2006:NDL},
we consider the system of chemical reactions introduced
by Schnakenberg \cite{Schnakenberg:1979:SCR}
\begin{equation}
\mbox{ \raise 1mm \hbox{%
$2 A + B \;\displaystyle\mathop{\displaystyle\longrightarrow}^{k_1}\; 3 A$,}}
\qquad\qquad
{\mbox{ \raise 0.851 mm \hbox{$\emptyset$}}}
\;
\mathop{\stackrel{\displaystyle\longrightarrow}\longleftarrow}^{k_2}_{k_3}
\;
{\mbox{\raise 0.851 mm\hbox{$A$,}}}
\qquad\qquad
\mbox{ \raise 1mm \hbox{%
 $\emptyset \;\displaystyle\mathop{\displaystyle\longrightarrow}^{k_4}\; B$,}}
\label{schnak}
\end{equation}
where we choose the rate constants as 
$k_1=4{\times}10^{-5} \; \mbox{sec}^{-1}$,
$k_2=50 \; \mbox{sec}^{-1}$, 
$k_3=10 \; \mbox{sec}^{-1}$ 
and 
$k_4 = 25 \; \mbox{sec}^{-1}$. 
We use the Gillespie SSA
(a5)--(d5) to simulate the time evolution of this system. 
To do that, let us note that the propensity function of
the first reaction is equal to $A(t) (A(t)-1) B(t) k_1$.
We also derive and solve the deterministic system of ODEs
corresponding to (\ref{schnak}). 
Using the same initial conditions $[A,B] = [10,10]$, we compare 
the results of the stochastic and deterministic models 
in Figure \ref{figstores}(a).
We plot the time evolution of $A(t)$. We see that the solution of the 
deterministic equations converges to a steady state while the stochastic 
model has oscillatory solutions. Note that there is a log scale on 
the $A$-axis
-- numbers of A given by the (more precise) SSA vary between zero 
and ten thousand. If we use a linear scale on the $A$-axis, then the
low molecular fluctuations would be invisible and the solution of the SSAs
would look as if there were ``almost deterministic oscillations", 
although it is the intrinsic noise which makes the oscillations possible. 
To understand this behaviour better, we plot the stochastic and deterministic
trajectories in the $(A,B)$-plane in Figure \ref{figstores}(b). We include
the nullclines of the deterministic system of ODEs (green lines). We see that
the deterministic system follows a stable nullcline into the steady
state (red circle). The stochastic model also initially ``follows" 
this nullcline
(with some noise) but it is the intrinsic noise which makes it
possible for the stochastic model to leave the stable nullcline
and oscillate (again we use a log scale on the $A$-axis).

\subsection{Biological pattern formation}

\label{secpattern}

\begin{figure}
\picturesAB{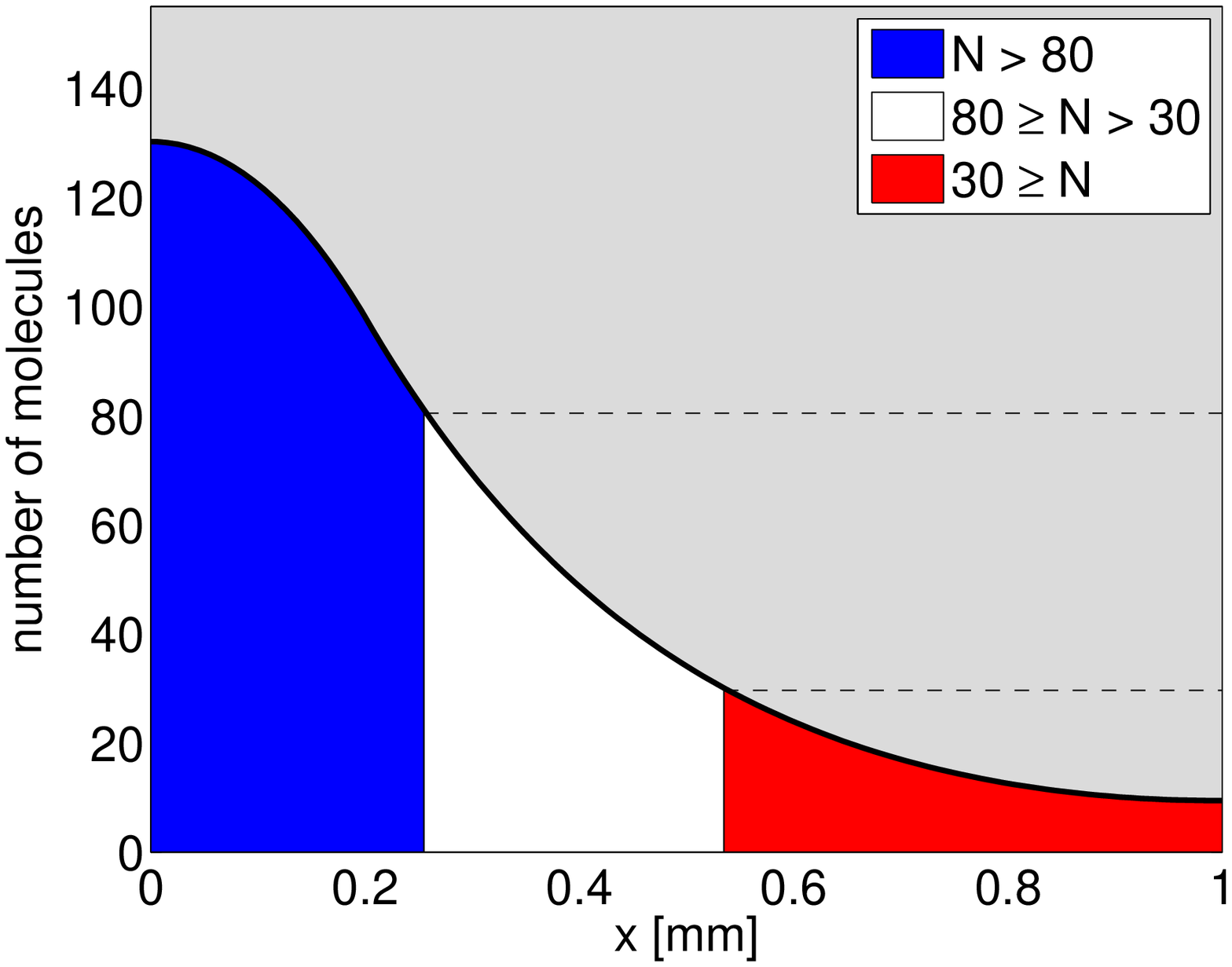}{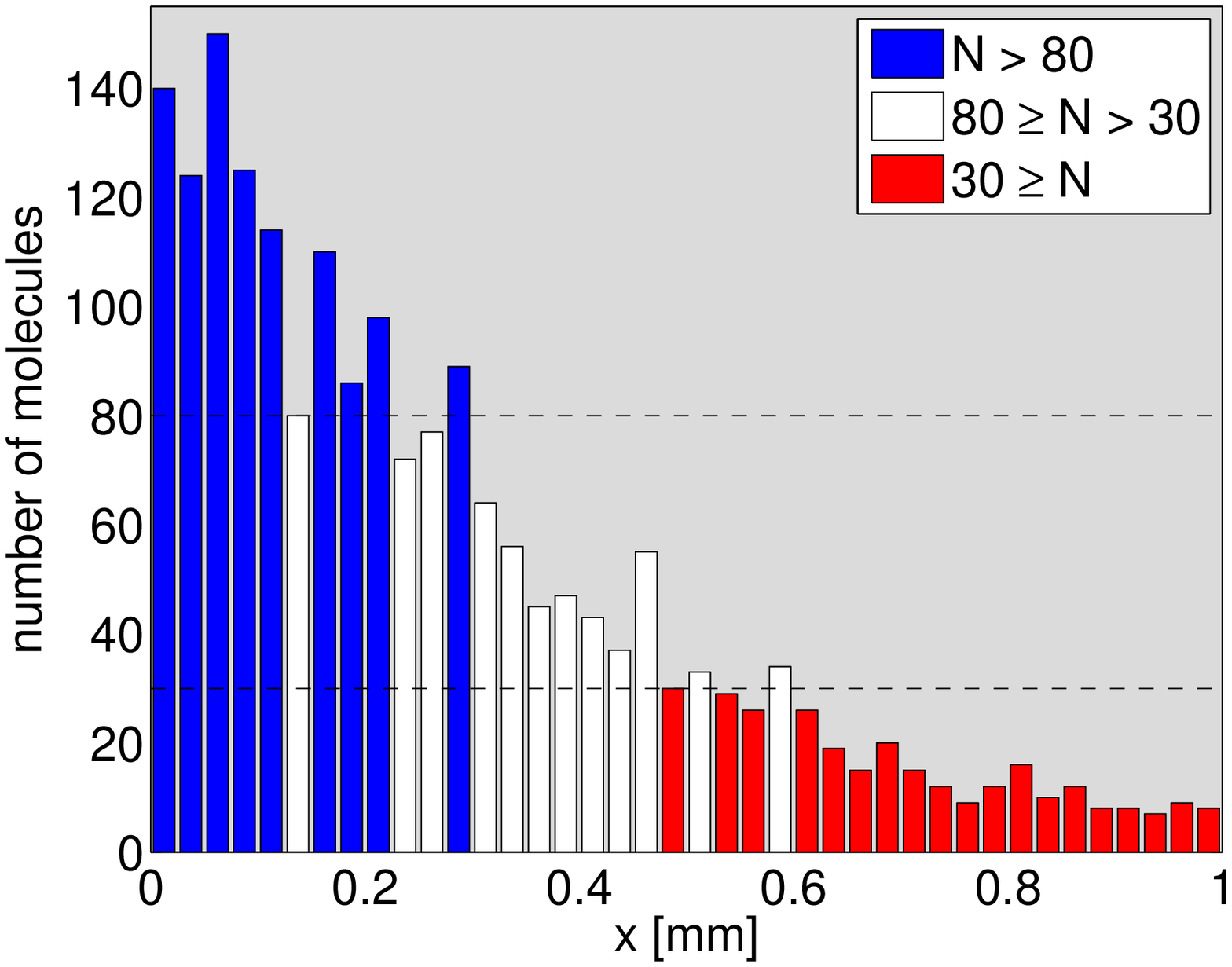}{2in}{5mm}
\caption{French flag problem. {\rm (a)} Deterministic model. {\rm (b)} 
Stochastic model.}
\label{figfrenchflag}
\end{figure}

Reaction-diffusion processes are key elements of 
pattern forming mechanisms in developmental biology. 
The illustrative example from Sections \ref{seccompartmentRD} 
and \ref{secRDsmol} was a caricature of more complicated 
morphogenesis applications \cite{Shimmi:2005:FTD,Reeves:2005:CAE}
where one assumes that some prepatterning in the domain 
exists and one wants to validate the reaction-diffusion
mechanism of the next stage of the patterning of the embryo.
In our example, we considered a chemical $A$ which is produced
in part $[0,L/5]$ of domain $[0,L]$. Hence, the domain $[0,L]$
was divided into two different regions (prepatterning) $[0,L/5]$ 
and $[L/5,L]$. The simplest idea of further
patterning is the so-called French flag problem 
\cite{Wolpert:2002:PD}. We assume that the interval $[0,L]$
describes a layer of cells which are sensitive to the
concentration of chemical $A$. Let us assume that a
cell can have three different fates (e.g. different genes
are switched on or off) depending on the concentration of
chemical $A$. Then the concentration gradient of $A$
can help to distinguish three different regions
in $[0,L]$ -- see Figure \ref{figfrenchflag}.
If the concentration of $A$ is high enough (above a certain threshold),
a cell follows the first possible program (denoted blue
in Figure \ref{figfrenchflag}). The ``white program" (resp. ``red
program") is followed for medium (resp. low) concentrations
of $A$. The deterministic version of the French flag problem
is presented in Figure \ref{figfrenchflag}(a). We consider
a solution of  $(\ref{rdPrDeDi})$--$(\ref{boPrDeDi})$
at time $30$ minutes which is the red curve in Figure
\ref{figrdgill}(b) or Figure \ref{figrdsmol}(b). 
The solution of $(\ref{rdPrDeDi})$--$(\ref{boPrDeDi})$
is decreasing in space. Introducing two 
thresholds, we can clearly obtain three well-defined regions
as seen in Figure \ref{figfrenchflag}(a).
The stochastic version of the French flag problem
is presented in Figure \ref{figfrenchflag}(b).
We take the spatial histogram presented in 
Figure \ref{figrdsmol}(b). We introduce two thresholds
(80 and 30 molecules) as before and replot the histogram 
using the corresponding colours. Clearly, the resulting
``French flag" is noisy. Different realizations of the SSA
would lead to different noisy French flags.
The same is true for the SSA from Figure \ref{figrdgill}(b).
 
\begin{figure}
\picturesAB{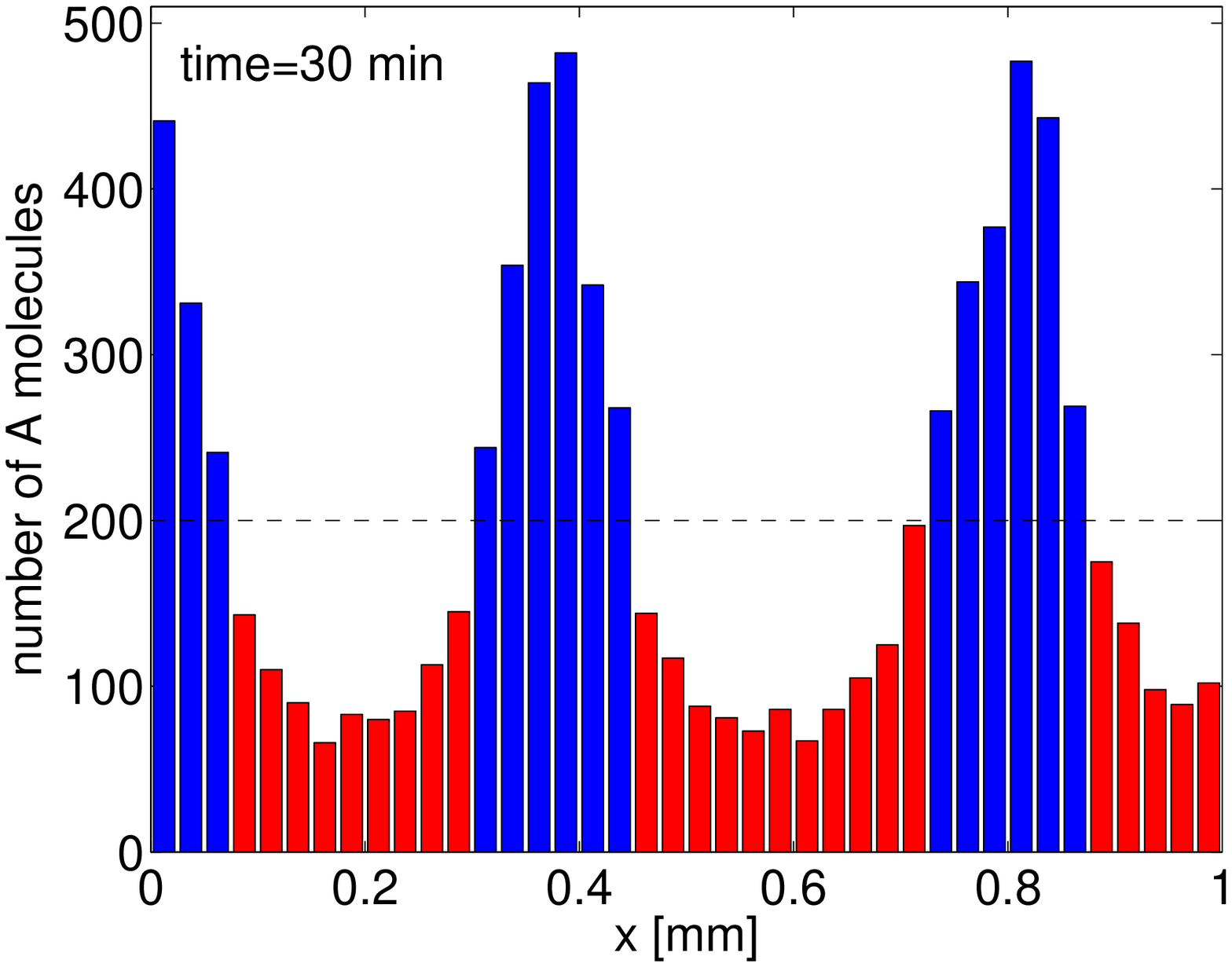}{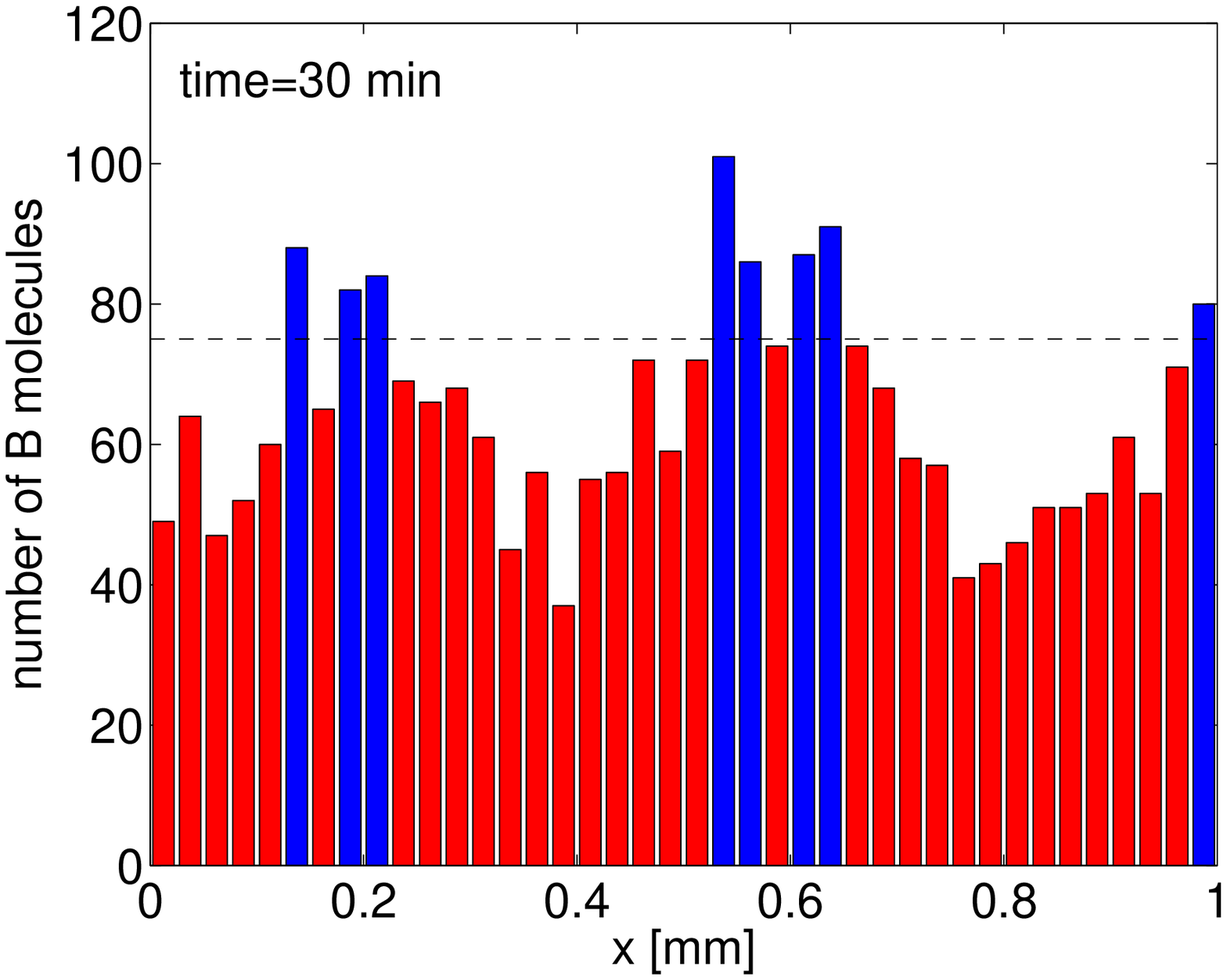}{2in}{5mm}
\caption{Turing patterns. {\rm (a)}
Numbers of molecules of chemical species $A$ in 
each compartment at time 30 minutes;
{\rm (b)} the same plot for chemical species $B$.}
\label{figturing}
\end{figure}

Our second example of patterning in developmental
biology are the so-called Turing patterns 
\cite{Turing:1952:CBM,Gierer:1972:TBP,Murray:2002:MB,Sick:2006:WDD}. 
They do not require any prepatterning. Molecules
are subject to the same chemical reactions 
in the whole domain of interest. For example, let
us consider a system of two chemical species $A$ and
$B$ which react according to the Schnakenberg system 
of chemical reactions (\ref{schnak}). Let us choose 
the values of rate constants as 
$k_1 = 10^{-6} \; \mbox{sec}^{-1}$,
$k_2 = 1 \; \mbox{sec}^{-1}$,
$k_3 = 0.02 \; \mbox{sec}^{-1}$
and $k_4 = 3 \; \mbox{sec}^{-1}$. The corresponding 
deterministic system 
of ODEs for (\ref{schnak}) has one nonnegative
stable steady state equal to $a_s=200$ and $b_s=75$
molecules. Introducing diffusion to the model,
one steady state solution of the spatial problem 
is the constant one ($a_s$, $b_s$) everywhere. 
However, such a solution might not be stable (i.e.
might not exist in reality) if the diffusion constants
of $A$ and $B$ differ significantly. We choose 
$D_A=10^{-5} \; \mbox{mm}^2 \, \mbox{sec}^{-1}$
and $D_B=10^{-3} \; \mbox{mm}^2 \, \mbox{sec}^{-1}$,
i.e. $D_B/D_A = 100.$ To simulate the reaction-diffusion 
problem with the Schnakenberg system of
chemical reactions (\ref{schnak}), we follow the 
method of Section \ref{seccompartmentRD}. We divide 
the computational domain $[0,L]$ into $K=40$ compartments 
of length $h = L/K=25 \,\mu$m. We denote the number of molecules 
of chemical species $A$ (resp. $B$) in the $i$-th compartment 
$[(i-1)h,ih)$ by $A_i$ (resp. $B_i$), $i=1,\dots,K$.
Diffusion is described by two chains of chemical
reactions (\ref{diffGillRDnonlinA})--(\ref{diffGillRDnonlinB})
where the rates of ``chemical reactions" are equal
to $d_A = D_A/h^2$ for chemical species $A$ and
$d_B = D_B/h^2$ for chemical species $B$. Chemical 
reactions (\ref{schnak}) are considered in every compartment
(the values of rate constants in (\ref{schnak}) are already 
assumed to be expressed in units per compartment).
Starting with a uniform distribution of chemicals 
$A_i(0)=a_s=200$ and $B_i(0)=b_s=75$, $i=1,2, \dots, K$, 
at time $t=0$, we
plot the numbers of molecules in each compartment at time $t=30$ 
minutes computed by SSA (a5)--(d5) in Figure \ref{figturing}.
To demonstrate the idea of patterning, compartments
with many molecules (above steady state 
values $a_s$ or $b_s$) are plotted as blue; 
other compartments are plotted as red. We see in Figure
\ref{figturing}(a) that chemical $A$ can be clearly used to
divide our computational domain into several regions.
There are two and half blue peaks in this figure. 
The number of blue peaks depends on the size of the
computational domain $[0,L]$ and it is not a unique number
in general. The reaction-diffusion system has several favourable 
states with a different number of blue peaks. As discussed 
in Section \ref{secssr}, the solution 
of the corresponding deterministic model converges to one 
of the favourable (stable steady) states of the system. 
The stochastic model 
enables stochastic switching between the favourable states,
i.e. between the states with a different number of blue
peaks.

\section{Discussion}

\label{secdiscussion}

We presented SSAs for systems of chemical reactions and
molecular diffusion. Then we presented methods for 
simulating both reactions and diffusion at the same time.
The algorithms for simulating (spatially homogeneous)
systems of chemical reactions were based on the work
of Gillespie \cite{Gillespie:1977:ESS}. We did not focus
on the computer implementation of the algorithms. We chose
simple examples which can be simulated quickly. If one
considers systems of many equations, there are ways
to make the Gillespie SSA more efficient \cite{Gibson:2000:EES}. 
For example, it would be a waste of time to recompute all 
the propensity functions at each time step. We simulate one 
reaction per one 
time step. Therefore, it makes sense to update only
those propensity functions which are changed by the 
chemical reaction which was selected in step (d5) of
SSA (a5)--(d5).

We only briefly touched on the concept of the Fokker-Planck
equation \cite{Risken:1989:FPE} when we discussed the 
Smoluchowski description
of diffusion. It is worth noting that there are
interesting connections between the chemical master
equation (which is equivalent to the Gillespie SSA)
and the Fokker-Planck equation which gives the time
evolution of the probability distribution.
Such connections are discussed (through the 
so-called chemical Langevin equation) in \cite{Gillespie:2000:CLE}.
The Smoluchowski equation is actually the same
mathematical object as the chemical Langevin equation,
i.e. the stochastic differential equation \cite{Arnold:1974:SDE}.
An algorithmic introduction to stochastic differential equations
can be found in \cite{Higham:2001:AIN}.

We presented two models of diffusion in this paper. One was based 
on the chain of ``chemical reactions" (\ref{diffGill}) computing 
the time evolution of the numbers of molecules in compartments. 
Coupling this model with the modelling of chemical reactions is 
straightforward and presented in Section \ref{seccompartmentRD};
such a compartment-based approach is used e.g. 
in \cite{Stundzia:1996:SSC,Isaacson:2006:IDC,Hattne:2005:SRD}. 
The second model for molecular diffusion was based on the
Smoluchowski equation (\ref{x5equation}). It was an example
of the so-called position jump process, that is,
a molecule jumps to a different location at each time step.
As a result, the trajectory of a molecule is discontinuous.
The individual trajectories of diffusing molecules
can be also modelled using the so-called velocity
jump processes \cite{Othmer:1988:MDB}, that is, 
the position of a molecule $x(t)$ follows the deterministic 
equation $\mbox{d}x/\dt=v$ where $v(t)$, the velocity of 
the molecule, changes stochastically. Such 
stochastic processes can be used not only for the
simulation of diffusing molecules but also for 
the description 
of movement of unicellular organisms like bacteria
\cite{Erban:2004:ICB,Erban:2005:STS} or amoeboid
cells \cite{Erban:2007:TAE}. Velocity jump processes
can be also described in terms of PDEs for the
time evolution of the probability distributions to find 
a particle (molecule or cell) at a given place.
Such equations are not exactly equal to the diffusion
equation. However, they can be reduced in the appropriate
limit to the diffusion equation 
\cite{Zauderer:1983:PDE,Erban:2007:RBC}. A classical
review paper on diffusion and other stochastic
processes was written by Chandrasekhar
\cite{Chandrasekhar:1943:SPP}, a nice introduction to
random walks in biology is the book by Berg \cite{Berg:1983:RWB}.

In this paper, we used only reflective boundary conditions,
that is, particles hitting the boundary were
reflected back. Such boundary conditions are suitable 
whenever there is no chemical interaction between molecules 
in the solution and the boundary of the domain. Considering 
biological applications, it is often the case that molecules
(e.g. proteins) react with the boundary (e.g. with receptors
in the cellular membrane). Then the boundary conditions
have to be modified accordingly. It has to be assumed
that some molecules which hit the boundary 
are reflected and some molecules are adsorbed by 
the boundary (e.g. become bound to the receptor or take 
part in membrane-based chemical reactions). 
The probability that a molecule is adsorbed rather than
reflected depends on the chemical properties of the
boundary and also on the SSA which is used for modelling
(further details are given in  
\cite{Erban:2007:RBC,Erban:2007:TSR}).

Our analysis of SSAs was based on the chemical master
equation. We successfully derived equations for
the means and variances in illustrative examples 
which did not include second-order reactions.
Other first-order reaction networks can be also 
analysed using this framework \cite{Gadgil:2005:SAF}.
The nonlinear chemical kinetics complicates the mathematical
analysis significantly. We can write a deterministic
description but it might be too far from the correct
description of the system \cite{Samoilov:2006:DEM}.
A review of more computational approaches for the analysis
of SSAs can be found in \cite{Kevrekidis:2003:EFM}.
Applications of such methods to stochastic reaction-diffusion
processes is presented in \cite{Qiao:2006:SDS}.

\section*{Acknowledgements}
This work was supported by the Biotechnology and Biological
Sciences Research Council (grant ref. BB/C508618/1), 
St. John's College, Oxford and Linacre College, Oxford (RE).
Authors would like to give thanks for helpful suggestions and 
encouraging comments during the preparation of this manuscript
to Ruth Baker, Hyung Ju Hwang, Chang Hyeong Lee, Hans Othmer, 
Jan Rychtar and Aidan Twomey.
 
\bibliographystyle{amsplain}
\bibliography{bibrad}

\end{document}